\documentclass[onefignum,onetabnum]{arxiv_style}
\usepackage[margin=1.2in]{geometry}

\usepackage{tikz}
\usetikzlibrary{arrows.meta}
\usepackage[version=4]{mhchem}

\ExplSyntaxOn
\keys_define:nn { mhchem }
 {
  arrow-min-length .code:n =
   \cs_set:Npn \__mhchem_arrow_options_minLength:n { {#1} } 
 }
\ExplSyntaxOff

\mhchemoptions{arrow-min-length=1.1em}
\usepackage{xfrac}
\usepackage{textcomp}

\renewcommand{\textrightarrow}{$\to$}

\usepackage{bbm}
\usepackage[export]{adjustbox}

\newcommand{\ssMHE}{sMHE}

\newcommand{\lM}{{M^\ell_i}}
\newcommand{\lC}{{C^\ell_i}}
\newcommand{\lT}{{T^\ell_i}}

\newcommand{\lN}{\mathcal N^\ell_i}

\usepackage{amssymb}
\usepackage{amsfonts}
\usepackage{amsmath}
\usepackage{enumitem}
\usepackage{tikz-cd}
\newcommand{\bigslant}[2]{{\raisebox{.2em}{$#1$}\left/\raisebox{-.2em}{$#2$}\right.}}
\usepackage{caption}
\usepackage{subcaption}

\newcommand{\R}{\ensuremath{\mathbb{R}}}
\newcommand{\Z}{\ensuremath{\mathbb{Z}}}
\newcommand{\N}{\ensuremath{\mathbb{N}}}
\DeclareMathOperator{\Image}{Im}
\DeclareMathOperator{\Ker}{Ker}

\usepackage{algorithm,algcompatible,amsmath}
\algnewcommand\INPUT{\item[\textbf{Input:}]}%
\algnewcommand\OUTPUT{\item[\textbf{Output:}]}%

\usepackage{multirow}

\usepackage{xpatch}

\usepackage[intoc]{nomencl}
\makeatletter
\patchcmd{\thenomenclature}{\section*}{\section}{}{}
\makeatother

\makenomenclature
\RequirePackage{ifthen}

\renewcommand{\nompreamble}{We provide a list of the symbols used in the Main Text and Supplementary Information}
\usepackage{etoolbox}
\renewcommand\nomgroup[1]{%
  \item[\bfseries
  \ifstrequal{#1}{M}{\ssMHE{} model}{%
  \ifstrequal{#1}{T}{Topological Data Analysis}{%
  \ifstrequal{#1}{G}{General}{}}}%
]}

\usepackage{placeins}

\headers{Zigzag persistence for coral reef resilience using a stochastic spatial model}{R.A.~McDonald, R.~Neuhausler, M.~Robinson,  L.G.~Larsen, H.A.~Harrington,  M. Bruna}

\title{Zigzag persistence for coral reef resilience using a stochastic spatial model}
\author{
Robert A. McDonald\footnotemark[1]\ \footnotemark[2]
\and
Rosanna Neuhausler\footnotemark[1]\ \footnotemark[3]
\and
Martin Robinson\footnotemark[4]
\and
Laurel G. Larsen\footnotemark[3]
\and
Heather A. Harrington\footnotemark[2]\ \footnotemark[5]
\and
Maria Bruna\footnotemark[6]
}

\begin{document}
\maketitle

\renewcommand{\thefootnote}{\fnsymbol{footnote}}

\footnotetext[1]{These authors contributed equally.}
\footnotetext[2]{Mathematical Institute, University of Oxford, Oxford, OX2 6GG, UK.}
\footnotetext[3]{Department of Geography, University of California, Berkeley, CA, 94720 USA.}
\footnotetext[4]{Computer Science Department, University of Oxford, Oxford OX1 3QG, UK.}
\footnotetext[5]{Wellcome Centre for Human Genetics, University of Oxford, Oxford OX3 7BN, UK (\email{harrington@maths.ox.ac.uk}).}
\footnotetext[6]{Department of Applied Mathematics and Theoretical Physics, Centre for Mathematical Sciences, University of Cambridge, Cambridge CB3 0WA, UK (\email{bruna@maths.cam.ac.uk}).}

\renewcommand{\thefootnote}{\arabic{footnote}}

\begin{abstract}
A complex interplay between species governs the evolution of spatial patterns in ecology.
An open problem in the biological sciences is characterising spatio-temporal data and understanding how changes at the local scale affect global dynamics/behaviour. 
Here, we extend a well-studied temporal mathematical model of coral reef dynamics to include stochastic and spatial interactions and generate data to study different ecological scenarios. 
We present descriptors to characterise patterns in heterogeneous spatio-temporal data surpassing spatially averaged measures.
We apply these descriptors to simulated coral data and demonstrate the utility of two topological data analysis techniques--persistent homology and zigzag persistence--for characterising mechanisms of reef resilience. 
We show that the introduction of local competition between species leads to the appearance of coral clusters in the reef. 
We use our analyses to distinguish temporal dynamics stemming from different initial configurations of coral, showing that the neighbourhood composition of coral sites determines their long-term survival. 
Using zigzag persistence, we determine which spatial configurations protect coral from extinction in different environments. 
Finally, we apply this toolkit of multi-scale methods to empirical coral reef data, which distinguish spatio-temporal reef dynamics in different locations, and demonstrate the applicability to a range of datasets.
\end{abstract}



\section{Introduction}

Spatial patterns arise in many natural systems, from systems of chemical species or morphogens \cite{Gray:1984cn}, cells in developing embryos \cite{pourquie2003segmentation,gregor2005diffusion}, skin patterns on fish and mammals \cite{Volkening:2015dj,murray2001mathematical}, and coral colonies in coral reefs \cite{adjeroud1997factors}. 
Alan Turing explained the mechanisms behind the spatial patterns observed in morphogenesis--the interplay of diffusion and reactions \cite{Turing:1952ja}. Recent work has shed light on the importance of early segregation, or spatial patterning in embryos for successful development \cite{warmflash2014method}. A common denominator of such systems is their complexity: they are dynamic, involve large numbers of particles or agents (e.g., molecules, cells, animals), and are inherently noisy. To elucidate the role of spatial patterns in such systems' function and spatial evolution requires quantitative tools that can cope with such complexity. In recent years, the area of topological data analysis (TDA) has blossomed to offer multiple promising methods \cite{hoef2022primer}. TDA can provide multiscale summaries of complex data.
Here we dive into the mechanisms of spatial patterning with TDA and other topological descriptors, with shallow-water coral reefs as a case study.

\begin{figure*}[t!]
\centering
\includegraphics[width=\linewidth]{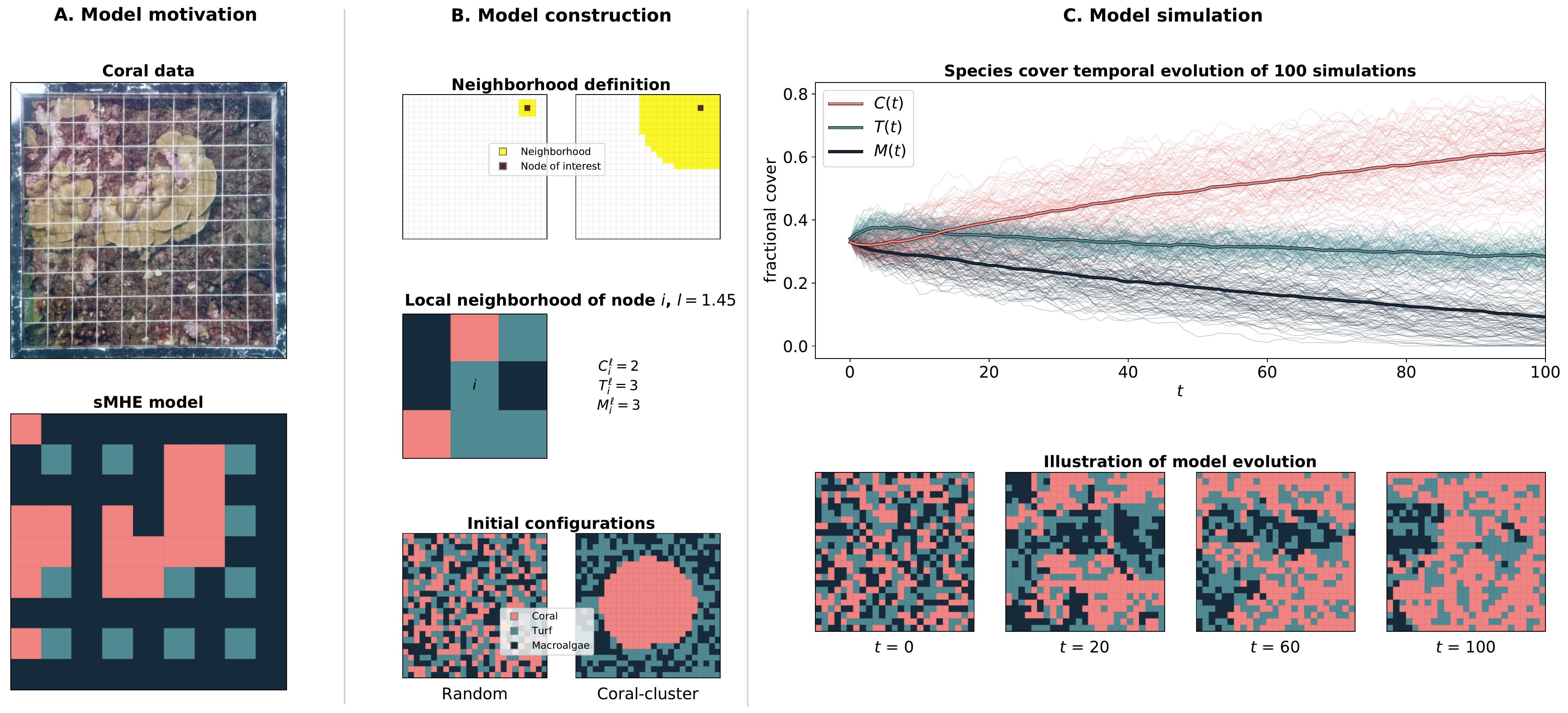}
\caption{
\textbf {Stochastic spatial \ssMHE{} model of a coral reef.} 
\textbf{A.} Model motivation. Quadrat photograph of 1m$^2$ taken on the coral reef in Nuku Hiva (Marquesas Islands, French Polynesia) in 2014, kindly provided by the Service National d'Observation CORAIL from CRIOBE.  A $10\times10$ grid mesh is placed over a section of reef, photographed, and the species at each grid intersection, i.e., every $10$cm, is noted. Motivated by the image, the \ssMHE{} model represents the most prevalent species within each sub-square. 
\textbf{B.} Model construction.  
The local neighbourhood of radius $\ell$ determines a node's area of influence in the evolution of the model. We show the neighbourhoods of influence corresponding to $\ell = 1.45$ and $\ell = 11$. 
The local neighbourhood of node $i$ (occupied by turf in the example plotted) has $C_i^\ell = 2$, $T_i^\ell =3$, and $M_i^\ell = 3$.
Given initial covers $C(0), T(0), M(0)$, the \ssMHE{} model is initialised either with a \emph{random} configuration, with a uniform distribution representing a spatially mixed reef (any node is initialised as coral, turf or macroalgae, with probability $C(0), T(0), M(0)$ respectively), or with a \emph{coral-cluster} configuration, where the coral is placed in a connected patch at the centre of the domain, with the other two species uniformly distributed in the remaining space.
\textbf{C.} Model simulation. Temporal evolution of the model with initial random configuration, $\ell$ = 1.45, grazing $g=.58$, and $C(0) = T(0) = M(0) = .33$, averaged over 100 realisations. Spatial snapshots of the \ssMHE{} model are shown at times $t=0$, $t = 20$, $t = 60$, and $t=100$ for one realisation.}
\label{fig:fig1}
\end{figure*}

Coral reefs provide a tremendous range of ecosystem services, including biodiversity, fishing, and tourism \cite{moberg1999ecological, bellwood2004confronting}. Due to the complexity and stability of their calcium carbonate structure, specifically in shallow waters, coral reefs supply the optimal foundation for various photosynthetic benthic organisms to settle and grow upon. As such, there is relentless competition for space on the reef. Under human or natural disturbances--which are becoming ever more common with climate change and coastal development--coral reefs have been observed to shift from coral- to algae-dominated states \cite{hughes1994catastrophes}. Many mechanistic models have been proposed for hypothesis-testing about mechanisms that drive spatial patterning and long-term resilience \cite{patternformationecosystems,gorospe2013genetic,fricke1986diversity,bradbury1983coral,maguire1977spatial,mumby2005patch}. In particular, Mumby, Hastings, and Edwards (MHE) developed a simple three-species temporal model for Caribbean coral reefs accounting for corals and two types of algae, algal turfs and macroalgae \cite{mumby2007thresholds}. The MHE model demonstrated that there are coral- and algae-dominated alternative states and established critical thresholds of fish grazing and coral cover delineating the resilience of each state. 

A natural question is how coral resilience is affected by the spatial distribution of coral and other species within the reef. To address this question, we build on the MHE model \cite{mumby2007thresholds} to develop a stochastic and spatial lattice-based model (\ssMHE{}) of a coral reef. Stochasticity is essential to model the unpredictable disturbances that affect coral reefs and enables transitions between the alternative stable states identified by the MHE model. 
Spatial models, both deterministic (based on partial differential equations) and stochastic (e.g. agent-based and cellular automata models), have been used to study pattern formation \cite{murray2001mathematical,Volkening:2015dj,Pearson:1993wk}. Due to stochasticity, multiple realisations of the \ssMHE{} model are required to produce statistical results and develop insight into the reef-level dynamics while retaining the spatial information. To this end, we use topological descriptors suited to averaging over realisations. We first consider neighbourhood descriptors that quantify the clustering of coral throughout the reef and then appeal to TDA.

TDA is a branch of computational mathematics that summarises the shape of data through topological invariants \cite{carlsson2009topology,edelsbrunner2022computational}.
Persistent homology (PH), a prominent tool in TDA, takes in data and outputs a multiscale topological summary of features, such as connected components, loops, and cavities \cite{ghrist}.
Depending on the type of data being studied, PH offers a flexible suite of methodologies which may be adapted to address the research question at hand \cite{roadmap}. 
PH has previously generated insight into many biological applications \cite{fishPHpaper, stolz2022multiscale, taylor2015topological}.
While competition for space is known to affect reef dynamics \cite{mumby2005patch}, much conventional analysis of reef data focuses on the prevalence of different species (measured through percentage cover).
We use TDA to analyse simulated and empirical data, due to its ability to capture spatial features and patterning not detected by standard analyses. 
To perform statistics, classification, comparison, and averages of topological features, vectorisation methods of PH have been developed \cite{persland,wasserman2018topological,adams2017persistence}.
Such statistical techniques allow the computation of robust spatial properties of complex, possibly noisy coral reef data.

We compute PH of coral data extracted from photographs of underwater reefs and from snapshots of the \ssMHE{} model.
However, standard PH is limited to studying static, non-dynamical data. 
Advances in TDA have enabled the analysis of data that evolves non-monotonically over time \cite{crockerplots, vineyards, formigrams, kimmemoliDMS}, including the generalisation of PH to zigzag persistence \cite{originalzigzag, zigzagapps,zigzagstability}.
As with standard PH, zigzag persistence detects topological features in data such as components, loops and cavities. However, zigzag persistence allows such features to be tracked over multiple time-snapshots, which is not possible with standard PH.
We therefore use zigzag persistence to analyse how the spatial composition of reefs changes over time. Vectorisation methods created for PH are also applicable to zigzag persistence, and we use these to describe average spatio-temporal properties of coral under different conditions.

Dynamic TDA methods have been previously applied to analysing aggregation models, fish swarms, and temporal networks \cite{fishswarm,crockerplots,myers2022temporal}. However, zigzag persistence of dynamic data for larger systems was computationally out of reach until recently \cite{bradzzalgorithm}. 
Here, we show that standard and zigzag PH provide complementary information on species competition in our model and zigzag PH reveals pathways to coral decline.

\begin{figure*}[h!]
\centering
\includegraphics[width=\linewidth]{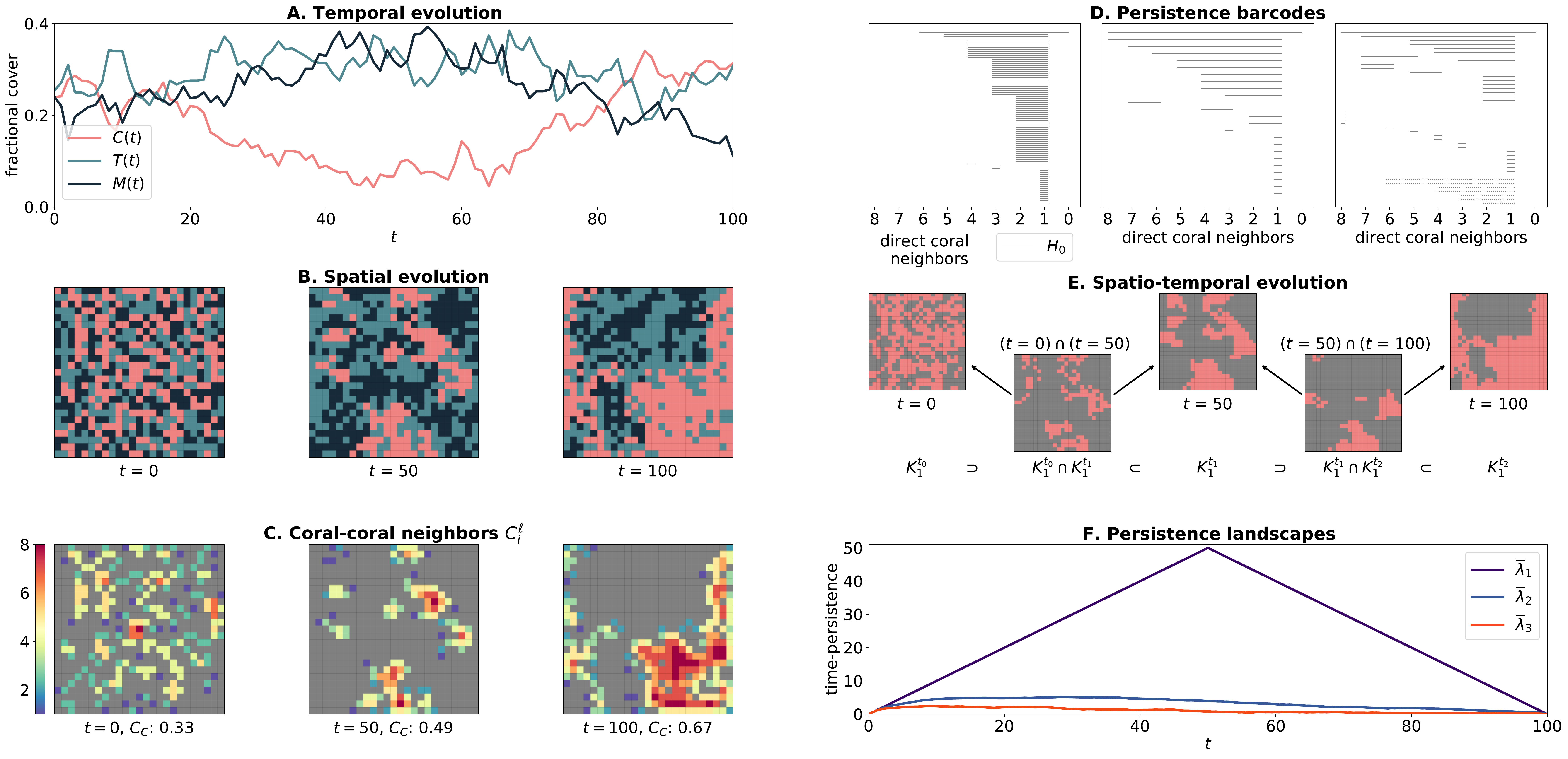}
\caption{\textbf {Data analysis of the \ssMHE{} model.} 
\textbf{A.} Temporal evolution of fractional covers $C(t)$, $T(t)$, $M(t)$  over 100 timesteps. 
\textbf{B.} Snapshots of the spatial evolution of the \ssMHE{} model at three different times. Colours represent the three species as in Fig.~\ref{fig:fig1}. 
\textbf{C.}  Coral-coral neighbours: heatmaps of $C_i^\ell$ for each node $i$ corresponding to the data in \textbf{B}. 
\textbf{D.} Persistence barcodes of the three snapshots in \textbf{B}. Solid bars represent clusters; dotted bars represent enclosed loops. The length of the bars represents the sizes of each. The filtration parameter is the number of direct coral neighbours. 
\textbf{E.} Spatio-temporal evolution: illustration of the zigzag sequence from a single simulation of the \ssMHE{} model. We show the three time snapshots in \textbf{B} (after pre-processing) and their intersections. 
\textbf{F.} Average persistence landscapes $\bar \lambda_1$, $\bar \lambda_2$, $\bar \lambda_3$, describing the three most time-persistent features in an average simulation of the \ssMHE{} model. The first landscape shows there is a single component that lasts throughout all times in all simulations. The second and third landscapes show that, on average, two other components appear early in the simulation and last for a short time.}
\label{fig:fig2}
\end{figure*}

\section{Stochastic spatial model and data generation}

The coral ecosystem is driven through complex interactions between its components and environmental variables, which arise from competition for space and resources \cite{mumby2005patch} (see Fig.~\ref{fig:fig1}A, top). In \cite{mumby2007thresholds}, Mumby, Edwards, and Hastings proposed the following model (MHE model) to describe such interactions in a simple three-component ODE system:
\begin{subequations} \label{mumby}
\begin{align} 
\dot{C} &= rCT - dC - aCM, 
\label{mumbyA}
\\ 
\dot{M} &= aCM - \frac{gM}{M+T}
+ \gamma MT.
\label{mumbyB}
\end{align}
\end{subequations}
The MHE model describes the temporal evolution of the fraction of a reef covered by either coral ($C$), macroalgae ($M$), or algal turf ($T$), where $T = 1- C - M$ (assuming the seabed area is fully covered). Previous work has extended the MHE model to account for the effect of ocean acidification \cite{MHEoceanacidification}, natural disasters \cite{MHEhurricanes} and more complex fishing dynamics \cite{MHEparrotfish1, MHEparrotfish2}. Other models consider multiple competing coral and macroalgae types \cite{eLifemodel} with complex intra-species dynamics \cite{multimodel}. We use the original MHE model as our starting point due to its simplicity and well-understood dynamics.

In the MHE model (\eqref{mumby}), competition between species is represented by the nonlinear terms, which are agnostic of spatial location. One possible modelling framework which would represent the spatial dependence of coral growth would be a system of partial differential equations (PDE) describing $C$, $M$ and $T$ as functions of time and space. However, to account for the complexity of the environment, such inter-species interaction should be stochastic, which would not be reflected in a deterministic PDE. Furthermore, we wish to construct a model which simulates data of the same form as photographed coral reefs overlayed with a grid mesh as in Fig.~\ref{fig:fig1}A.

In the stochastic spatial MHE model (\ssMHE{} model), we make inter-species competition location-dependent by considering a square-grid discretisation of the seabed domain with $25\times 25$ nodes, where each node $i$ is spatially embedded in 2D and occupied by either $C_i, T_i, M_i \in \{0,1\}$ (see Fig.~\ref{fig:fig1}B, bottom). The number of nodes (625) was chosen to balance simulation time (which increases exponentially with the number of nodes) with the effectiveness of TDA computations (which distinguish more spatial behaviour at finer resolutions). We introduce a local neighbourhood of radius $\ell$ within which interactions can occur (see Fig.~\ref{fig:fig1}B, top). 
For example, the first term in \eqref{mumbyA} describes the interaction between $C$ and $T$, namely the recruitment of coral through the overgrowth of turf at a rate $r$. In the \ssMHE{}, this interaction can only occur if there is a node $i$ with turf (for coral to overgrow) within a radius $\ell$ of a node $j$ with coral, that is, $T_i = 1, C_j = 1$ and $|i-j|\le \ell$. The interaction is then represented as the ``reaction'' $T_i + C_j \to C_i + C_j$ with rate $r$ (meaning that the transition $T_i\to C_i$ occurs with probability $r\Delta t$ within a time interval $\Delta t$). The full set of reactions of the \ssMHE{} is
\begin{subequations} \label{stoch_spatial_model}
\begin{align}
& T_i + C_j \xrightarrow{\makebox[1.75mm]{$r$}} C_i + C_j, & & C_i \xrightarrow{\makebox[7mm]{$d/\nu_1$}} T_i, \quad C_i + M_j \xrightarrow{\makebox[1.75mm]{$a$}} M_i + M_j,\\
& M_i \xrightarrow{\makebox[7mm]{$g /\nu_2$}} T_i, &  & T_i + M_j \xrightarrow{\makebox[1.75mm]{$\gamma$}} M_i + M_j,
\end{align}
\end{subequations}
where $\nu_1, \nu_2$ are neighbourhood-dependent functions (see Appendix \ref{SIsec:sMHE} for the full specification of Eqs.~\eqref{stoch_spatial_model}).

Throughout this work, we keep all the model parameters fixed to $r =1$, $d = 0.4$, $a = 0.2$, $\gamma = 0.75$ except the neighbourhood radius $\ell$ and the grazing rate $g$ at which fish graze on macroalgae. The fixed parameters are taken from \cite{mumby2007thresholds}. We initialise the model to initial global densities $C(0), T(0), M(0)$, either using a random initial configuration (uniformly distributed) or a coral-cluster initial configuration (where all coral nodes are placed together in the centre of the domain, and $M$ and $T$ are uniformly distributed around in the remaining space, see Fig.~\ref{fig:fig1}B, bottom). The \ssMHE{} model \eqref{stoch_spatial_model} is simulated as follows: at fixed time steps $\Delta t$, each node $i$ is considered in turn and is allowed to react with a probability according to the neighbourhood-dependent rates (see Appendix \ref{SIsec:sMHE} for details on the stochastic simulation algorithm). 
A simulation of the \ssMHE{} model is a sequence of ternary matrices (see Fig.~\ref{fig:fig1}C, bottom), which we call snapshots, in which each entry indicates which of the three species occupies that location within the reef. From the matrix, we can extract the global fractional cover of each species at any time, e.g., $C(t) = \sum_i C_i(t)/625$.

\section{Descriptors for coral data analysis}
A simple, effectively non-spatial descriptor is the fractional cover, given by the proportion of nodes of a given type in the reef (Fig.~\ref{fig:fig1}C, \ref{fig:fig2}A). 
We explore the spatial dynamics of the \ssMHE{} model using a collection of \emph{neighbourhood descriptors}, PH, and zigzag persistence.

\subsection{Neighbourhood descriptors}
We introduce nine neighbourhood descriptors to quantify the average neighbourhood composition of nodes in the reef. Let node $i$ be of type $C$ ($C_i = 1$). The neighbours of node $i$ are those nodes within a radius $\ell$ of $i$ (Fig.~\ref{fig:fig1}B, middle).
We count the number of neighbours $\sigma_i^\ell$ of node $i$ that are of type $\sigma$ 
for $\sigma \in \{C, T, M\}$. We have $\sigma_i^\ell \in \{0, \dots, n_i^\ell\}$, where $n_i^\ell$ is the total number of neighbours of node $i$ (e.g., $n_i^{1.45} = 8$ for an internal node $i$). We then define the neighbourhood descriptors $\sigma_C$ as the average of $\sigma_i/n_i^\ell$ across all nodes $i$ of type $C$. The descriptor $C_C$, for instance, gives the average fraction of coral neighbours that are coral (see Fig.~\ref{fig:fig2}C).
The other six descriptors are defined similarly, considering nodes $i$ of type $T$ or $M$ to give nine descriptors in total (see Appendix \ref{SIsec:sMHE} for details).
We may average the fractional coral cover over many realisations (see Fig.~\ref{fig:fig1}C) to give a non-spatial summary of the model's behaviour. However, the same coral fractional cover can take many different spatial patterns. For example, the random or coral-cluster initial configurations can be set with the same fractional cover (0.33), yet the local neighbourhood information differs significantly ($C_C (\text{random})<0.2)$ whereas $C_C (\text{coral-cluster})>0.8)$, (Fig.~\ref{fig:fig1}B, bottom). This local neighbourhood descriptor highlights that coral occupies more than 80\% of the neighbours of coral nodes in the coral-cluster configuration but less than 20\% for the random configuration, and therefore distinguishes spatially inhomogeneous reefs with the same fractional covers.\\

\subsection{Persistence}
PH offers an algorithmic way to quantify the connectivity of multiscale data. 
We represent a snapshot of the \ssMHE{} model by a sequence of \textit{cubical complexes}. A cubical complex is a data structure that represents nodes on a grid by vertices and connects adjacent vertices with edges and squares. 
To encode information about the density of coral nodes, 
we assign an integer to each vertex $i$ with $C_i = 1$ based on the number of direct coral neighbours. Mathematically, we define a \textit{density filtration function} $f:\mathcal{I} \rightarrow N$, where $\mathcal{I}$ is the set of nodes in the reef and $N = \{0, \ldots, 8\}$ \cite{TDAmnist,carlssonTDAdatamodelling}.
We use cubical complexes and $f$ to create a multi-scale lens called a \textit{filtration}.

In the first filtration step, we create a cubical complex using only those coral nodes with eight direct coral neighbours (i.e., completely surrounded by coral nodes), adding edges and squares between adjacent coral nodes. At each subsequent step, we include coral nodes with 7, 6,~\dots coral neighbours to the cubical complex. We define a nested sequence of cubical complexes according to the number of direct coral neighbours, $K_8 \subset K_7 \subset \dots \subset K_2 \subset K_1.$ 

For a given time-snapshot of the \ssMHE{} model, we build the filtration and then compute standard PH (see Appendix \ref{SIsec:PH}) for full details). 
PH quantifies the topological features, such as clusters $H_0$ (i.e., connected components) or loops $H_1$ (i.e., one-dimensional holes) across the filtration. The appearance and disappearance of components and loops across the filtration can be visualised as a multiset of intervals called a \textit{barcode}. Here, barcodes quantify features according to their size since large numbers of direct coral neighbours indicate large clusters (see Fig.~\ref{fig:fig2}C, D). 
In Fig.~\ref{fig:fig2}D, the barcodes capture the temporal evolution of the data, from the random spatial structure described by many short bars at $t=0$ to the single long bar at time $t=100$ representing one large coral cluster.

PH considers specific times of the \ssMHE{} data independently, making it difficult to decide whether a single component of coral persists or whether different coral clusters appear at each timestep.
We want to trace the time evolution of specific spatial features in the \ssMHE{} model. Due to the non-monotonicity of coral dynamics (i.e., coral locations can be occupied by other species and then return to coral), standard PH is not suitable, since coral clusters are not nested over time.

\begin{figure}[hb!]
\centering
\includegraphics[width=.6\linewidth]{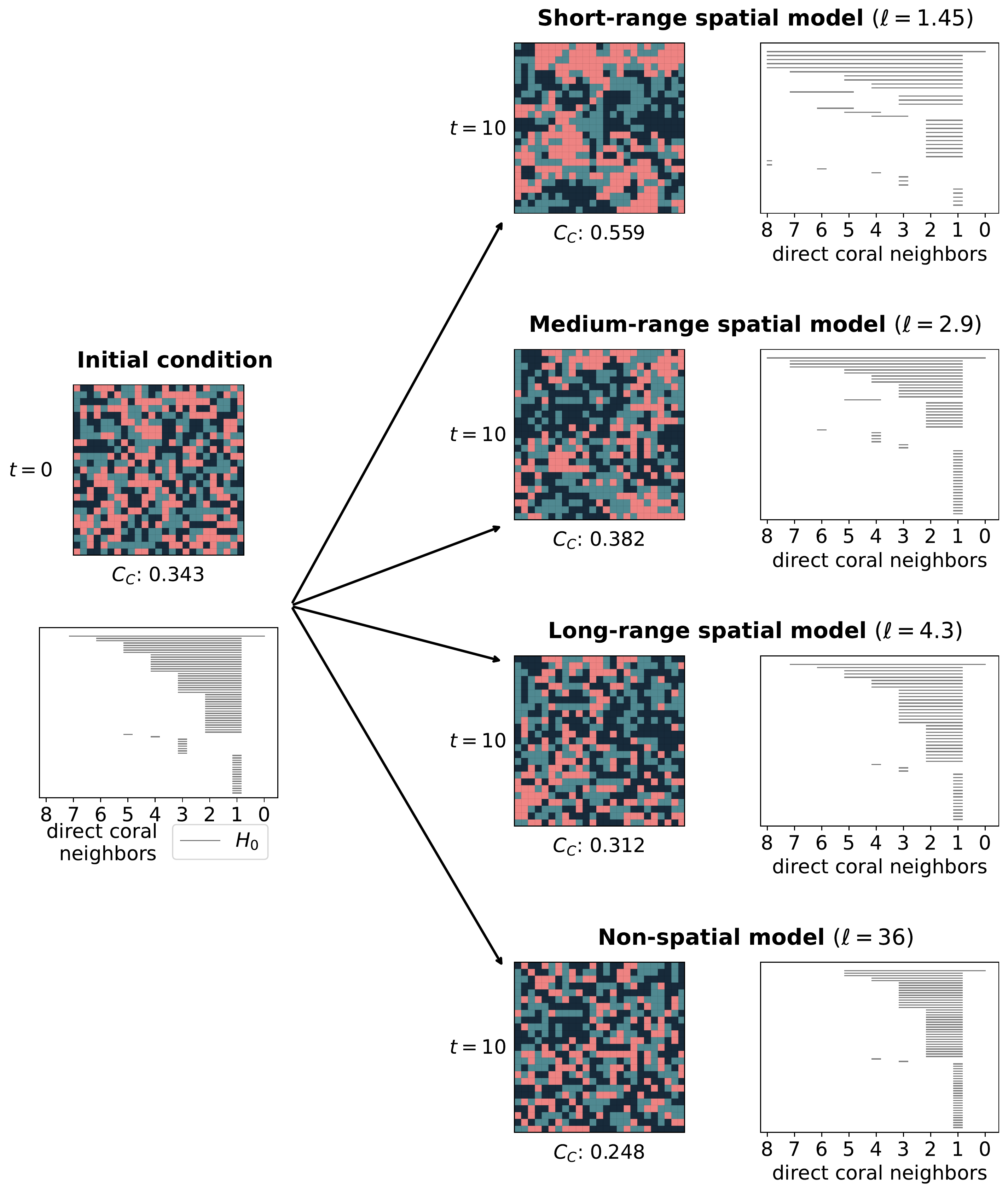}
\caption{\textbf{Effect of neighbourhood threshold} Spatial and non-spatial versions of the \ssMHE{} model are compared by running simulations up to time $t=10$. The grazing rate is kept constant at $g=0.53$, the random initial configuration is used with equal initial fractional covers $C(0)$, $T(0)$, and $M(0)$. The neighbourhood size is varied: $\ell=1.45, 2.9, 36$. As described in Fig.~\ref{fig:fig1}, a neighbourhood size of $\ell=1.45$ gives a spatial model, where only the immediate eight neighbours of each node affect its transition as the model is updated. The values $\ell=2.9, 4.3$ give spatial models where a larger grid of nodes affects the reaction rates. When $\ell=36$, all nodes are considered neighbours of all others, giving a non-spatial model. The snapshots of the \ssMHE{} model are printed at time $t=0$ and at time $t=10$ for each of these scenarios. The coral-coral neighbourhood descriptor, $C_C$, distinguishes the three cases. The value $C_C=0.559$ for the $\ell=1.45$ spatial model indicates that coral clusters more tightly together in this case, whereas the value $C_C=0.248$ shows that this does not happen for the non-spatial model. The adjacent persistence barcodes give further details of this difference. For the $\ell=1.45$ spatial model, there are a few large components of coral (indicated by solid bars) and some enclosed loops (indicated by dotted bars). For the non-spatial case, there are many smaller components of coral and no loops. The $\ell=2.9, 4.3$ plots give intermediate results in both $C_C$ and the persistent barcode.}
\label{fig:fig3}
\end{figure}

\subsection{Zigzag persistence}
To track the evolution of topological features over many time steps, we propose to use zigzag persistence \cite{originalzigzag}, which generalises the notion of filtration to a 
\textit{zigzag diagram}. At each time point $t_m$, we choose one of the complexes from the filtration described above to represent a snapshot of the \ssMHE{} model. We choose $K_1^{t_m}$, the cubical complex obtained by including coral nodes with at least one coral neighbour. 
We then insert the intersection of every pair of successive complexes into this sequence, giving:
\begin{equation}
K^{t_0}_1 \supset (K^{t_0}_1 \cap K^{t_1}_1)  \subset K^{t_1}_1 \supset (K^{t_1}_1 \cap K^{t_2}_1) \cdots K^{t_M}_1.
\label{zzfiltration}
\end{equation}
A diagrammatic representation of \eqref{zzfiltration} with three time points is given in Fig.~\ref{fig:fig2}E. We use this sequence to compute connected components (i.e., $H_0$ of \eqref{zzfiltration}), which emerge and disappear over many timesteps in a simulation of the \ssMHE{} model. We perform a pre-processing step to reduce the noise of the \ssMHE{} data before the computation of zigzag persistence (see Appendix \ref{SIsec:zigzag}).

Due to the stochastic nature of the \ssMHE{} model, the zigzag barcode of a single simulation would not provide an accurate summary of the model's behaviour. Persistence landscapes offer a way to average the spatial information encoded in barcodes \cite{persland}. We create a persistence landscape from a simulation of the \ssMHE{} model as follows. For each component of coral that is detected in a simulation of the model, we plot a landscape peak at $(t_l, s_l)$, where $t_l$ is halfway through the component's life span, and $s_l$ is half of the length of time for which it exists. For example, if a component is born at time $t=50$ and dies at time $t=100$, a landscape peak is placed at $(75, 25)$. A `tent' is then constructed by connecting $(t_l, s_l)$ linearly down to the birth and death times on the $t$-axis. The $k$th persistence landscape $\lambda_k$ takes, at every time, the $k$th largest value among all the `tents' (see Appendix \ref{SIsec:zigzag} for a full explanation of this process). 

The landscape $\bar \lambda_k$ is the average of $\lambda_k$ over many simulations, and the maximum of each $\bar \lambda_k$ may be interpreted as the average half-lifetime of the $k$th most persistent cluster. Therefore, zigzag persistence landscapes quantify the typical spatial information across multiple simulations of the \ssMHE{} model and rank which features are most significant over time.

 The first landscape ($\bar \lambda_1$) in Fig.~\ref{fig:fig2}F highlights that a single cluster of coral dominates the domain from $t=0$ through $t=100$. We can observe that the time-persistence of the second and third landscapes ($\bar \lambda_{2}, \bar \lambda_{3}$) decreases as time increases, suggesting that smaller components either disappear or join with the main one as time increases. Zigzag persistence, therefore, confirms that, on average, a single component is emerging over time, which PH alone cannot conclude.

We use these tools to explore the model dynamics and spatial structure. Specifically, we consider the effect of local neighbourhood radius $\ell$, the initial configuration of species (random or coral-cluster configurations), and the rate $g$ that fish graze on macroalgae, and quantify how changing these three parameters affects the \ssMHE{} model.

\section{Results}

\subsection{Spatial interactions lead to coral clustering}
The range of the spatial interactions in the \ssMHE{} model is controlled by the neighbourhood radius $\ell$ (see Fig.~\ref{fig:fig1}B). 
For large values of $\ell$, the neighbourhood of interaction spans the whole domain, and the \ssMHE{} model reduces to a non-spatial stochastic model. However for small values of $\ell$ the rates of reactions \eqref{stoch_spatial_model} are highly dependent on a node's immediate neighbours.  
The neighbourhood descriptor $C_C$ and PH describe the spatial patterning observed in simulations for different values of the neighbourhood radius $\ell$ (see one realisation in Fig.~\ref{fig:fig3}).
In particular, larger clusters of coral appear when the interaction range $\ell$ is small, whereas we observe little spatial patterning when $\ell$ is large. 
Zigzag persistence determines over multiple timesteps that a stable cluster persists over time in the \ssMHE{} model with $\ell=1.45$, whereas no such cluster persists in the non-spatial model (see Supplementary Material Fig.~\ref{SIfig:fig3supplement}).
\clearpage

\begin{figure}[h!]
\centering
\includegraphics[width=.525\linewidth]{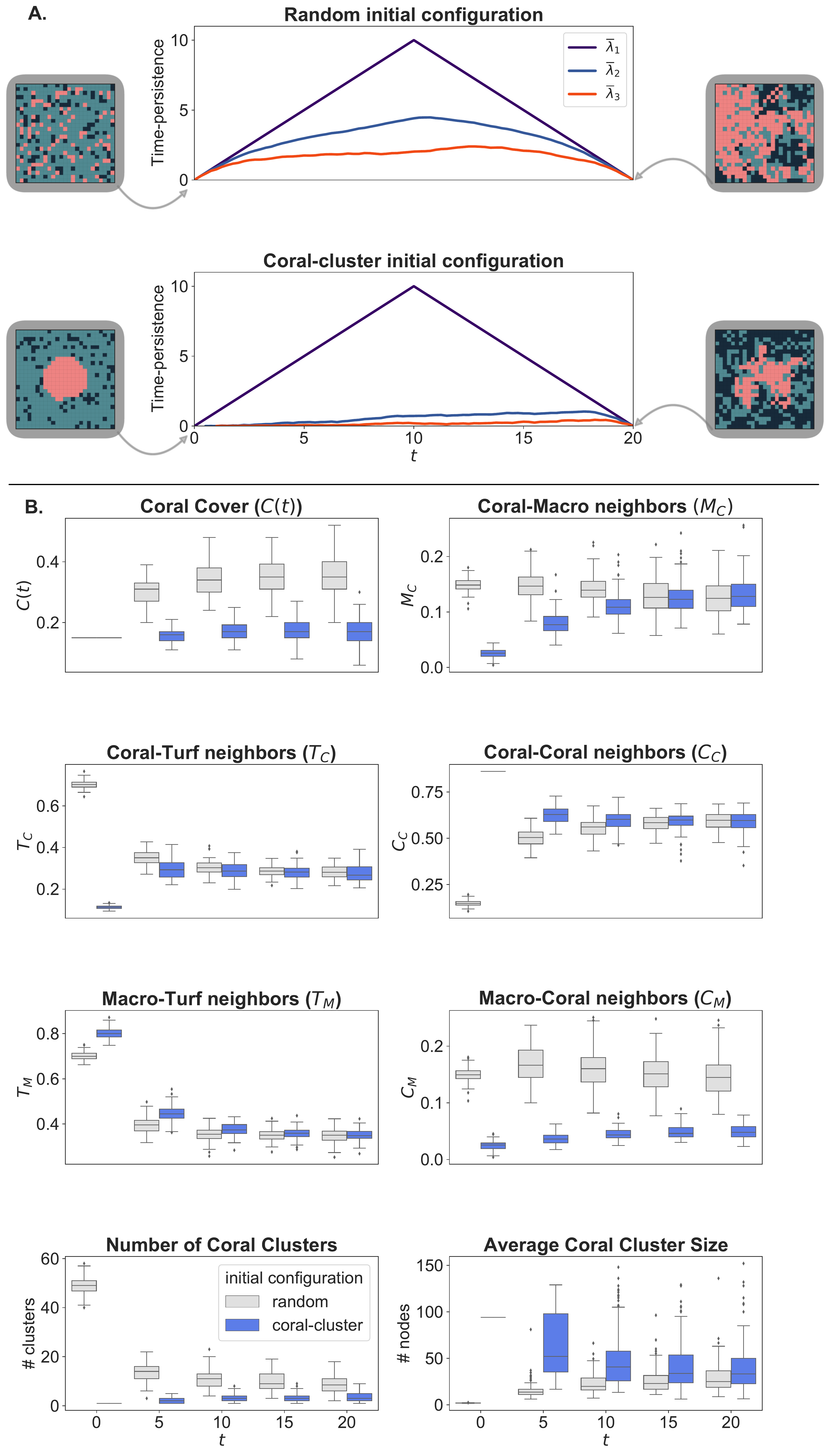}
\caption{\textbf{Effect of initial configuration} Comparison of 100 realisations of the \ssMHE{} model initialised with either the random or coral-cluster initial configurations from Fig.~\ref{fig:fig1}B, with grazing fixed to $g= 0.53$ and initial fractional covers $C(0) = M(0) = 0.15$. \textbf{A.} Average persistence landscapes $\bar \lambda_1, \bar \lambda_2, \bar \lambda_3$. While the random initial configuration leads to three non-trivial landscapes, the coral-cluster initial configuration gives a significant first landscape $\bar \lambda_1$. These average landscapes indicate that the random initial configuration leads to coral clustering in many different components, which persist as separate patches up to at least time $t=20$. In contrast, the coral-cluster initial configuration begins with a single connected component of coral. The analysis shows that few other components form throughout the simulation and that coral remains clustered in the initial central patch. Examples of model snapshots at $t=0$ and $t=20$ are printed at the side of each landscape plot. \textbf{B.} Coral fractional cover $C(t)$ and neighbourhood descriptors. The plot of $C(t)$ shows that the random initial configuration favours coral growth, since the fractional coral cover increases in these simulations. In contrast, coral appears to die out when simulations start with the coral-cluster initial configuration. The $T_C$, $M_C$, and $C_C$ plots show that these descriptors clearly distinguish the two initial configurations at time $t=0$ but cannot tell them apart after a small number of timesteps. Neighbourhood descriptors give similar values for the two initial configurations from time $t=5$ onwards.}
\label{fig:fig4}
\end{figure}
\clearpage
\subsection{Zigzag persistence distinguishes initial configurations}
Intuitively, we may think that a single large coral cluster might offer the best conditions for coral resilience over time. To test this hypothesis, we initialise the \ssMHE{} model to two initial conditions with identical fractional covers ($C(0), M(0), T(0)$) but very different spatial configurations, namely the random and coral-cluster initial configurations (see Fig.~\ref{fig:fig1}B). Perhaps surprisingly, depending on the grazing conditions, the analysis suggests that coral can be more robust over time when initialised with the random configuration, whereas it stagnates or even becomes extinct when initialised with the coral-cluster configuration. 
The neighbourhood descriptors, as well as the average cluster size and total number of clusters, distinguish the two initial configurations, showing significant differences between the two cases at $t=0$. 
However, these descriptors can not discern the two cases as time progresses (see Fig.~\ref{fig:fig4}B). 
On the other hand, zigzag persistence landscapes $\bar \lambda_2, \bar \lambda_3$--averaged over many simulations--significantly differ between the two cases over the whole simulation time (see Fig.~\ref{fig:fig4}A). 
We can understand this result by noting that the random initial configuration yields a more significant average number of coral-turf neighbours ($T_C(0) = 0.765$ and $0.113$ for random and coral-cluster initial conditions, respectively), increasing the locations where coral growth can occur (via the interaction $T_i+C_j \to C_i+C_j$) when compared to the coral-cluster initial configuration. In this way, under certain parameter regimes, the spatial configuration of nodes in the \ssMHE{} model may determine the long-term behaviour of the system.

\subsection{Zigzag persistence describes coral extinction pathway} \label{sec:results3}
In many reefs, coral's spatio-temporal dynamics critically depends on the fish population feeding on the macroalgae \cite{muthukrishnan2014multiple}. Thriving fish populations (i.e., high grazing rate $g$) keep macroalgae levels low, which leads to better conditions for coral to flourish. In contrast, overfishing and natural disasters shrink the fish population and hence $g$, which may result in macroalgae overgrowth and coral decay.

The MHE model (\eqref{mumby}) captured these coral-reef interactions and established that the grazing rate $g$ is a bifurcation parameter, where low grazing leads to coral extinction, and intermediate grazing drives the system to display two alternative states \cite{mumby2007thresholds}.

The \ssMHE{} model (\eqref{stoch_spatial_model}) reproduces this behaviour: for $g<0.52$ the system evolves to a macroalgae-dominated reef; for $g>0.55$ the system evolves to a coral-dominated reef. In the region of multistability ($g \in [0.52, 0.55]$), we observe that simulations can evolve to either the coral- or macroalgae-dominated state. However, in contrast to original MHE model, where the long-term behaviour of simulations in the metastable region are determined by their initial condition, the stochastic nature of the \ssMHE{} model means that identical initial conditions may lead to different outcomes. At $g=0.53$, we observed that around half of all simulations (initiated with equal numbers of coral, turf and macroalgae nodes in the random configuration) evolved to the macroalgae-dominated state, with the other half converging to a coral-dominated state.

Metastability implies that the system takes a long time to converge to such a state (up to $1000$ timesteps in some realisations). We explore whether summaries of the early species' behaviour (we choose $t\in[0,100]$) predict the system outcome.
Non-spatial descriptors (e.g., the fractional covers $C(t), M(t)$, see Supplementary Material Figure 11) can give an early indication of the system outcome in the metastable region. 

But what makes coral die out in some runs of the \ssMHE{} in the metastable region and resist in others?

\begin{figure}[h!]
\centering
\includegraphics[width=.75\linewidth]{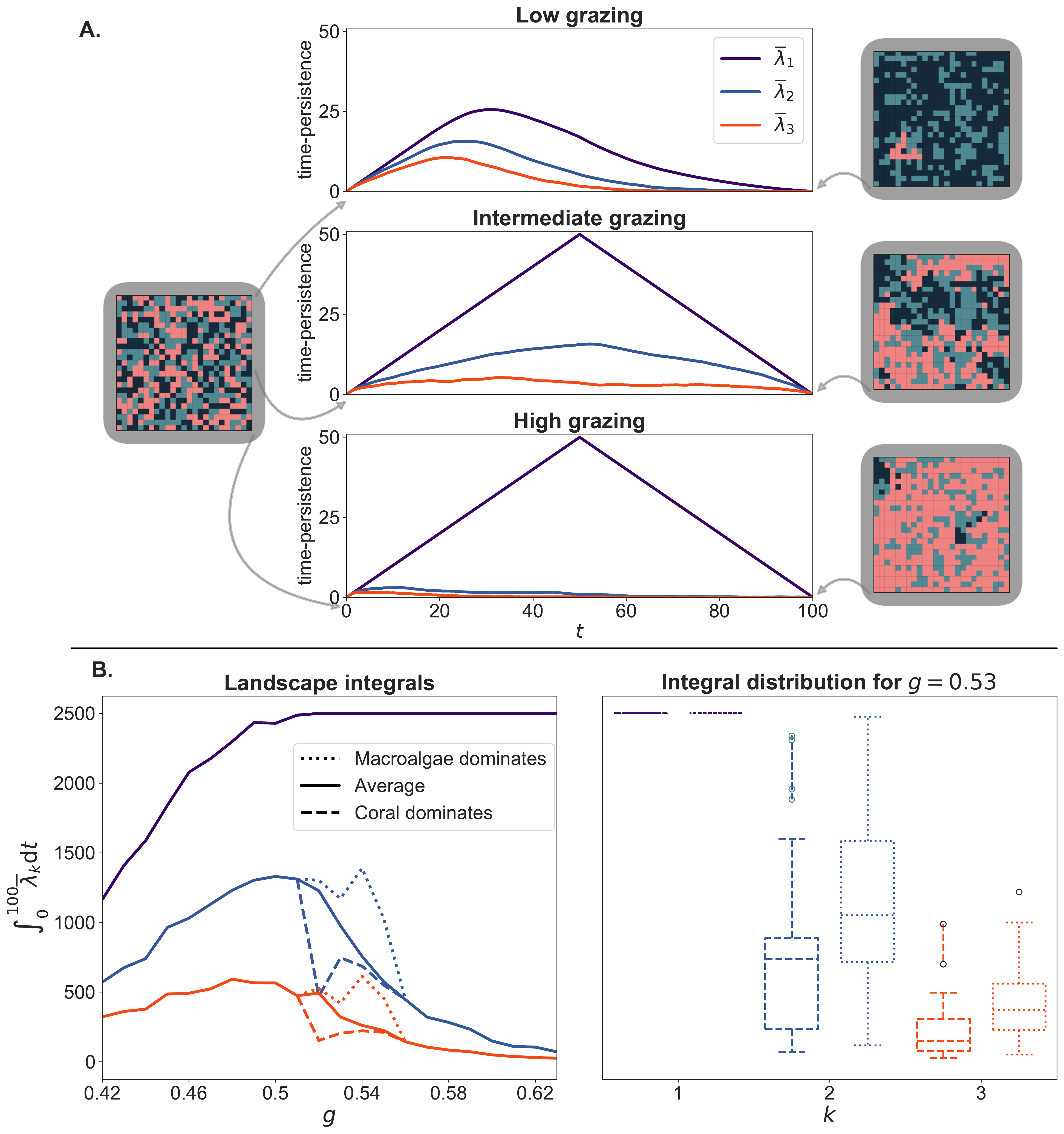}
\caption{\textbf{Effect of grazing parameter.} \textbf{A.} Average persistence landscapes using 100 realisations of the \ssMHE{} model up to $t=100$ for three grazing rates: $g=0.42$ (low grazing), $g=0.53$ (intermediate grazing) and $g=0.62$ (high grazing). All realisations are initialised with the same initial configuration (pictured near $t=0$). Representative outputs of the \ssMHE{} model at $t=100$ for each grazing rate are also shown. The integrals of $\bar \lambda_k, k=1, 2, 3$ are computed for each grazing rate. The peak of the first landscape increases as the grazing rate increases. The increasing size of the first landscape indicates that the most resilient component of coral lasts for a longer time for higher, coral-favouring grazing rates. The second landscape is small for low $g$ since all components of coral are short-lived in these conditions, as macroalgae quickly overgrows them. On the other hand, the second landscape is small for large $g$ as well, because smaller components of coral are quickly amalgamated into the main component in these conditions, as coral begins to dominate the entire domain. These phenomena are balanced for $g$ in the middle of this range, where either coral or macroalgae may be successful. Zigzag persistence, therefore, not only describes the change in spatial behaviour as $g$ is increased, it indicates that the ``tipping point''--where coral and macroalgae are evenly matched--corresponds to the peak of the second landscape $\bar \lambda_2$, which occurs near $g=0.51$. \textbf{B.} The average integrals, between $t=0$ and $t=100$, of the $k$th landscapes for different grazing rates. The maximum integral of $\bar \lambda_2$ coincides with the metastable region. Values of these integrals are plotted separately for simulations where coral ended up dominating the reef from those where macroalgae dominated. The integrals of $\bar \lambda_2$ and $\bar \lambda_3$ are greater in cases where coral dies out, indicating that it does so through many short-lived components rather than the principal component.}
\label{fig:fig5}
\end{figure}

To address this question, we turn to zigzag persistence. 
Zigzag persistence can predict the system's outcome (Fig.~\ref{fig:fig5}A) but also characterises how the coral's spatial clustering affects coral outcome (Fig.~\ref{fig:fig5}B). 
In particular, we find that $\bar \lambda_2$ and $\bar \lambda_3$ (corresponding to the second and third most dominant topological features) are higher in simulations where coral eventually dies out. The difference in landscape size for different outcomes indicates that under these reef conditions (in particular, low coral-turf neighbours, $T_C(0) = .361$), coral extinction occurs through multiple small clusters (as opposed to one cluster that shrinks to nothing), while coral domination establishes itself as one persistent cluster. 

\clearpage
\clearpage

\subsection{Topology distinguishes empirical data with same coral cover}
\begin{figure*}[ht]
\centering
\includegraphics[width=.675\linewidth]{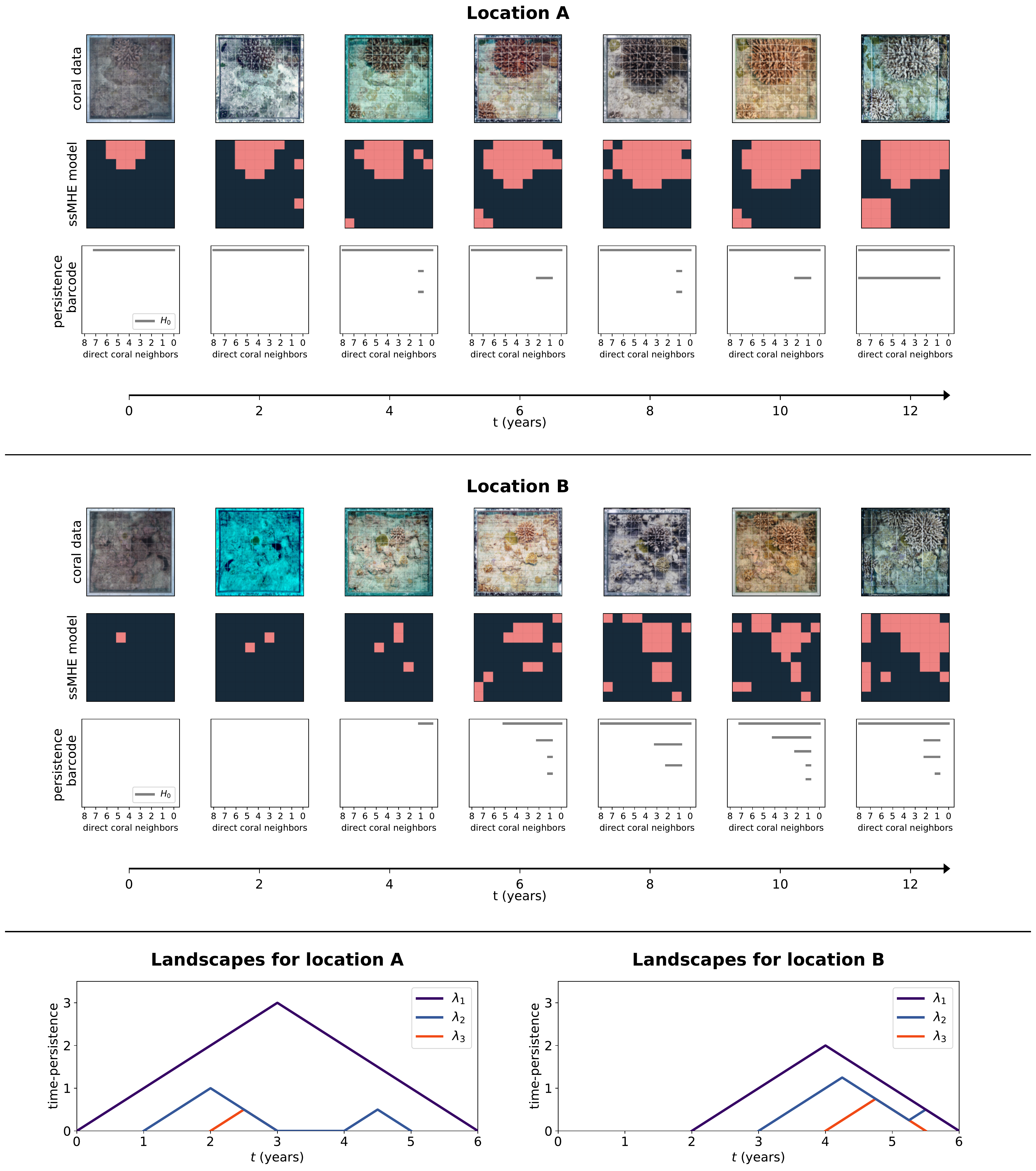}
\caption{\textbf{Analysis of empirical data}. Comparison of two multi-year coral datasets at different locations within the Rarotonga reef (Cook Islands) from 2009 to 2022. Data was kindly provided by the Service National d'Observation CORAIL from CRIOBE. \textbf{Locations A and B.} For each location we show a 1m$^2$ section of reef with a $10 \times 10$ grid placed on top, photographed on seven different occasions over a 12-year period. We estimate (by eye) the grid intersections where coral is present and use this to represent each photograph by a snapshot of the \ssMHE{} model. Pink squares denote locations where we see coral and all other squares are coloured blue. We then compute persistence barcodes of the model snapshots corresponding to each photograph at each location. In location A, all barcodes contain one persistent bar, with some containing additional short bars. This indicates that one large component of coral is present within each photograph of location A, as well as some small components at certain timesteps. On the other hand, the barcodes corresponding to the photographs of location B show many short bars and one long bar--with the longest bar increasing in length as time goes on. This suggests that there is principal component of coral in location B which is getting larger. \textbf{Landscapes for both locations.} The zigzag landscapes for each location provide additional information to the persistence barcodes and distinguish the spatio-temporal evolution in each location. In location A, the first landscape indicates that a single component of coral exists throughout all photographs. The second and third landscapes show short-lived components of coral which either disappear or join to the principal component. The peak of the first landscape is smaller and occurs later for location B, indicating that the principal component emerges towards the end of the study period. While the percentage cover of coral ($\approx 40\%$) and the $C_C$ neighbourhood descriptor ($\approx 0.75$) are similar at the two locations at the final timestep, the zigzag persistence landscapes clearly distinguish the spatio-temporal evolution of coral at the two locations.}
\label{fig:fig6}
\end{figure*}

We next  use the longitudinal spatial data from the Moorea IDEA project \cite{davies2016simulating} to exemplify the use of the proposed topological descriptors. 
Our methods would be ideally suited for a grid with finer spatial resolution, but data of coral reefs via drones \cite{drone_data_coral} which offer higher spatial resolution currently lack multi-year collection.
Here, we apply our topological methods to two multi-year datasets from the Rarotonga reef at two different locations (see Figure~\ref{fig:fig6}, locations A and B). While the percentage cover of coral is similar at the start ($t=0$) and at the end ($t=12$) of the twelve-year period for both locations, persistence barcodes and zigzag landscapes distinguish the spatio-temporal evolution (see Figure~\ref{fig:fig6}, landscapes).

The coral configurations in the two locations (A and B) resemble the coral-cluster and random configurations of the \ssMHE{} model respectively. 
The percent Similar to the model analysis in Figure~\ref{fig:fig4}, the coral-coral neighbourhood descriptor ($C_C$) is the same for both locations at the end of the time-course. However, the persistence barcodes and zigzag landscapes differ. Barcodes from the empirical coral-cluster (Figure~\ref{fig:fig6}, location A) have one persistent bar for each time iterate, whereas the dataset analogous to the random configuration (Figure~\ref{fig:fig6}, location B) has many short bars in the early iterates and one long bar at the final timestep.
The spatial-temporal evolution at different locations can be distinguished by the zigzag landscapes  (Figure~\ref{fig:fig6}, landscapes). In location A, the first zigzag landscape is larger than for location B, and the peaks of landscapes for location B occur later than for location A. 
The topological analysis therefore reveals that, in location A, one large cluster persists across twelve years, whereas in location B, many small clusters join together over time.
Our model predicts that configurations similar to location A are more likely to lead to coral extinction than those similar to location B.
To justify this prediction, we would require empirical reef data with better resolution (spatial and temporal).
In future, we aim to compare data with coral evolution models such as the \ssMHE{} model.

\section{Discussion}

Motivated by ecology and evolution, as well as the increasing availability of spatial data of such processes, we introduced a stochastic spatio-temporal lattice-based model (\ssMHE{} model \eqref{stoch_spatial_model}) of coral reefs. We collected data through computational experiments and proposed topological descriptors to quantify coral behaviour and predict mechanisms. Specifically, we explored these descriptors on multiple realisations of coral reef dynamics under changes to the initial configurations and the values of key model parameters.

The evolution of the \ssMHE{} model depends on two factors. First, species evolve based on the rate values in Eqs.~\ref{stoch_spatial_model} (which model internal and external factors affecting the reef as in the original MHE model \cite{mumby2007thresholds}) and, second, on the spatial arrangement of nodes in the reef. A combination of both factors determines which reactions are possible (depending on the make-up of the neighbourhood) and more likely to occur. The fish grazing rate $g$ is a critical parameter in both the MHE and \ssMHE{} models,
with macroalgae dominating for low enough $g$ and coral dominating for high enough $g$ (regardless of the spatial arrangement of species in the reef). 
In contrast, in the intermediate metastable region, the coral- and macroalgae-favouring reactions balance out so that the spatial arrangement becomes a deciding factor. 
Zigzag persistence can discern these multiple pathways, intricately dependent on species' competition. 
For example, we found that the random initial configuration yields higher coral growth than the coral-cluster initial configuration (Fig.~\ref{fig:fig4}) under turf abundance. Conversely, when macroalgae growth overwhelms coral, zigzag persistence suggests that coral goes extinct by becoming dissected into many components (Fig.~\ref{fig:fig5}B). This indicates that, when macroalgae prevalence is high, coral survives better when clustered together, thus limiting the macroalgae-overtaking-coral interaction in ~\eqref{stoch_spatial_model}. 
Together, these insights show the potential use of our model in helping assess the vulnerability of reefs and better design artificial reefs. By tweaking our model to the conditions of a reef of interest, one can theorise coral's ideal spacing for survival.
\clearpage
\clearpage

The region of metastability of the \ssMHE{} model (in which different outcomes are possible)
results from including spatial and stochastic dynamics in our model. This variation in outcomes motivates questions such as: are there disturbances or interventions that could ``shock'' reefs into one state or another? Could this region make the fate of reefs more ``reversible'' than one would assume in the analysis of deterministic models? Could this behaviour explain some of the ``noise'' observed in real data, with high coral morality in some locations and coral survival in others with similar conditions?

The advantage of topological spatial descriptors is their versatility in analysing spatial and temporal structure in complex data. With the flexibility of PH to propose and adapt different filtrations, the standard PH pipeline can be tailored to study a wide range of other spatially patterned systems.
Here we showcased the power of zigzag persistence, a topological measure that has recently benefited from improved computation \cite{bradzzalgorithm}. We combined zigzag persistence with statistical landscapes \cite{persland} to enhance the identification of the geometry of the initial configuration of species as well as the geometric mechanisms of species competition and, ultimately, coral extinction in the metastable parameter region. 

Since zigzag persistence quantifies spatial patterns in images, it can be used as a pre-processing step prior to classification. For example, if using data sources such as drones or satellite images, one may vectorise persistence using persistence landscapes, which can then be fed into a machine learning algorithms. Based on the positive findings from analysing short-time duration data generated by the \ssMHE{} model, a promising future direction is to apply topological descriptors to such high-resolution, short-time duration drone data. Given such highly resolved data, we expect the classification, annotation, and localisation of species in images with machine learning and topological statistics will become automatic.

We have found that persistent homology and zigzag persistence provides valuable insights into the spatial dynamics of the \ssMHE{} model. Topological quantification has been used to describe complex biological models, with different filtrations and topological descriptors used depending on the specific model in question \cite{nardini, tdadorsogna, fishPHpaper}. Such topological analyses may, in future, enable the comparison and validation of complex mechanistic models by adapting topological approximate Bayesian computation \cite{TABC}.

\section{Materials and Methods}
All simulations of the \ssMHE{} model were performed in Python: \url{https://github.com/rneuhausler/coralModel-TDA-study}. The neighbourhood descriptors are calculated while running the model, as the neighbourhood information is used in the model's evolution (see \eqref{stoch_spatial_model}). 
Computation of PH and zigzag persistence was implemented in Python using the BATS package:  \url{https://github.com/CompTop/BATS.py}. Code for all figures is available at \url{https://github.com/rmcdomaths/zigzagcoralmodel}.
All data used in this project may be simulated by running scripts available at \url{https://github.com/rneuhausler/coralModel-TDA-study}. Alternatively, the data may be downloaded directly from \url{https://doi.org/10.6084/m9.figshare.23717409.v1}.
A list of symbols used in this work can be found in Appendix E.

\section{Acknowledgements}

We thank M.~Porter for insightful comments on this manuscript and R.~Bardenet, M.~Bonsall, R.~Neville, and J.~Page for valuable discussions that contributed to the early stages of this work. The authors thank B.J.~Nelson for extending the zigzag code to include cubical complexes. RAM thanks the EPSRC and Lord Crewe's Charity. MB acknowledges funding from St John’s College, Oxford, and the Royal Society (URF$\backslash$R1$\backslash$180040). HAH acknowledges funding from EPSRC EP/R018472/1, EP/R005125/1 and EP/T001968/1, the Royal Society RGF$\backslash$EA$\backslash$201074 and UF150238. RN acknowledges funding from National Science Foundation DGE-1450053 and the National Aeronautics and Space Administration 80NSSC19K1378.
For the purpose of Open Access, the authors have applied a CC BY public copyright license to any  Author Accepted Manuscript (AAM) version arising from this submission.
\FloatBarrier

\appendix

\section{The \ssMHE{} lattice-based coral model}
\label{SIsec:sMHE}
\subsection{Construction}
\label{SIsubsec:modelconstruction}

We fully state the spatial, stochastic, lattice-based version of the MHE model (the \ssMHE{} model), giving explicit formulae for the reaction rates. Coral reefs are generally analysed through digital images, split into a coarse grid of smaller squares. To report the percentage cover of coral, researchers judge which species exists at the intersection of these squares. The number of sites occupied by coral is counted and then divided by the total number of intersections (Fig. ~\ref{fig:fig1}A of the main text, top). An area of the seabed is therefore described as a pixel-like structure, and it is this structure that inspired the \ssMHE{} model. We consider a two-dimensional domain (reef) subdivided into $J$ nodes such that each node $i$ is occupied by one species $C_i, T_i, M_i \in \{0, 1\}$ ($C_i + T_i + M_i = 1$ for each $i$). A non-spatial summary of the \ssMHE{} model at time $t$ is given by the total fraction of nodes occupied by a certain species: $C(t) = \sum_i C_i(t)/J$, $T(t) = \sum_i T_i(t)/J$, $M(t) = \sum_i M_i(t)/J$.

\subsection{Reactions}
\label{SIsubsec:reactions}

We introduce the local neighbourhood $\lN$ of node $i$ of radius $\ell$ as
$$
\lN = \left \{ j\ne i \ : \ \| i-j\|_2 \le \ell \right \},
$$
and define $n_i^\ell = \sum_{j\in \lN}$ to be the total number of neighbours of node $i$.
The makeup of the local neighbourhood of node $i$ is given by counting the number of coral, turf, and macroalgae neighbours of node $i$:
\begin{equation}
	\lC = \sum_{j \in \lN} C_{j}, \qquad \lT  = \sum_{i \in \lN} T_{j}, \qquad \lM = \sum_{j \in \lN} M_{j}.
	\label{SIeq:localN}
\end{equation}
For example, if $\ell = 1$, the neighbourhood has at most four nodes (up, down, left, right, less if it is a boundary node), and the above quantities add up to four. When $\ell = 1.45$, internal nodes have a neighbourhood of eight other nodes (as depicted in Fig.~\ref{fig:fig1}B of the main text, top left). We keep $\ell = 1.45$ for most of the simulations in this paper.

We may now define the \ssMHE{} model by the following set of reactions:
\begin{subequations}
\label{SI:stoch_spatial_model}
\begin{alignat}{3}
	&\ce{$T_i + C_j$ ->[$r$] $C_i + C_j$}, & \qquad &
	\ce{$C_i$ ->[$d /\nu_1 $] $T_i$}, & \qquad & \ce{$C_i + M_j$ ->[$a$] $M_i + M_j$},\\
	&\ce{$M_i$ ->[$g /\nu_2$] $T_i$}, & & \ce{$T_i + M_j$ ->[$\gamma$] $M_i + M_j$},& &
\end{alignat}
\end{subequations} 
Here $r, d, a, g, \gamma$ are parameters from the model of Mumby, Hastings and Edwards (MHE model), with $r =1$, $d = 0.4$, $a = 0.2$, $\gamma = 0.75$ fixed throughout this work while we allow for the grazing parameter $g$ to vary. The rates $\nu_1$ and $\nu_2$  
are neighbourhood-dependent functions given by
$$ \nu_1 = 1 + \frac{C_i^\ell}{n_i^\ell}, \qquad \nu_2 = 1+ \frac{M_i^\ell + T_i^\ell}{n_i^\ell}. $$
The neighbourhood-dependent rate $\nu_1$ is introduced to make natural coral mortality depend on its neighbourhood (in particular, the mortality rate halves when a coral node is surrounded by coral relative to an isolated coral node). Similarly, a macroalgae node is less likely to be grazed upon when it is surrounded by macroalgae and turf (which can also be grazed). The macroalgae overgrowth rate $g/\nu_2$ therefore decreases as $\lT$ and $\lM$ increase. This corresponds to a spatially-dependent version of the grazing rate in the MHE model ($gM/(M+T)$). All reaction rates depend on the local neighbourhood $\lN$ of node $i$, either explicitly through $\nu_1, \nu_2$, or implicitly, since the reactions on node $i$ depend on the species types of nodes $j\in \lN$.

\subsection{\ssMHE{} model data}
\label{SIsubsec:modeldata}
The \ssMHE{} model represents a coral reef by a grid of $J$ nodes, each occupied by one of three species: coral, turf, or macroalgae. A \textit{snapshot} of the \ssMHE{} model is, therefore, a ternary matrix whose entries encode which species occupies the corresponding node in the reef. We represent such snapshots by a three-coloured grid, as seen in Fig.~\ref{fig:fig1} in the main text. A \textit{simulation} of the \ssMHE{} model is then a series of snapshots representing the reef at different timesteps. Formally, a simulation is an ordered sequence of ternary matrices--one for each snapshot.

\subsection{Neighbourhood descriptors}
\label{SIsec:neighbourhooddescriptors}
Since many initializations of the \ssMHE{} model are possible for a single set of initial percentage covers, we define \textit{neighbourhood descriptors} to distinguish spatially distinct configurations.

Let the makeup of the local neighbourhood of node $i$: $\lC$, $\lT$, $\lM$, be defined as in Subsection \ref{SIsubsec:reactions}. The neighbourhood descriptors are the averages of these quantities over all nodes $i$ of each type. For example, $T_C$ gives the average number of the neighbours of a coral node that are occupied by turf: 
 \begin{equation}
 	T_C = \frac{\sum_i C_i (T_i^\ell / n_i^\ell)}{\sum_i C_i}.
 	\label{SIeq:Tc}
 \end{equation}
Our naming convention is such that the second species name refers to the reference species, and the first is the neighbour being counted. Note that Definition \ref{SIeq:Tc} is not symmetric, nor are other neighbourhood descriptors if the reference species differs from the type of neighbour being counted. For example, in a reef comprising a single internal coral node and otherwise all turf nodes, we will have $T_C = 1$, but $C_T \approx 0$, since the coral node has all turf neighbours, whereas most neighbours of turf nodes are not coral.

Each of the nine neighbourhood descriptors may be computed from a snapshot of the \ssMHE{} model at time $t$. The time evolution of the neighbourhood descriptors over a simulation of the \ssMHE{} model gives a spatio-temporal summary of its behaviour.

\subsection{Initialization of the \ssMHE{} model}
\label{SIsubsec:initialisation}
To initialise the \ssMHE{} model, we specify the total percentage cover of each species $C(0)$, $T(0)$, $M(0)$ and then choose the spatial structure of the initial configuration.  

\begin{figure}[htb]
	\begin{center}
	$(C(0), M(0)) = (0.5, 0.25)$ \hspace{2.5cm} $(C(0), M(0)) = (0.33, 0.33)$\\
		\includegraphics[width = 0.34\textwidth]{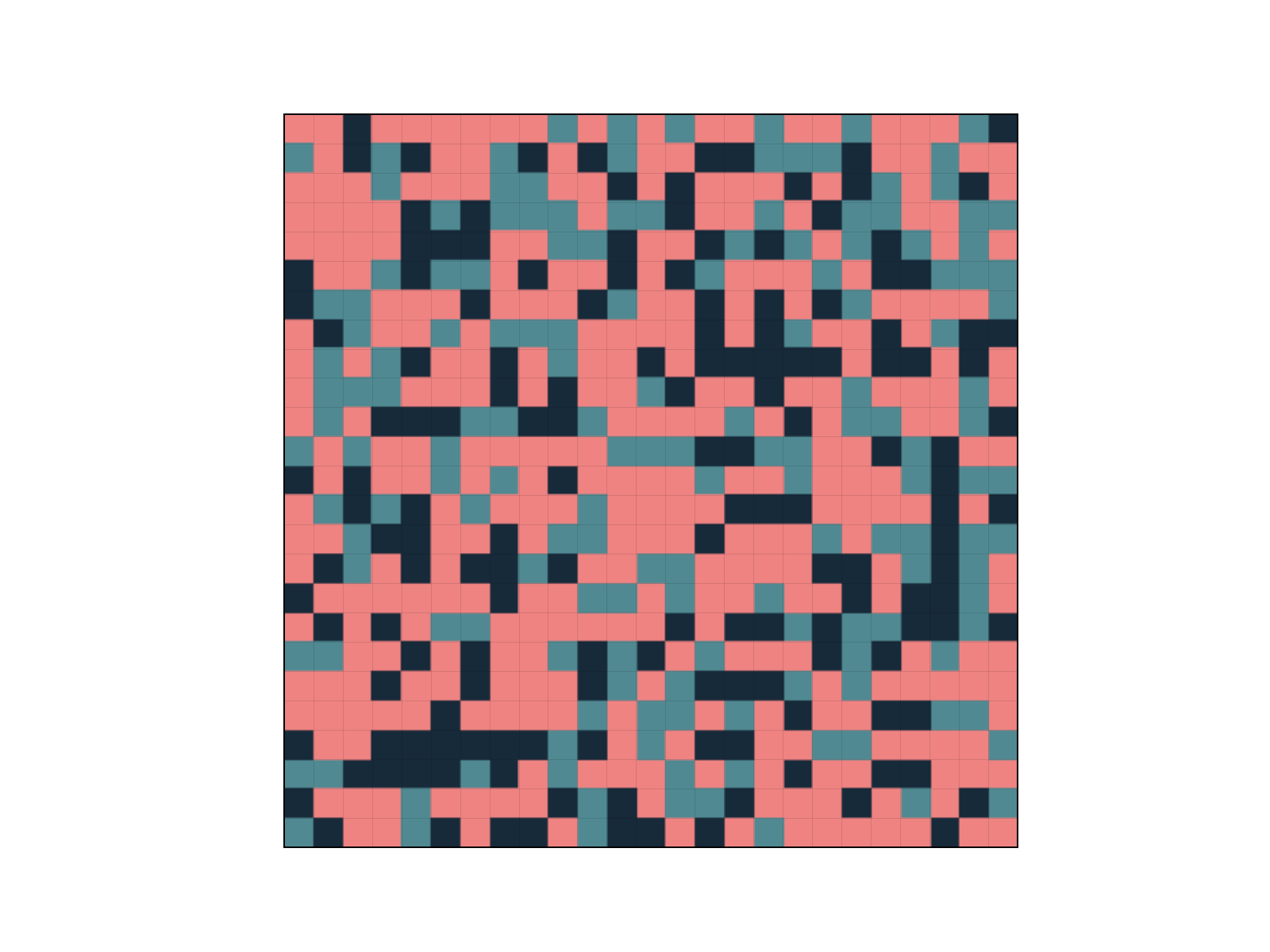} 
		\includegraphics[width = 0.34\textwidth]{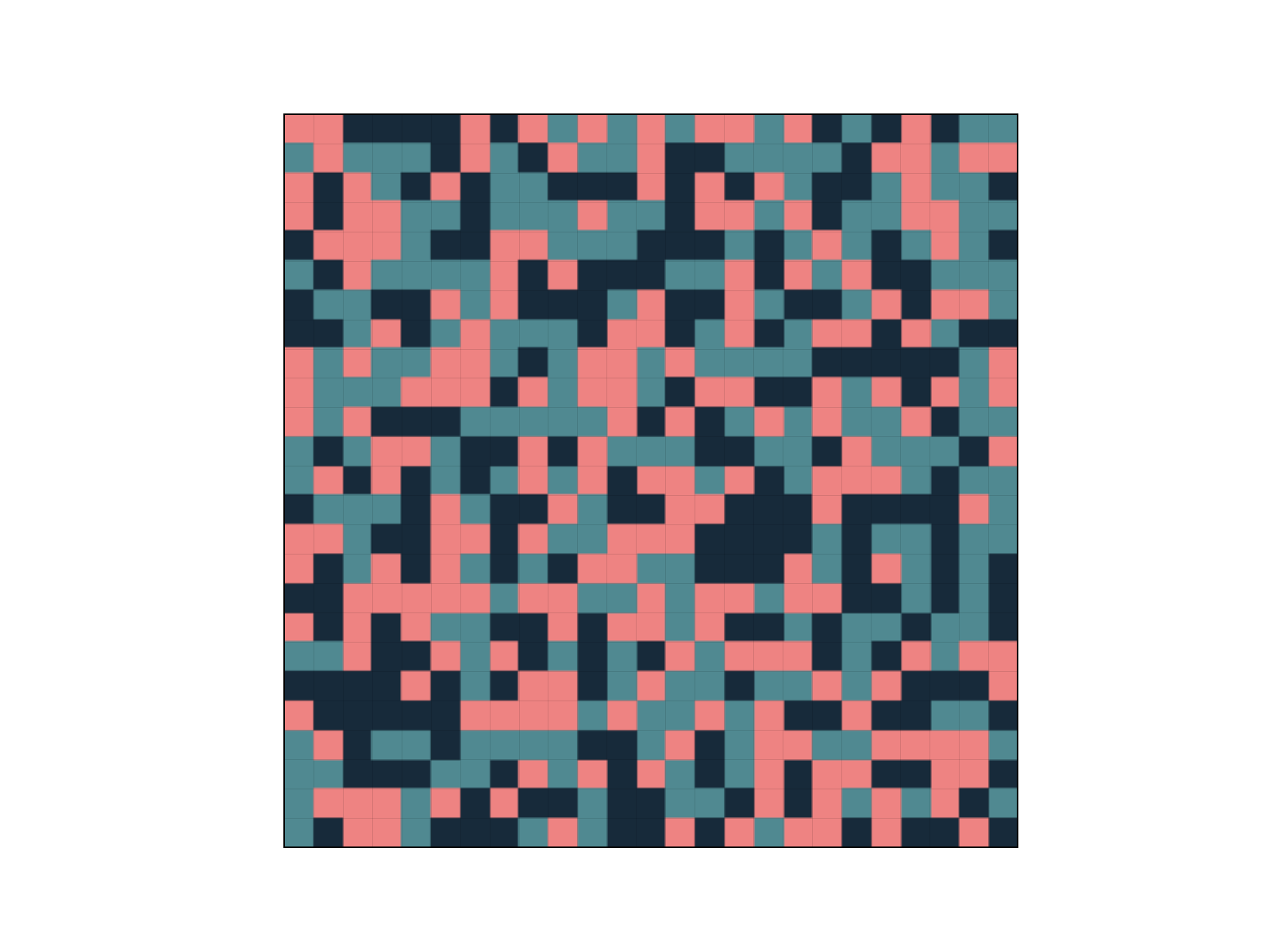} \\
		\includegraphics[width = 0.34\textwidth]{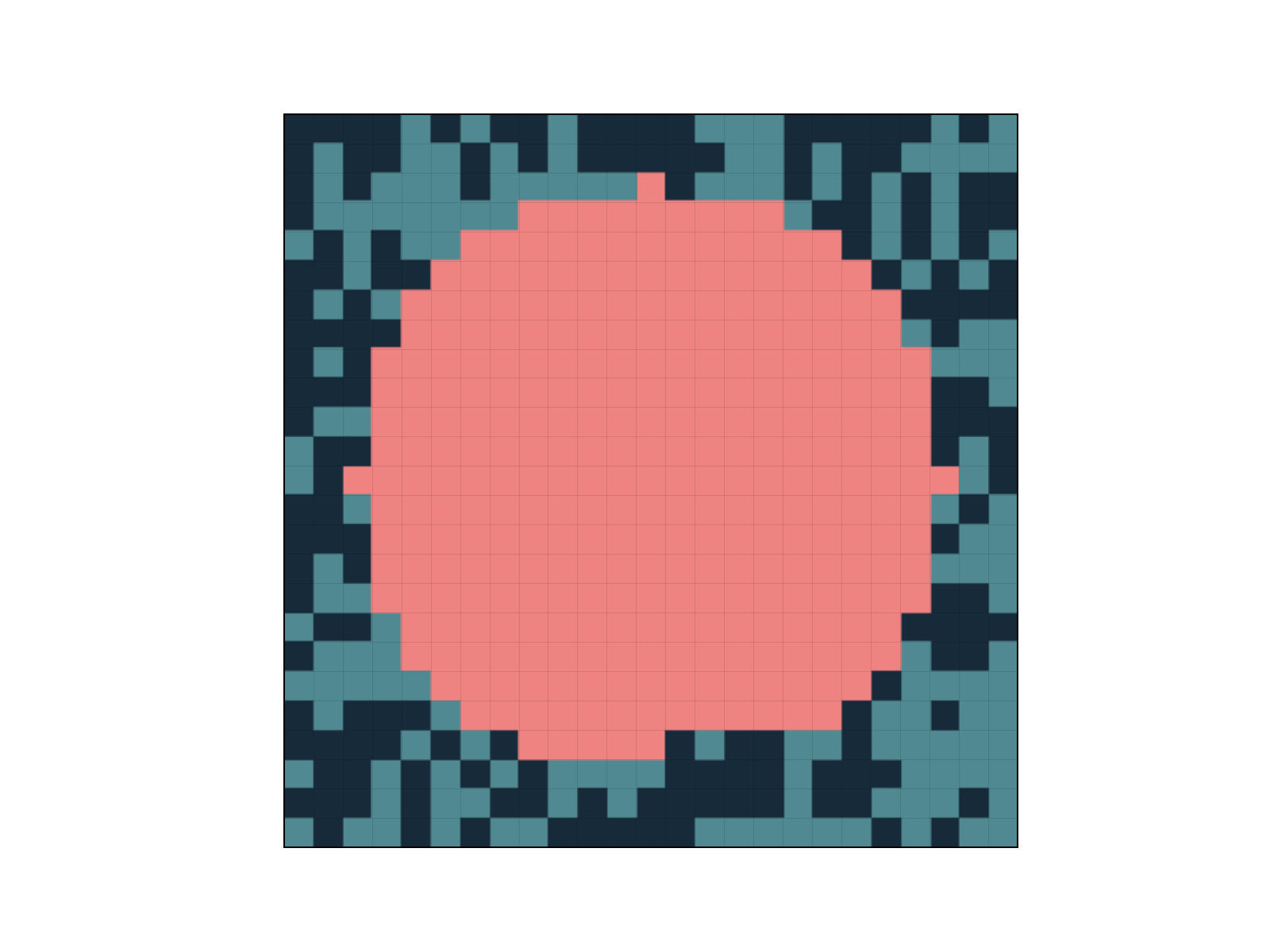} 
		\includegraphics[width = 0.34\textwidth]{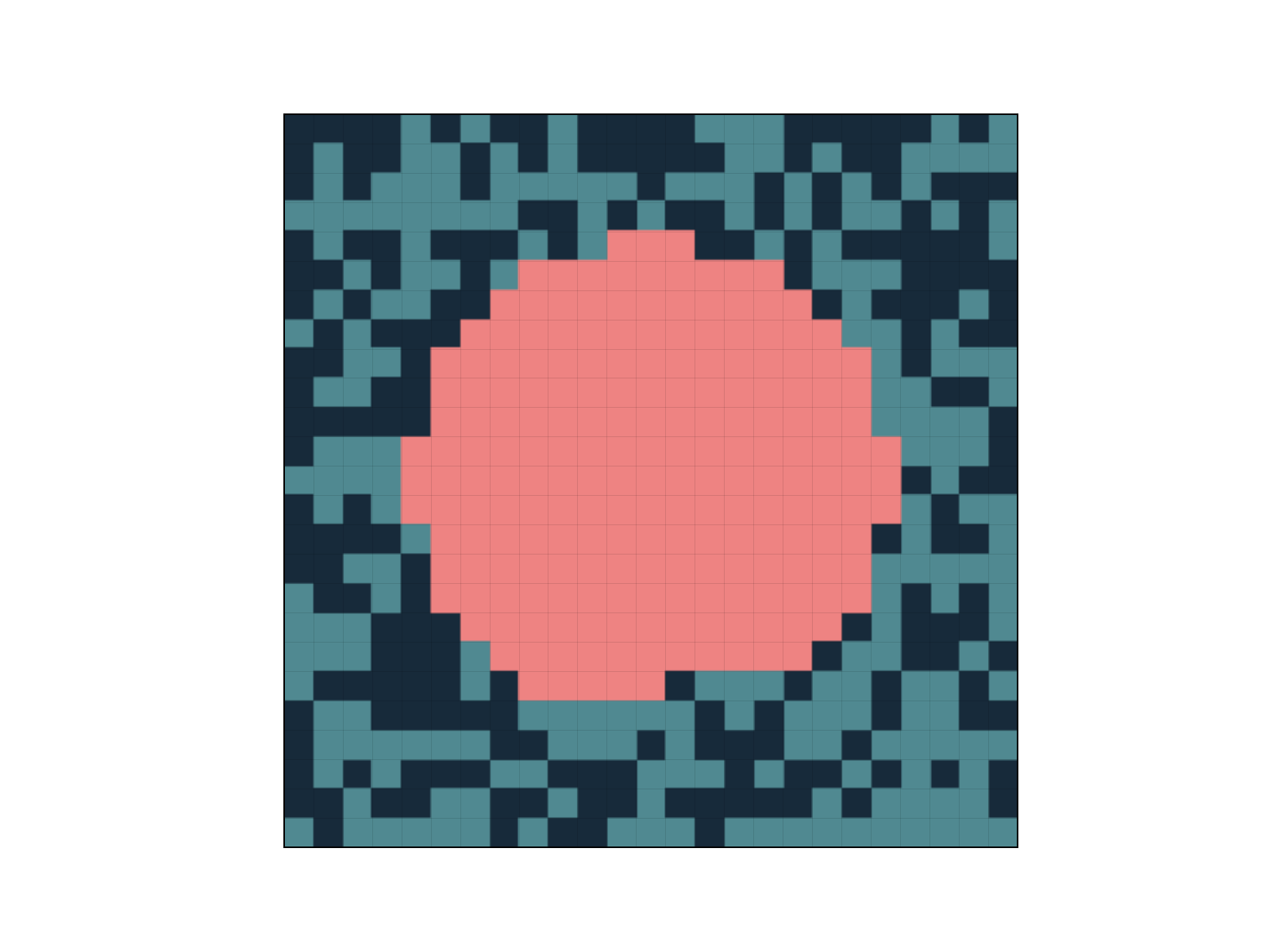}
	\end{center}
	\caption{Initial configurations of the \ssMHE{} model \eqref{SI:stoch_spatial_model} for two different initial fraction covers. Top row: random initial configuration. Bottom row: coral-cluster initial configuration.}
\label{SIfig:initial_profiles}
\end{figure}

We consider two different initial spatial configurations of a reef. 
\begin{enumerate}
    \item The \textit{random} initial configuration assigns a species to node $i$ with a probability according to $C(0)$, $T(0)$, $M(0)$, that is, for example $C_i(0) = 1$ with probability $C(0)$. 
    \item The \textit{coral-cluster} initial configuration assigns all coral nodes (according to $C(0)$) to a single cluster in the centre of the domain, and turf and macroalgae nodes randomly around the coral cluster according to $T(0)$ and $M(0)$.
\end{enumerate}
 For a fixed percentage cover $C(0)$, the two initial configurations achieve approximate minimal and maximal values of the neighbourhood descriptor $C_C$, since they either aim to cluster coral as little or as much as possible. Fig.~\ref{SIfig:initial_profiles} shows instances of the two initial configurations for different values of global fraction covers.
\newpage
\section{Persistent homology}
\label{SIsec:PH}
The \ssMHE{} model simulates spatial data. Standard analyses do not reveal structures such as connected components or enclosed loops within snapshot data. Yet, there is evidence from the contrast of the results obtained from the coral cluster and random initial configurations (Fig.~\ref{fig:fig1} of the main text) that such features may be important. Here we explain in detail the two topological tools we use to quantify the spatial information simulated by the \ssMHE{} model. This section briefly discusses the computation of persistent homology (PH) of a single-time snapshot of the \ssMHE{} model; the zigzag persistence of time-evolving simulations is discussed in the next section.

In Subsection \ref{SIsubsec:cubical}, we review cubical complexes, which are the algebraic objects that we build on the coral spatial data. We construct filtrations on snapshots of the \ssMHE{} model (Subsection \ref{SIsubsec:filtrations}) and then use these filtrations to compute PH barcodes (Subsection \ref{SIsubsec:PH}). We provide necessary details of cubical complexes  \cite{chomp} and standard persistence \cite{ghrist} by Kaczynski, Mischaikow, and Mrozek.

\subsection{Cubical complexes}
\label{SIsubsec:cubical}
The first step in analysing data from a topological perspective is to endow the data with an algebraic structure. A common approach is to approximate the shape and connectivity of the data by building \textit{simplicial complexes}, which represent the data by nodes, edges, and triangles. We do not take this approach. Instead, since coral data is on a grid, we use \textit{cubical complexes} \cite{chomp}, which represent the structure of coral data more faithfully than simplicial complexes would. 
\begin{definition}[Elementary interval] 
An elementary interval is a closed subset $E \subset \R$ of the form $E = [e,e+1]$ or $E = [e,e]$ for some integer $e \in \Z$. Depending on the form, intervals are called non-degenerate or degenerate, respectively.
\end{definition}
Elementary intervals are the building blocks of our algebraic structure and are used to define elementary cubes. Elementary cubes are analogous to simplices in TDA. 
\begin{definition}[Elementary cube] An elementary cube $q$ is a finite product of elementary intervals $q = E_1 \times E_2 \times \cdots \times E_{\iota} \subset \R^{\iota}$, where some or all of the elementary intervals may be degenerate. The value $\iota$ is sometimes referred to as the \textit{embedding number} of the elementary cube.
\end{definition}

\begin{figure}[tbh]
\textbf{A.} \hfill
\begin{center}
    
\includegraphics[width=0.9\textwidth]{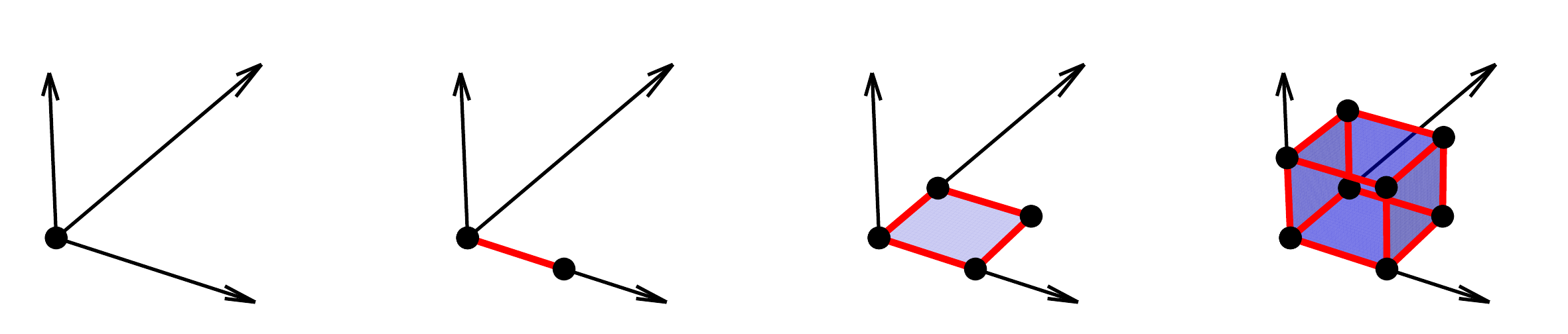}
\end{center}
    
\textbf{B.} \hfill
\begin{center}
          \includegraphics[width=0.6\textwidth]{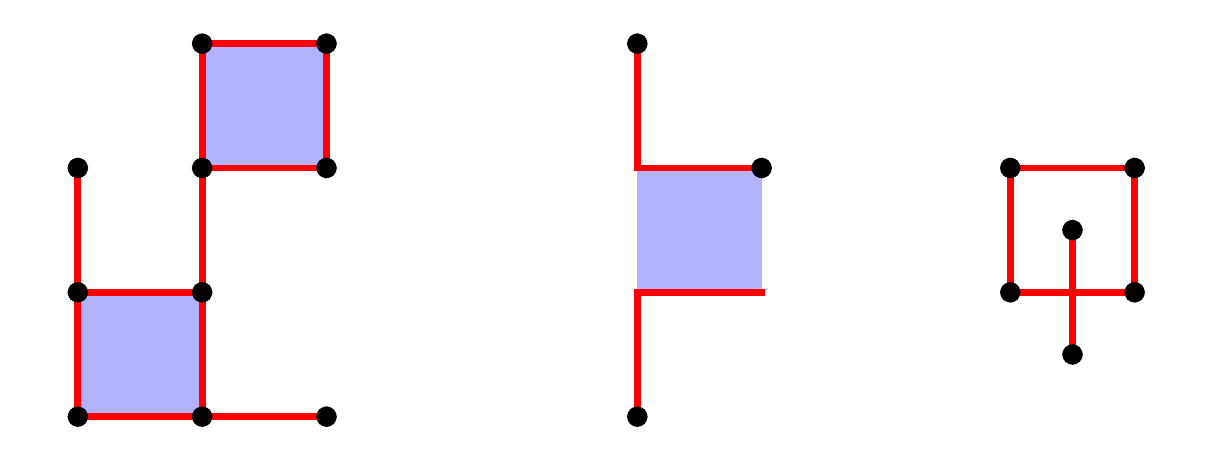}

    \end{center}
     \caption{Definition of cubical complexes. \textbf{A.} Elementary cubes, from left to right: a vertex, an edge, a square, and a cube. \textbf{B.} An example and two non-examples of cubical complexes. Observe that the structure in the middle is missing $0$ and $1$-dimensional elementary cubes on the boundary of the central $2$ dimensional elementary cube, which violates the first condition of Definition \ref{SIdef:cubcomp}. The structure on the right has an intersection of two $1$-dimensional elementary cubes that is not an elementary cube itself, violating the second condition.}
        \label{SIfig:cubicalDef}
\end{figure}

\begin{ex}
Three elementary cubes are given in Fig.~\ref{SIfig:cubicalDef}A. All three have an embedding number $\iota=3$ since they may be viewed as subsets of $\R^3$. Elementary cubes where each elementary interval is degenerate are called `vertices'. Elementary cubes where one, two, and three of its elementary intervals are non-degenerate will be called `edges', `squares', and `cubes', respectively. 
\end{ex}

A cubical complex is a collection of elementary cubes satisfying specific properties. These are analogous to simplicial complexes' usual `intersection' and `inclusion' properties.

\begin{definition}[Cubical complex]
A cubical complex is a collection of elementary cubes $\mathcal{Q} = \{ q_j \}_{j \in J}$ (where $J$ is some finite indexing set) satisfying the following two conditions:
\begin{enumerate}
    \item For each elementary cube $q \in \mathcal{Q}$, any elementary cube $\tilde{q}$ with $\tilde{q} \subseteq q$ must also be an elementary cube in $\mathcal{Q}$. 
    \item For any two elementary cubes $q, q' \in \mathcal{Q}$, the intersection $q \cap q'$ must be an elementary cube in $\mathcal{Q}$.
\end{enumerate}
\label{SIdef:cubcomp}
\end{definition}
\begin{ex}
Figure \ref{SIfig:cubicalDef}B (left) shows an example of a cubical complex. Two examples that fail each of the cubical complex conditions in Definition~\ref{SIdef:cubcomp} are shown in Figure \ref{SIfig:cubicalDef}B (middle and right).
\end{ex}

\clearpage

\subsection{Filtrations}
\label{SIsubsec:filtrations}
Here we build a filtration, a sequence of nested cubical complexes, on a fixed-time snapshot of the \ssMHE{} model.

On snapshot data of the \ssMHE{} model, we define a function $f:\mathcal{I} \rightarrow N$, where $\mathcal{I}$ are the nodes $i$ in the reef and $N = \{0, 1, \ldots, 7, 8\}$. If $C_i = 1$, $f(i)$ is the number of direct coral neighbours of node $i$. If $C(i)=0$ then $f(i)$ is automatically zero. 
\begin{ex}
Fig.~\ref{SIfig:coralPH}A gives an example of $f$ applied to a snapshot of the \ssMHE{} model. Each entry in the matrix is the number of direct coral neighbours the corresponding coral node has, with non-coral nodes $i$ having $f(i) = 0$.
\end{ex}
We use $f$ to define a filtration of a snapshot of the \ssMHE{} model.
\begin{definition}[\ssMHE{} snapshot filtration]
Given a snapshot of the \ssMHE{} model at time $t$, the \ssMHE{} snapshot filtration is a nested sequence of cubical complexes $\{K^t_8, K^t_7, \dots, K^t_1\}$, such that: \[K^t_8 \subset K^t_7 \subset \dots \subset K^t_1.\]
\label{SIdef:filtration}
\end{definition}

Given a snapshot of the \ssMHE{} model at time $t$, we build the cubical complex $K_\eta^t$ as follows. For each node $i = (i_1, i_2)$ with $f(i) \geq \eta$ (that is, for each coral node with at least $\eta$ direct coral neighbours) we create a $0$-dimensional cube. We then fill in $1$- and $2$-dimensional cubes according to whether such nodes $i$ are adjacent. Algorithm \ref{SIalg:convert} describes this process.

\begin{algorithm} [h!]

    \caption{Construct a filtration of cubical complexes from a snapshot of the \ssMHE{} model.}
  \begin{algorithmic}[1]
  
    \INPUT Snapshot of the \ssMHE{} model at time $t$
    \OUTPUT Filtration--sequence of nested cubical complexes \{$K_\eta^t$\} for $\eta = 8, 7, \dots, 1$.
    \STATE{Define $f:\mathcal{I} \rightarrow N$ based on the snapshot of the \ssMHE{} model as above}
    \FOR{$\eta = 8, 7, \dots, 1$}
    \FOR{each node $i = (i_1, i_2)$ of the reef in the snapshot}
    \IF{$f(i_1, i_2) \geq \eta$ }
    \STATE{
add the vertex $[i_1,i_1]\times[i_2,i_2]$ to the complex}
\IF{$f(i_1, i_2+1) \geq \eta$ }
    \STATE{
add the interval $[i_1,i_1]\times[i_2,i_2+1]$ to the complex}
     \ENDIF
     \IF{$f(i_1+1,i_2) \geq \eta$ }
    \STATE{
add the interval $[i_1,i_1+1]\times[i_2,i_2]$ to the complex}
\ENDIF
\ENDIF

   \ENDFOR
   
   \FOR{each set of 4 adjacent intervals in the complex}
   \STATE{add the square $[i_1,i_1+1]\times [i_2,i_2+1]$ to the complex}
   \ENDFOR
   
    \STATE{label the resulting cubical complex as $K^t_\eta$, and add it to the sequence}

   \ENDFOR
    
  \end{algorithmic}
      \label{SIalg:convert}

\end{algorithm}

Algorithm \ref{SIalg:convert} adapts the algorithm of Wagner, Chen, and Vu\c{c}ini \cite{vucini} for converting a greyscale image into a cubical complex filtration.
\begin{ex}
An example of Algorithm \ref{SIalg:convert} applied to a snapshot of the \ssMHE{} model is given in Fig.~\ref{SIfig:coralPH}B.
\end{ex}
\newpage
\subsection{Computation of persistent homology of the \ssMHE{} model}
\label{SIsubsec:PH}

To obtain a spatial summary of a snapshot of the \ssMHE{} model, we compute \textit{homology groups} corresponding to cubical complexes in the \ssMHE{} snapshot filtration. Homology groups are defined via chain complexes. 

\begin{definition}[Chain complex]
Given a cubical complex $\mathcal{Q}$ containing elementary cubes of embedding number $2$, define $D_0, D_1$ and $D_2$ as the free Abelian groups generated by the vertices, edges, and squares in $\mathcal{Q}$ respectively. An element of $D_1$, for example, is therefore a sum $\sum n_j^1 \epsilon_j$, where $n^1 \in \N$ and $\epsilon_j$ are edges in $\mathcal{Q}$. We may arrange these groups into a sequence connected by group homomorphisms $\partial_0$, $\partial_1$, $\partial_2$ and $\partial_3$:
\begin{equation}
\begin{tikzcd}
0 \arrow[r, "\partial_3"] & D_2 \arrow[r, "\partial_2"] & D_1 \arrow[r, "\partial_1"] & D_0 \arrow[r, "\partial_0"] & 0.
\end{tikzcd}
\label{SIeq:chaincomplex}
\end{equation}

The zeros in \eqref{SIeq:chaincomplex} represent trivial groups---with the trivial homomorphism connecting them to adjacent groups. The homomorphisms $\partial_1$ and $\partial_2$ are called boundary maps and are defined on edges and squares as follows.
\begin{itemize}
    \item $\epsilon = [v,v+1]\times[w,w] \implies \partial_1(\epsilon) = [v+1,w] - [v,w]$, $\epsilon = [v,v]\times[w,w+1] \implies \partial_1(\epsilon) = [v,w+1] - [v,w]$.
  
    \item $\zeta = [v,v+1]\times[w,w+1] \implies \partial_2(\zeta) = [v+1,v+1] \times [w, w+1] - [v, w] \times [w, w+1] + [v, v+1] \times [w, w] - [v, v+1] \times [w+1, w+1] $.
\end{itemize}
The formulae for individual elementary cubes are then extended linearly to be defined on $D_1$ and $D_2$: $\partial_1(\sum_j n_j^1 \epsilon_j) = \sum_j n_j^1 \partial_1(\epsilon_j)$ for edges $\epsilon_j$ and $\partial_2(\sum_j n_j^2 \zeta_j) = \sum_j n_j^2 \partial_2(\zeta_j)$ for squares $\zeta_j$.
A chain complex is the pair $(D,\partial)$, where $D = \{D_0, D_1, D_2\}$ and $\partial = \{\partial_0, \partial_1, \partial_2\}$ satisfy the condition that the boundary of a boundary is zero (e.g., $\partial_{1} \circ \partial_2 = 0$, see Hatcher \cite{hatcher} for a proof).
\label{SIdef:xcomplex}
\end{definition}
Chain complexes provide an algebraic structure to define homology groups.
\begin{definition}[Zero- and one-dimensional homology groups]
Given a cubical complex $\mathcal{Q}$ and the chain complex $(D,\partial)$ defined in Def.~\ref{SIdef:xcomplex}, the zero- and one-dimensional homology groups are defined as the quotient groups:
\begin{align*}
    H_0(\mathcal{Q}) := \bigslant{\Ker{\partial_0}}{\Image{\partial_1}}, \qquad
    H_1(\mathcal{Q}) := \bigslant{\Ker{\partial_1}}{\Image{\partial_2}}.
\end{align*}
\label{SIdef:homgroups}
\end{definition}

\begin{figure}
\hspace{10mm} \textbf{A. Direct coral neighbours filtration function} \hfill
    \begin{center}
    
    \includegraphics[valign = c, width = 0.3\textwidth]{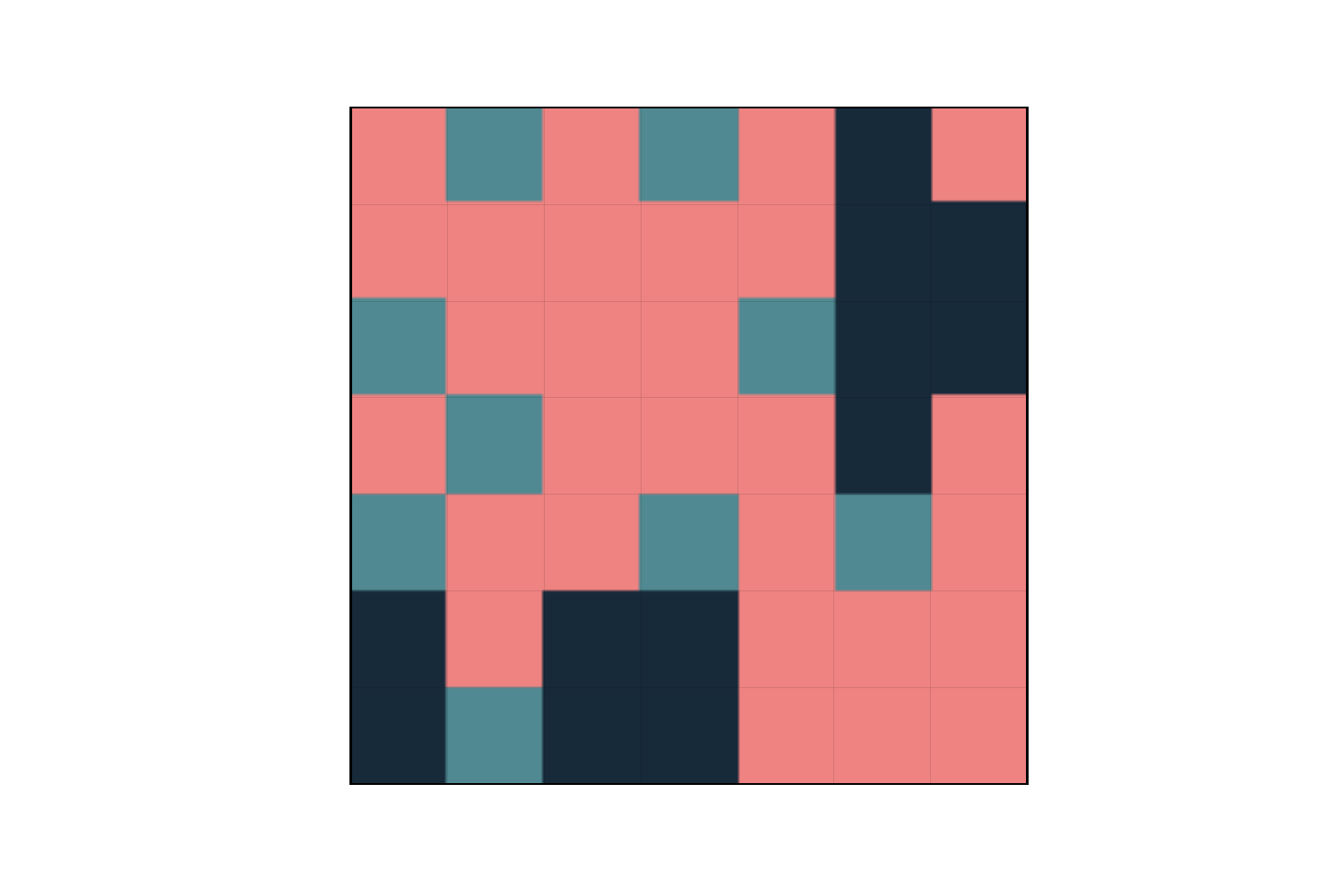}
    $
    \begin{pmatrix}
    0 & 4 & 5 & 5 & 3 & 0 & 0\\
    3 & 6 & 8 & 7 & 4 & 0 & 0\\
    0 & 6 & 7 & 7 & 0 & 0 & 0\\
    2 & 0 & 6 & 6 & 3 & 0 & 1\\
    0 & 4 & 4 & 0 & 4 & 0 & 3\\
    0 & 2 & 0 & 0 & 4 & 7 & 4\\
    0 & 0 & 0 & 0 & 3 & 5 & 3
    \end{pmatrix}
    $
    \end{center} 
    \begin{center}
       \pmb{ \Bigg\downarrow}
    \end{center}
    
\vspace{-2mm}  
\hspace{10mm} \textbf{B. \ssMHE{} snapshot filtration} \hfill
    \begin{center}
    \begin{tabular}{cccc}
    \hline
    \multicolumn{1}{|c|}{\includegraphics[width = 0.15\textwidth]{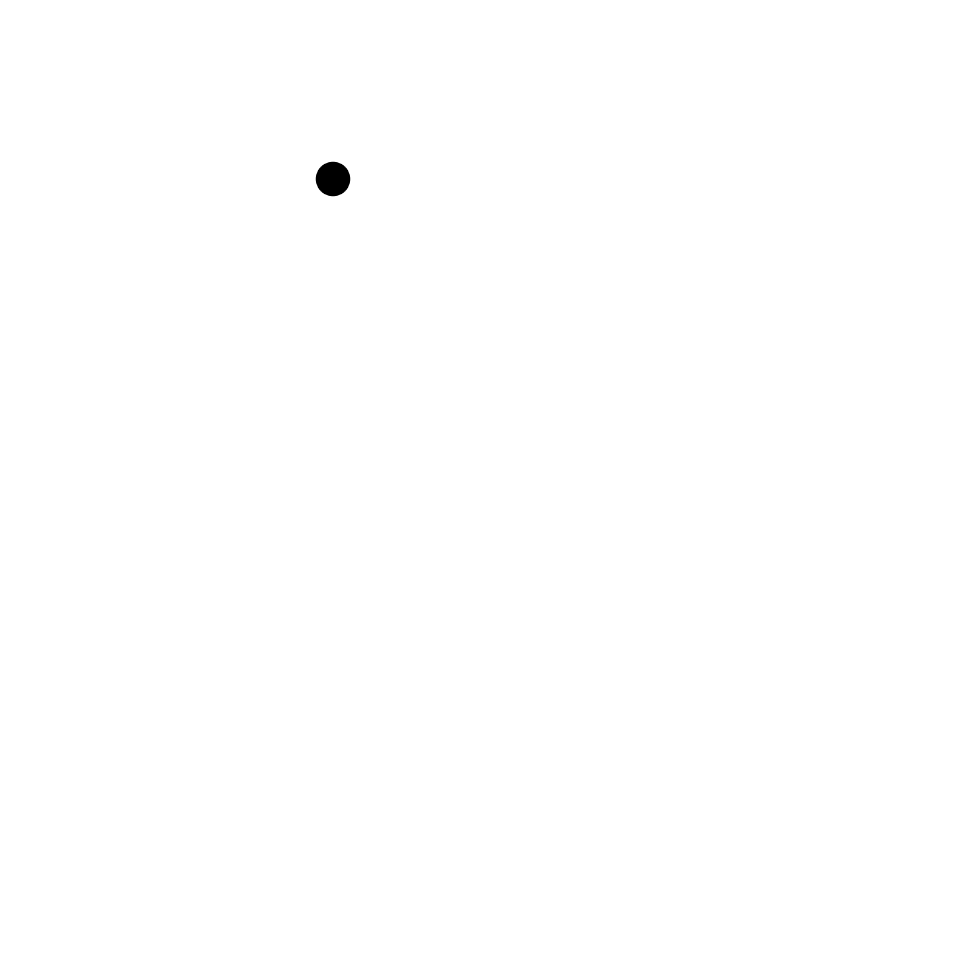}} &
    \multicolumn{1}{|c|}{\includegraphics[width = 0.15\textwidth]{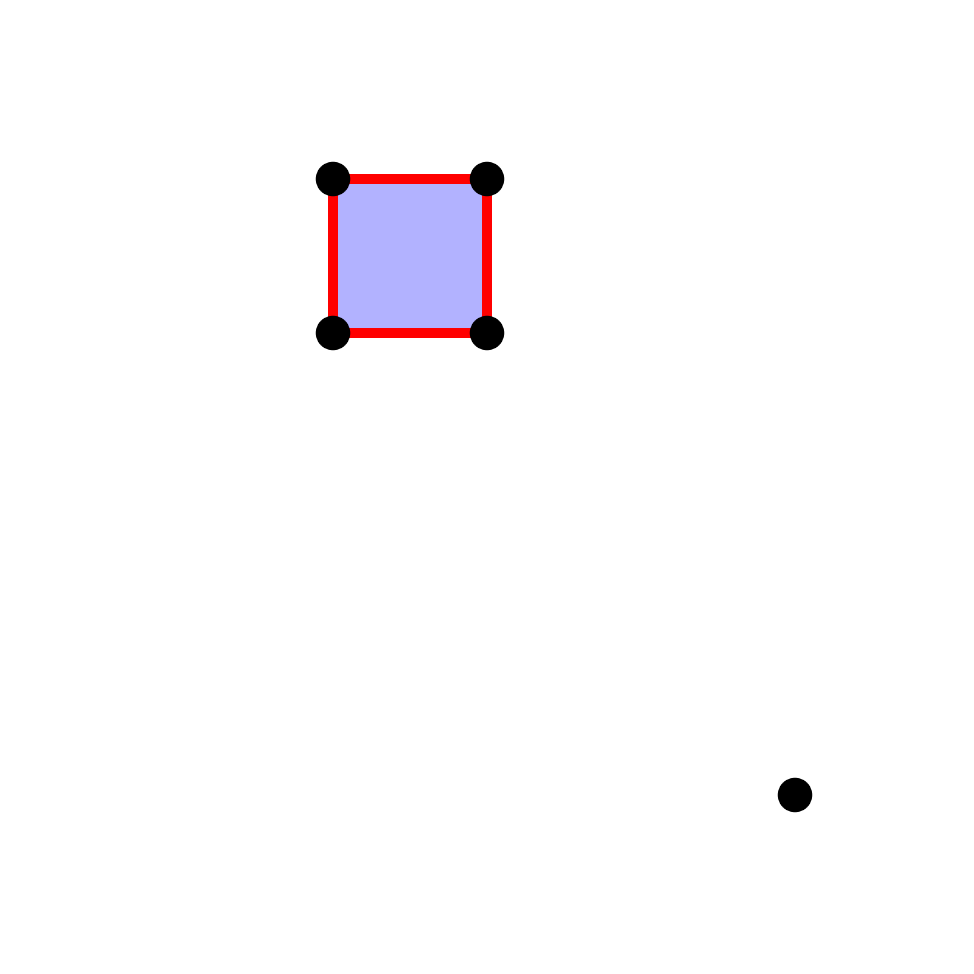}} &
    \multicolumn{1}{|c|}{\includegraphics[width = 0.15\textwidth]{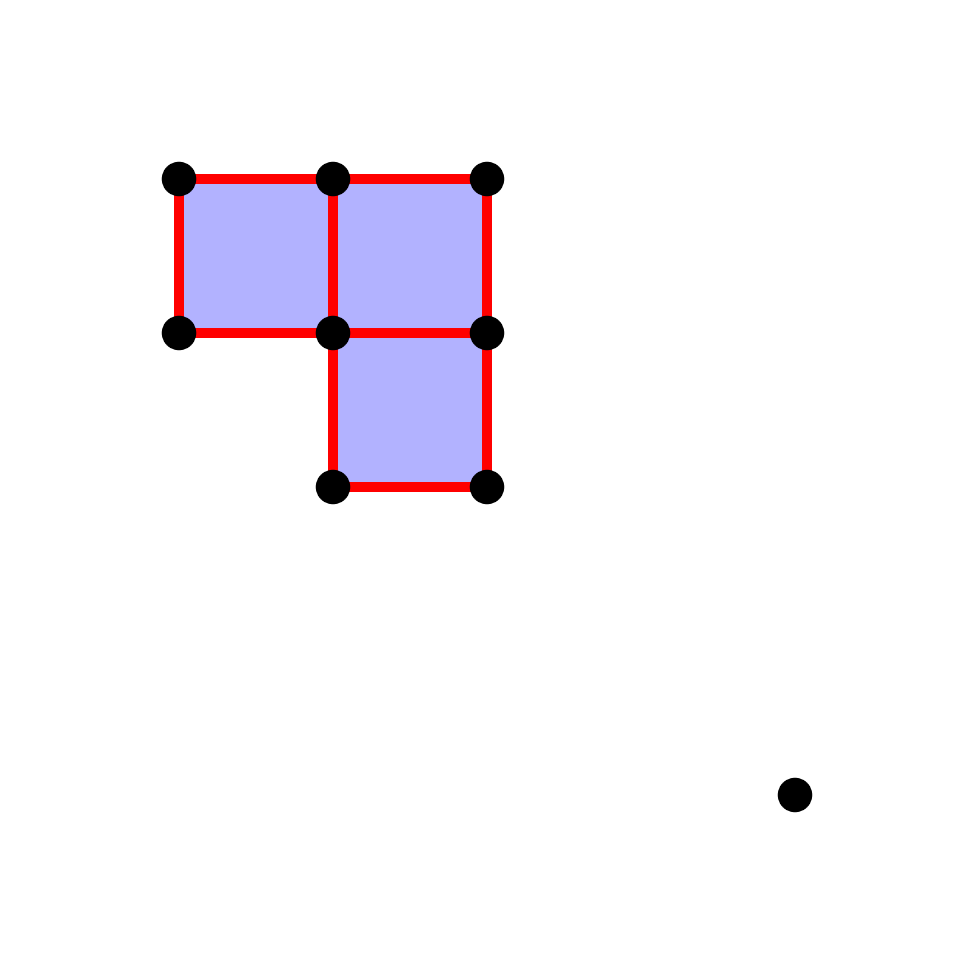}} & \multicolumn{1}{c|}{\includegraphics[width = 0.15\textwidth]{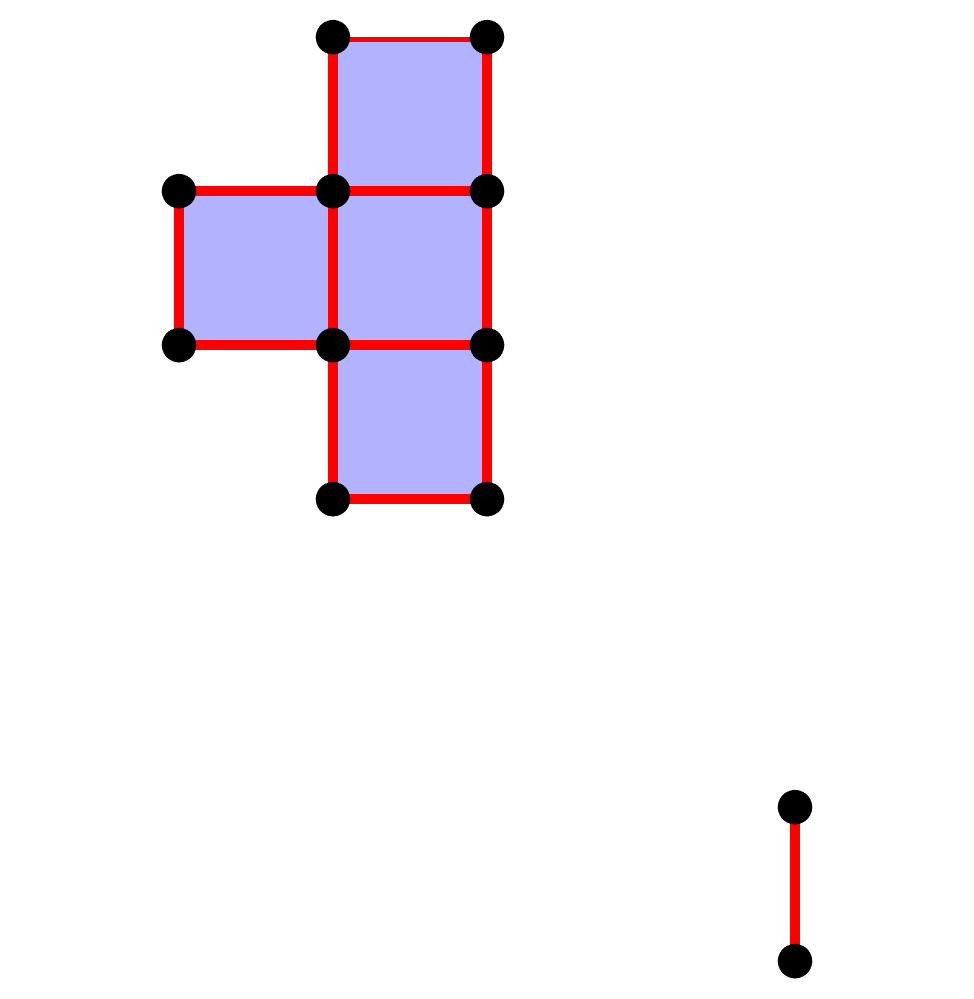}} \\ \hline
    $K_8^t$ & $K_7^t$ & $K_6^t$ & $K_5^t$ \\ \\ \\
    \hline
    \multicolumn{1}{|c|}{\includegraphics[width = 0.15\textwidth]{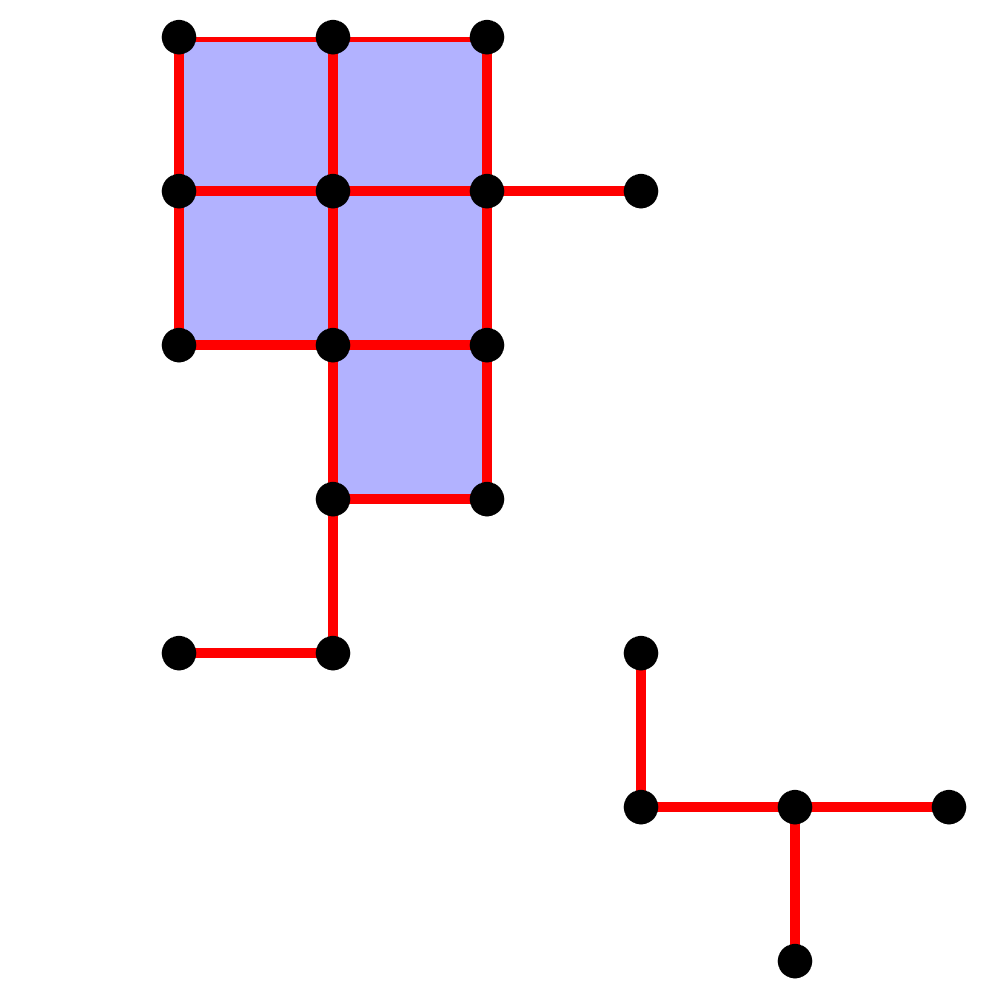}} & \multicolumn{1}{c|}{\includegraphics[width = 0.15\textwidth]{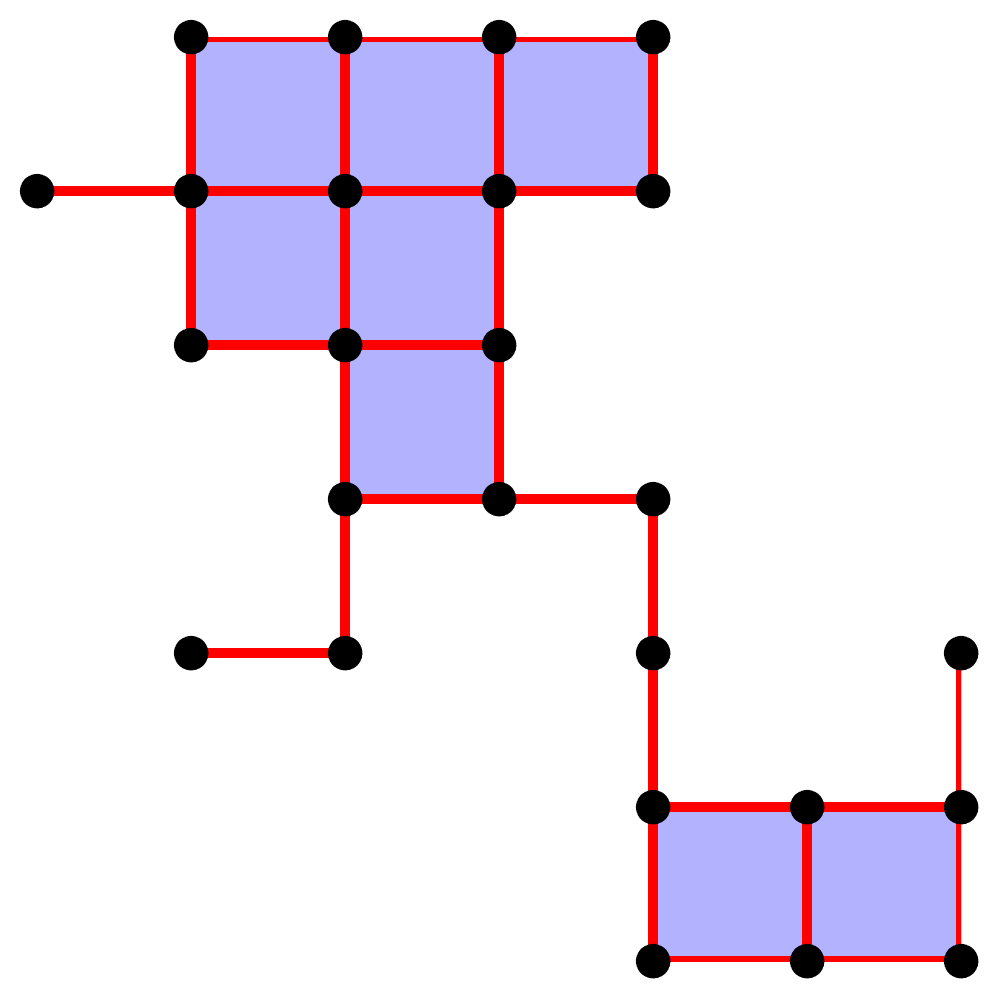}} & \multicolumn{1}{c|}{\includegraphics[width = 0.15\textwidth]{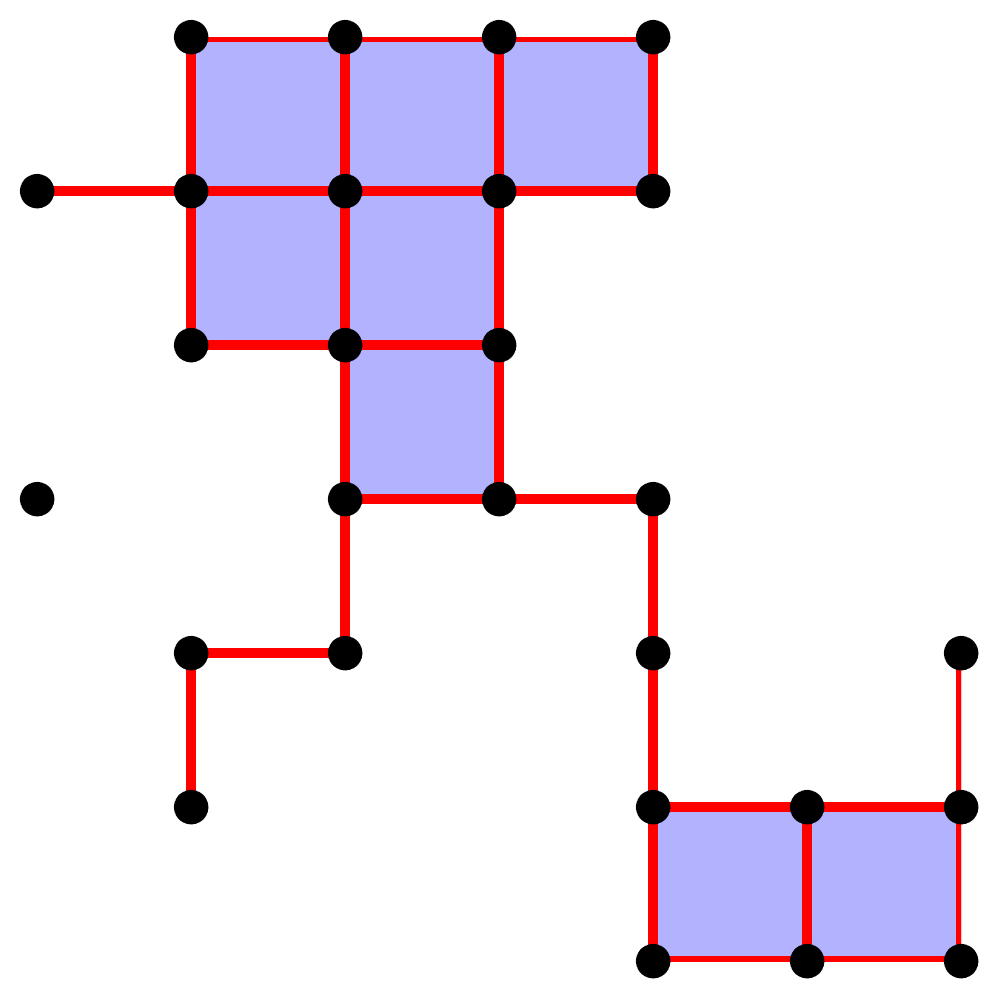}} & \multicolumn{1}{c|}{\includegraphics[width = 0.15\textwidth]{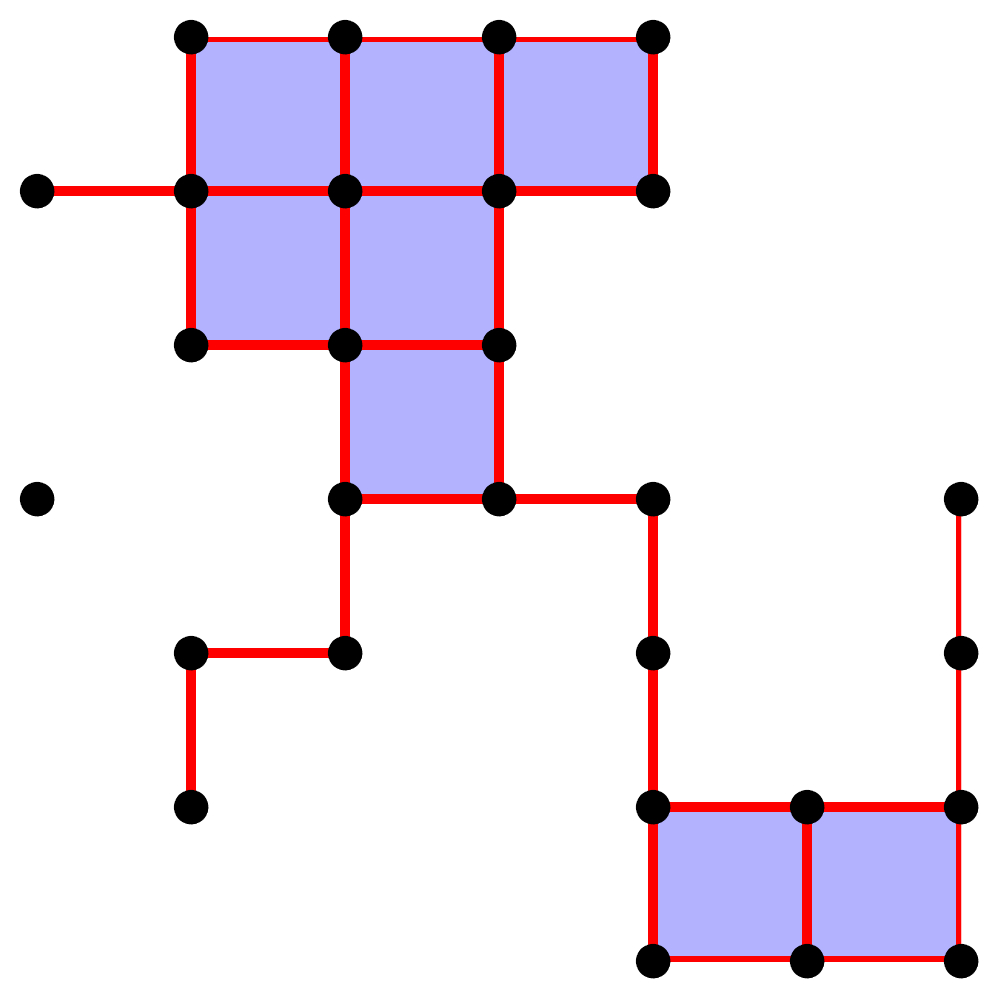}} \\ \hline
   $K_4^t$ & $K_3^t$ & $K_2^t$ & $K_1^t$
    \end{tabular}\  
        \end{center} 
          \begin{center}
       \pmb{ \Bigg\downarrow}
    \end{center}
\vspace{-2mm}
\hspace{10mm} 
\textbf{C. Persistence barcode} \hfill
\begin{center}
         \includegraphics[width = 0.3\textwidth]{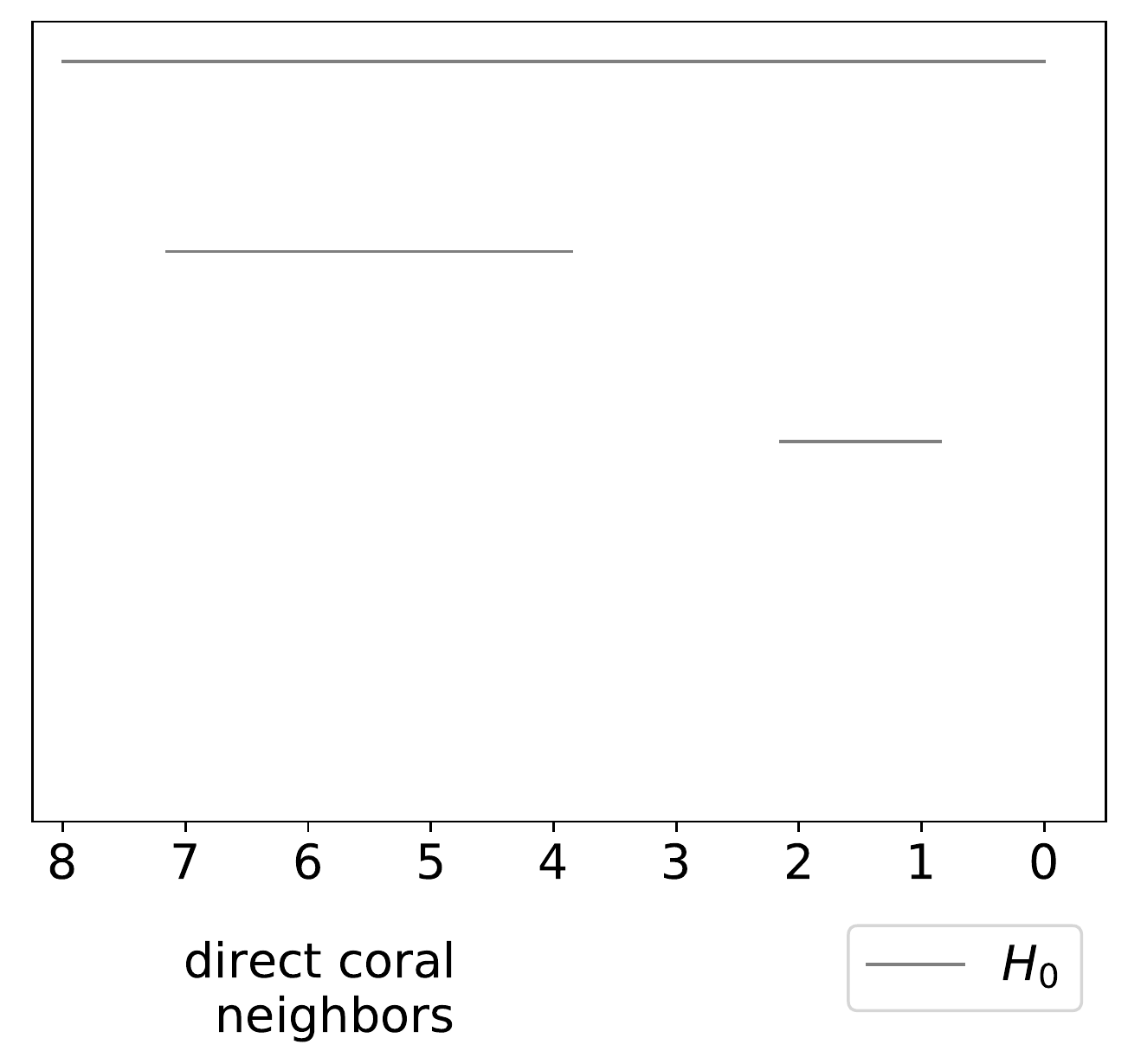}

\end{center}
\caption{The persistent homology pipeline used to find topological information about snapshots of the \ssMHE{} model. \textbf{A. Direct coral neighbours filtration function} Left: a snapshot of the \ssMHE{} model; right: a matrix representing the function $f$, whose values are the number of direct coral neighbours each coral node has. \textbf{B. \ssMHE{} snapshot filtration} Filtration of the snapshot of the \ssMHE{} model in \textbf{A}. Vertices are included wherever there are coral nodes with at least $\eta$ neighbours (for $\eta=8, 7, \dots, 1$), and intervals and squares are added between adjacent edges. \textbf{C. Persistence barcode} The persistence barcode generated by computing zero- and one-dimensional homology of the filtration in \textbf{B}. One long solid bar represents the large cluster of coral, and two shorter bars represent the two small components visible in \textbf{A}.} 
\label{SIfig:coralPH}
\end{figure}

Homology groups encode spatial information about the cubical complex $\mathcal{Q}$ and, in turn, the data from which $\mathcal{Q}$ was generated.
\begin{theorem}[Theorem 2.59 in \cite{chomp}]
The rank of the zeroth homology group $H_0(\mathcal{Q})$ gives the number of connected components in the cubical complex $\mathcal{Q}$.
\label{SIthm:hzero}
\end{theorem}
\begin{theorem}[Section 1.1 in \cite{chomp}]
The rank of the first homology group $H_1(\mathcal{Q})$ is equal to the number of enclosed regions or loops in the cubical complex $\mathcal{Q}$.
\label{SIthm:hone}
\end{theorem}
\begin{definition}[Betti numbers]
The ranks of the zeroth and first homology groups, $H_0(\mathcal{Q})$ and $H_1(\mathcal{Q})$, are called the zeroth and first Betti numbers, $\beta_0$ and $\beta_1$, respectively.
\label{SIdef:bettis}
\end{definition}
Homology groups can be computed algorithmically from data. Recall that we represent snapshot of the data simulated from the \ssMHE{} model by the \ssMHE{} snapshot filtration (\eqref{SIdef:filtration}). This filtration may be restated as a sequence of cubical complexes connected by inclusion maps.
\begin{equation}
    K^t_8 \longrightarrow K^t_7 \longrightarrow \dots \longrightarrow K^t_2\longrightarrow K^t_1.
    \label{SIeq:filtrationinclusion}
\end{equation}
We now compute the homology of each of these complexes and exploit the fact that homology is \textit{functorial}. When homology is taken of the sequence \eqref{SIeq:filtrationinclusion}, the inclusion maps \textit{descend} to give inclusion maps between homology groups.
\begin{equation}
    H_0(K^t_8) \longrightarrow H_0(K^t_7) \longrightarrow \dots \longrightarrow H_0(K^t_2) \longrightarrow H_0(K^t_1).
    \label{SIeq:persmod}
 \end{equation}
While $H_0$ measures the number of connected components in a single complex, the sequence \eqref{SIeq:persmod} tracks the number of connected components present at each filtration value $\eta$. Consider a feature that contributes to $\beta_0$--the rank of a zeroth homology group. The filtration value which corresponds to the first cubical complex in \eqref{SIeq:filtrationinclusion} at which the feature appears is called its \textit{birth}, and the value at which it disappears is its \textit{death}. The difference between death and birth times of a feature is called its \textit{persistence}. If a feature persists through many stages of \eqref{SIeq:persmod} then it represents a large component of coral in the snapshot. Those features with shorter persistences correspond to smaller components of coral.  The persistence of topological features is visualised through a \textit{barcode}. A barcode is a multiset of intervals containing the birth-death pairs of each feature (connected component or loop) that appears in \eqref{SIeq:persmod}. Intervals are plotted as a horizontal bar chart, where each bar represents a different topological feature.
\begin{ex}
Fig.~\ref{SIfig:coralPH}C gives a barcode of the \ssMHE{} model snapshot in Fig.~\ref{SIfig:coralPH}A.
\end{ex}

Persistent homology computed from the \ssMHE{} snapshot filtration may be used to describe the topology of a single snapshot of the \ssMHE{} model. We use the number of direct coral neighbours as the filtration parameter since this value is proportional to the size of the corresponding connected component. Other applications, such as \cite{crockerplots}, use Euclidean distance as a filtration parameter, which generates the more standard Vietoris--Rips complexes \cite{roadmap}.
One quantity we would not be able to use as a filtration parameter is time. As time increases, we cannot guarantee that cubical complexes generated from consecutive snapshots of the \ssMHE{} model will be nested. We can not, in general, define inclusion maps such as \eqref{SIeq:filtrationinclusion} between consecutive snapshots of the \ssMHE{} model.

\begin{figure}[htbp]
\begin{center}
{\renewcommand{\arraystretch}{0.1}	
\begin{tabular}{p{0.3\textwidth}p{0.3\textwidth}p{0.3\textwidth}}
\begin{center}
$t=0$\\
\includegraphics[width = 0.175\textwidth]{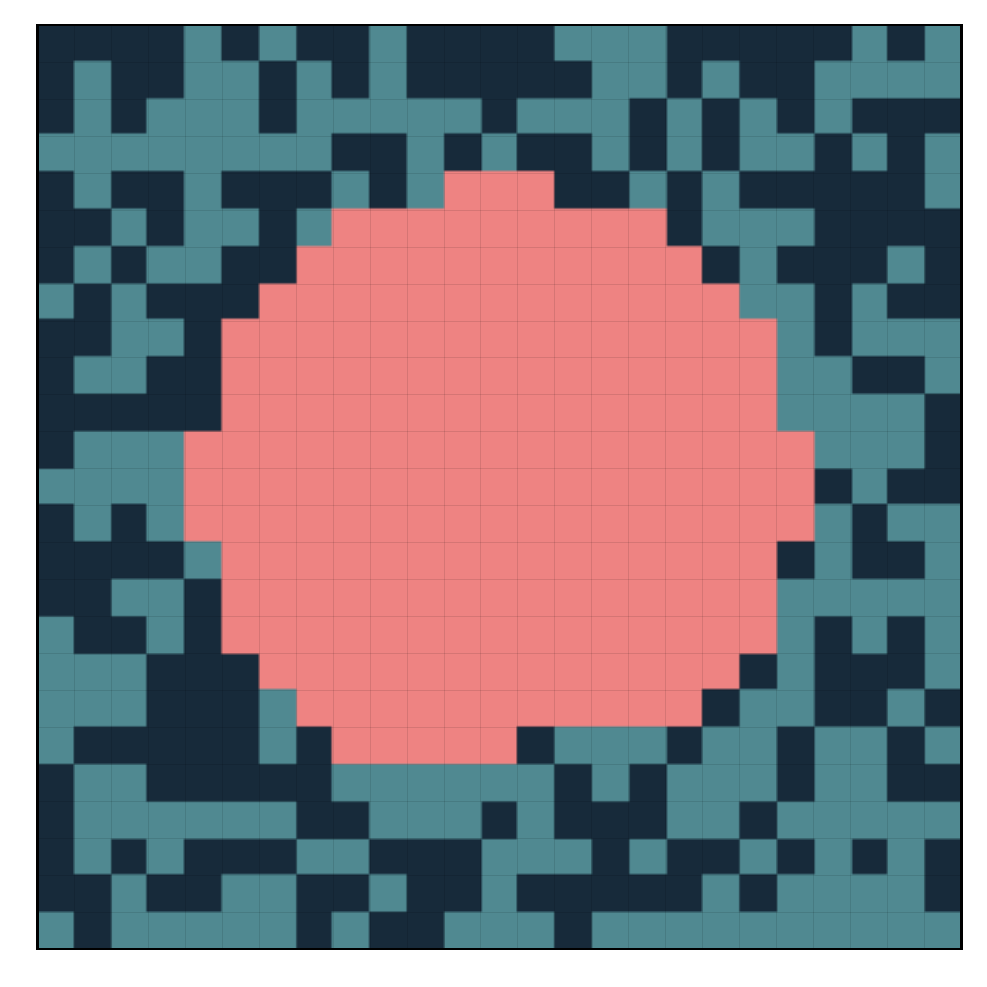}\\
\includegraphics[width = 0.175\textwidth]{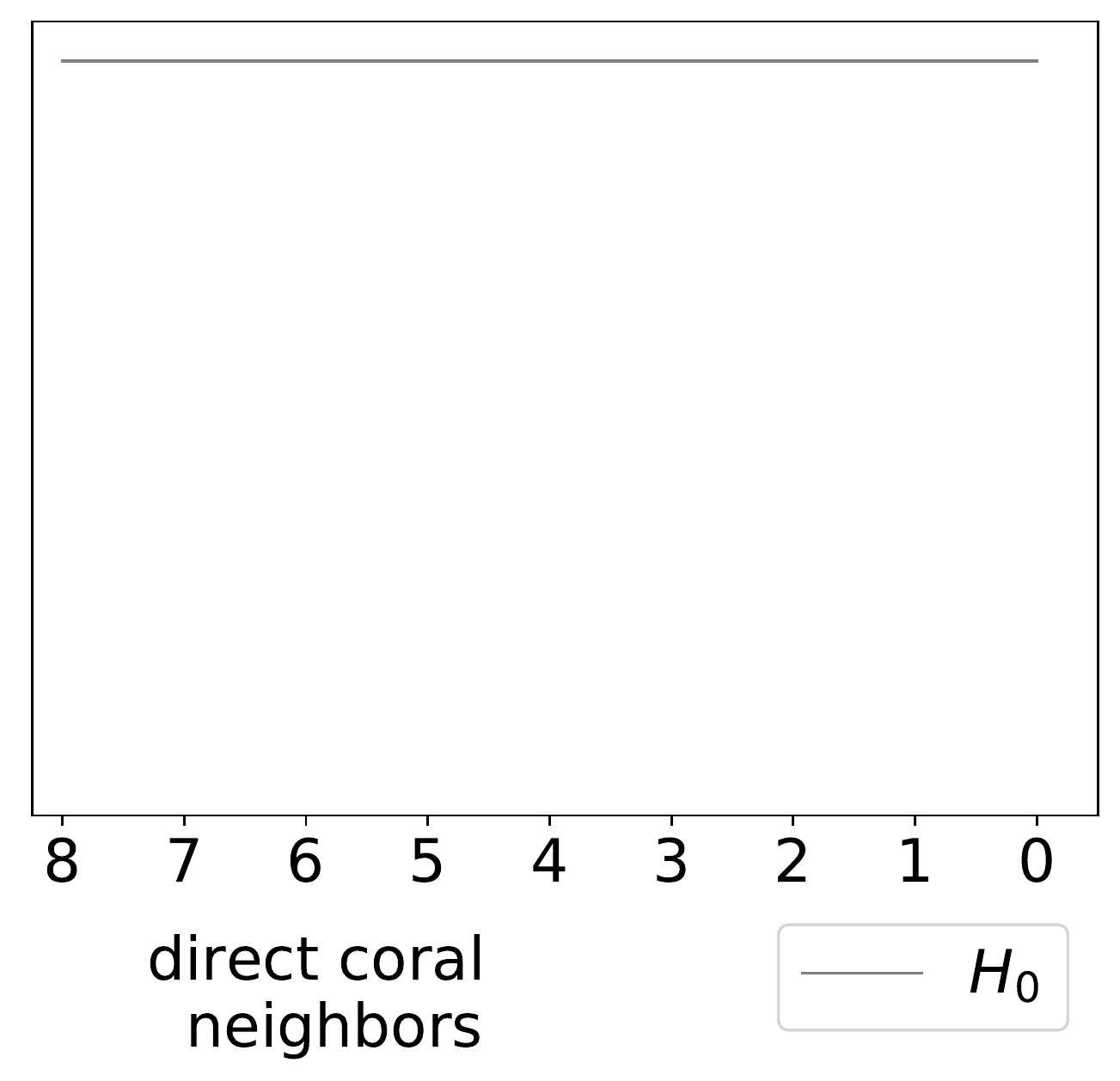}\\
\end{center} 
& \begin{center}
$t=20$\\
\includegraphics[width = 0.175\textwidth]{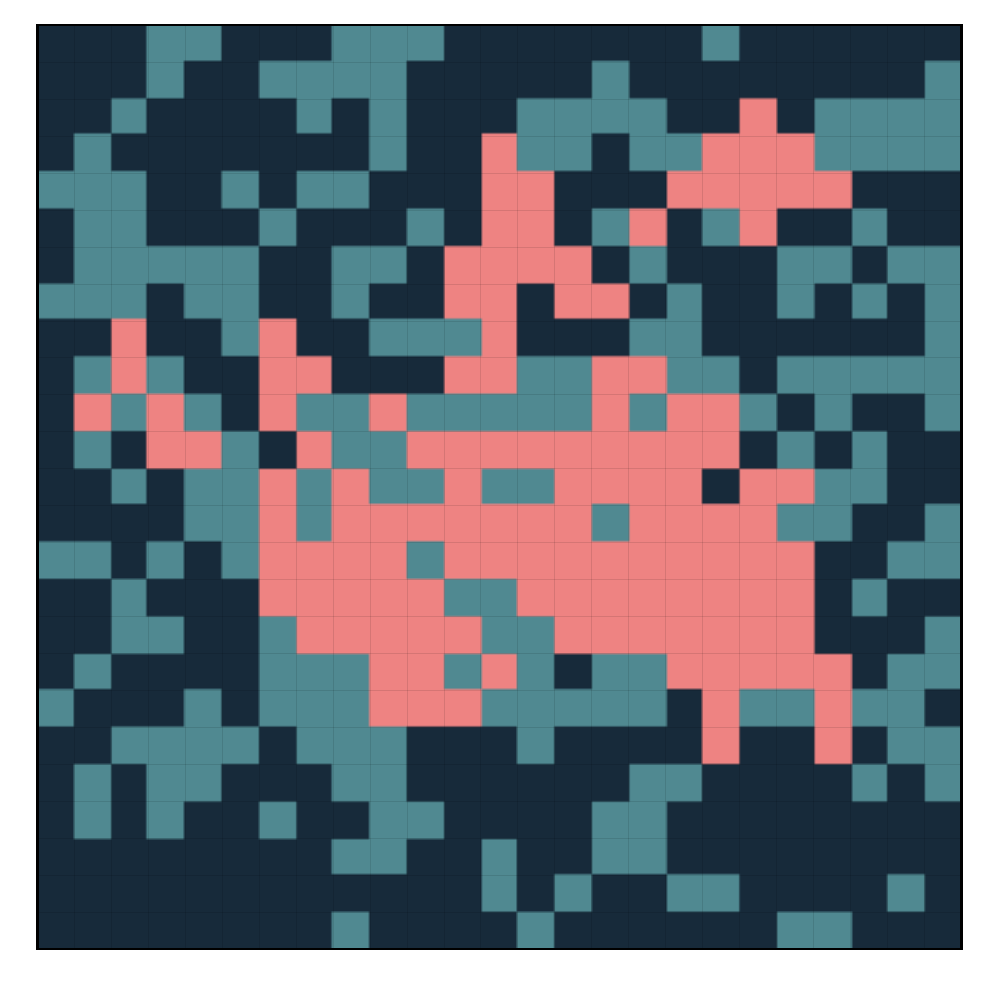} \\
\includegraphics[width = 0.175\textwidth]{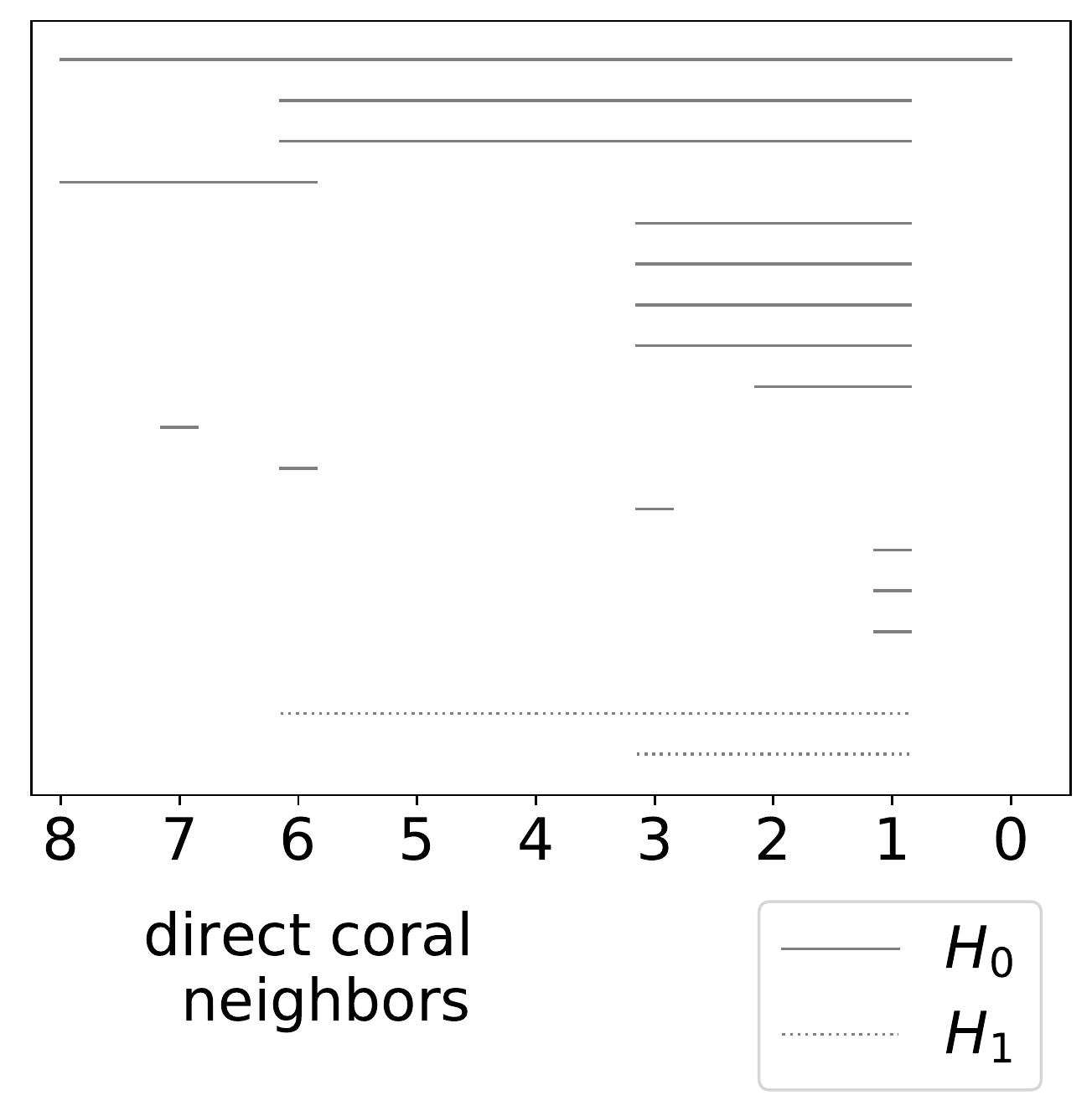}
\end{center}
& \begin{center}
$t=40$\\
\includegraphics[width = 0.175\textwidth]{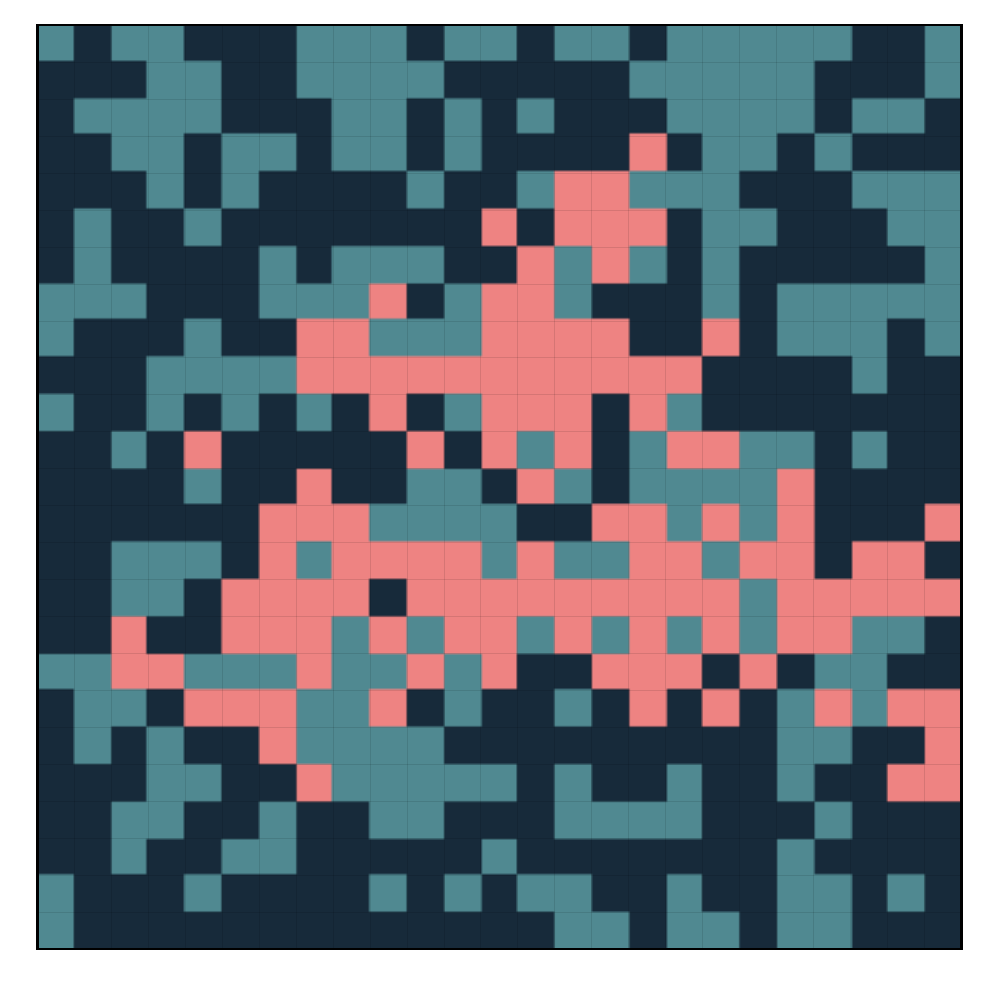}\\
\includegraphics[width = 0.175\textwidth]{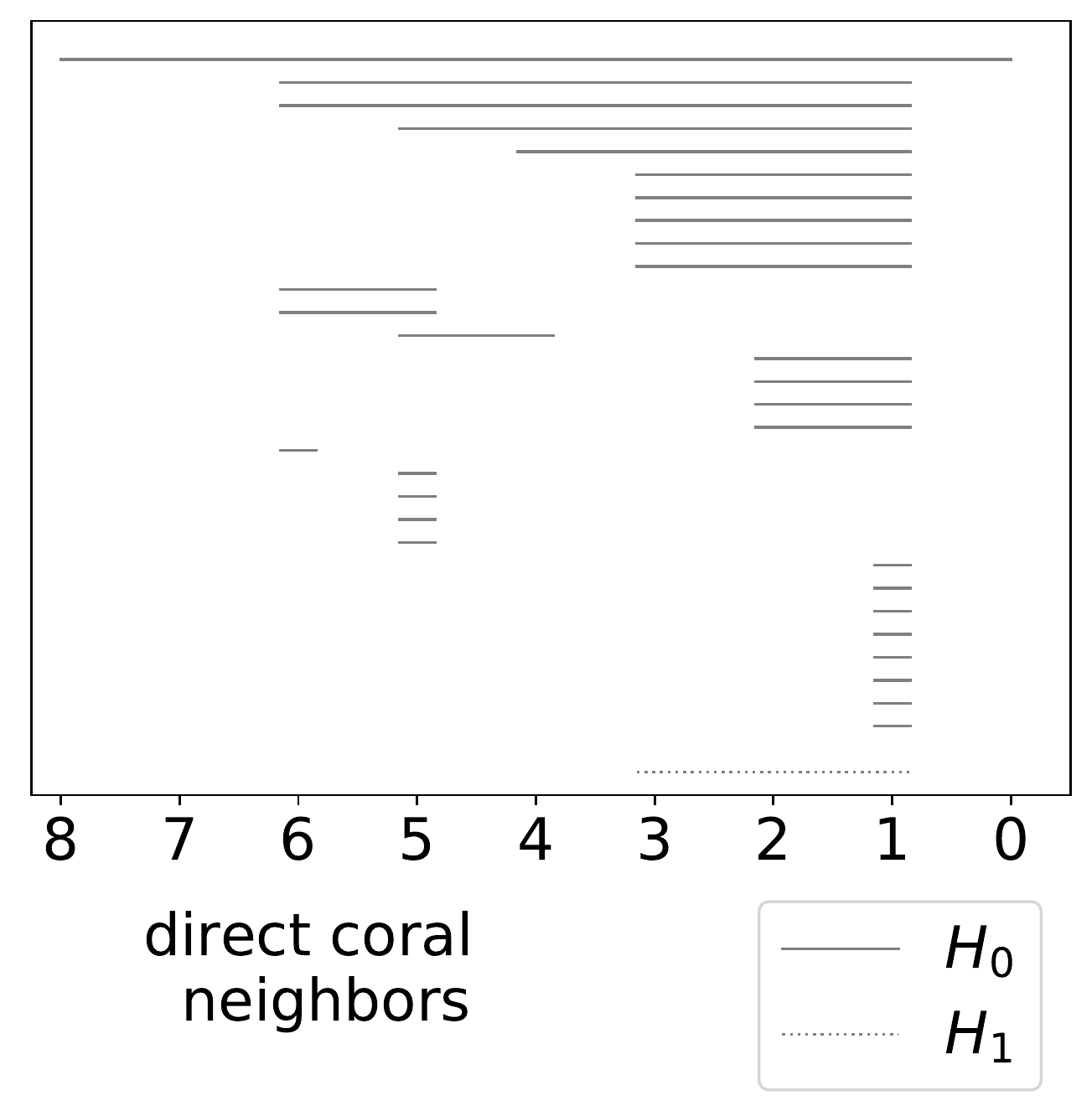}
\end{center} 
\\
\begin{center}
$t=60$\\
\includegraphics[width = 0.175\textwidth]{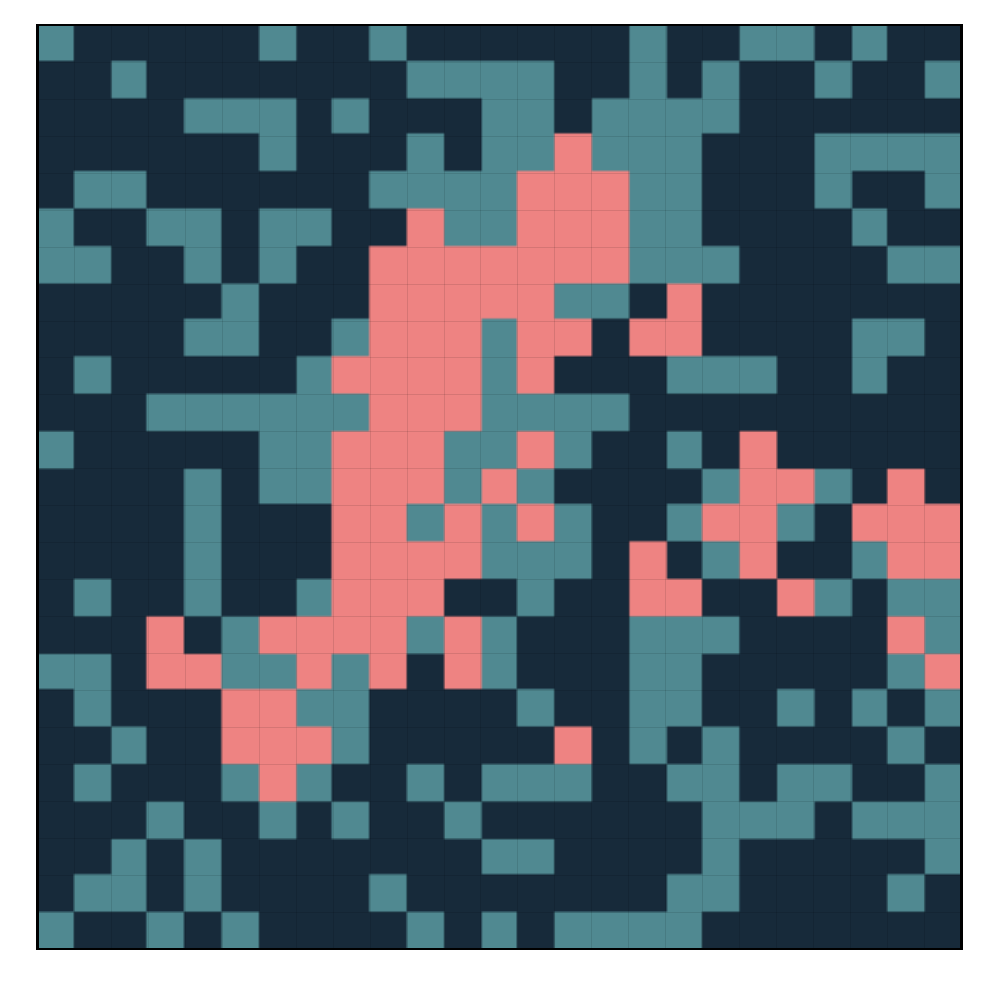}\\
\includegraphics[width = 0.175\textwidth]{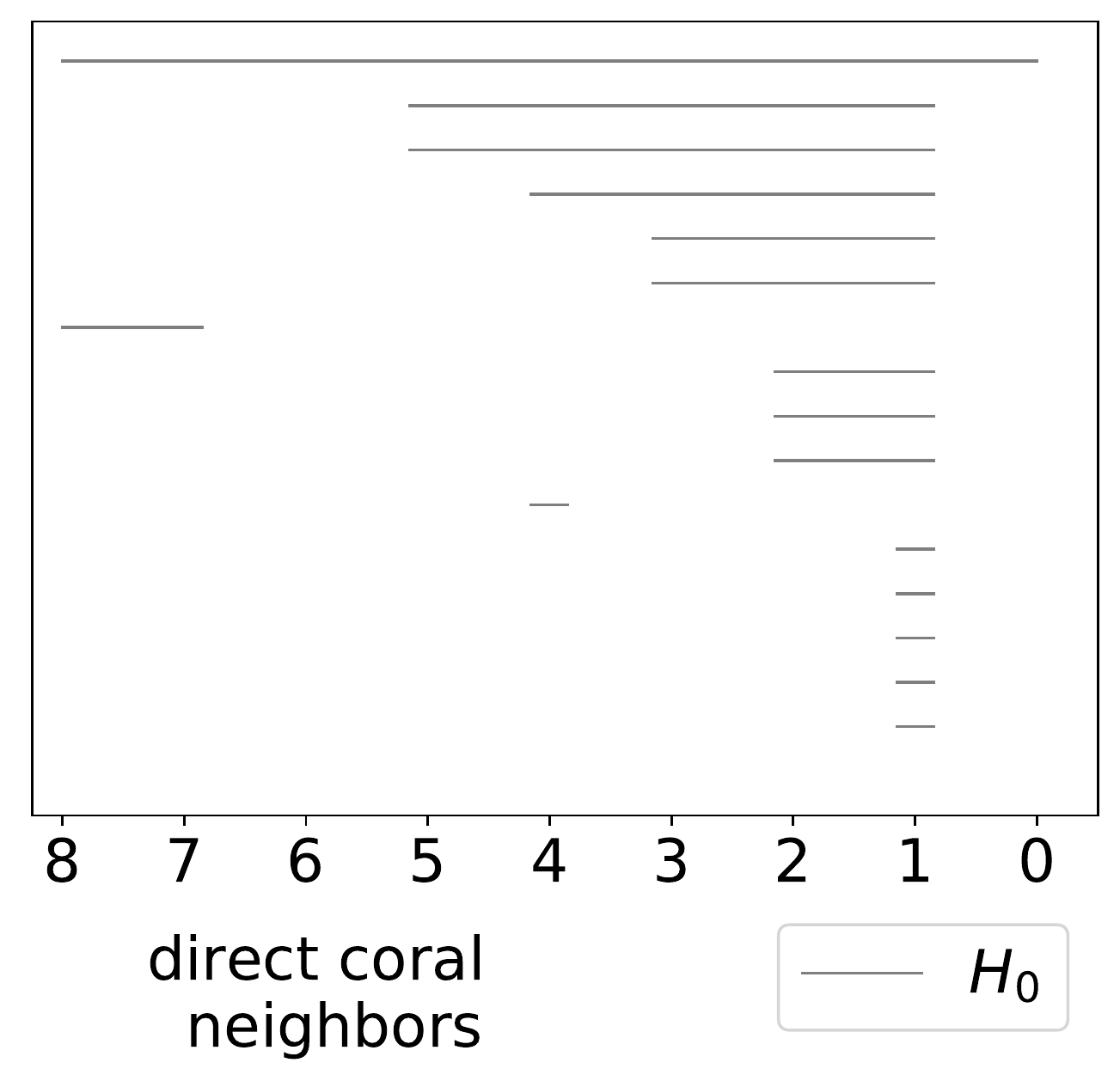}
\end{center} 
& \begin{center}
$t=80$\\
\includegraphics[width = 0.175\textwidth]{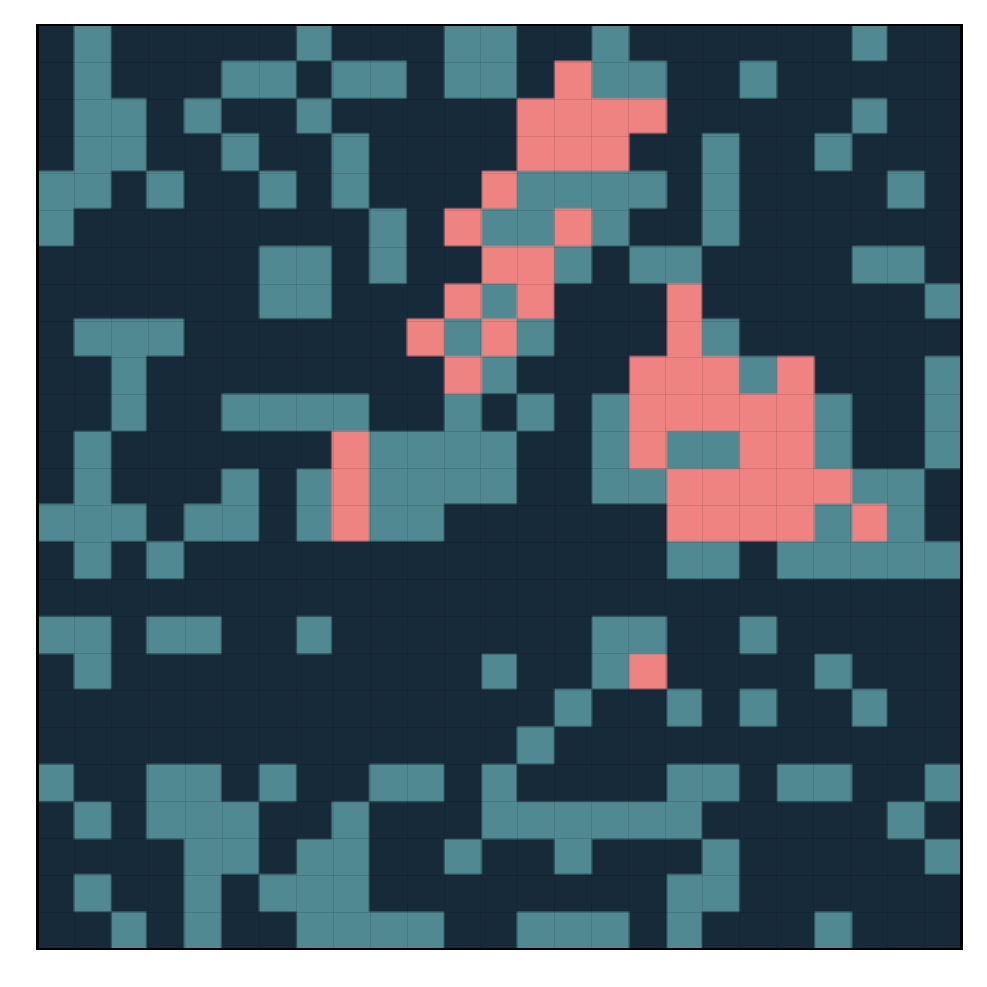} \\
\includegraphics[width = 0.175\textwidth]{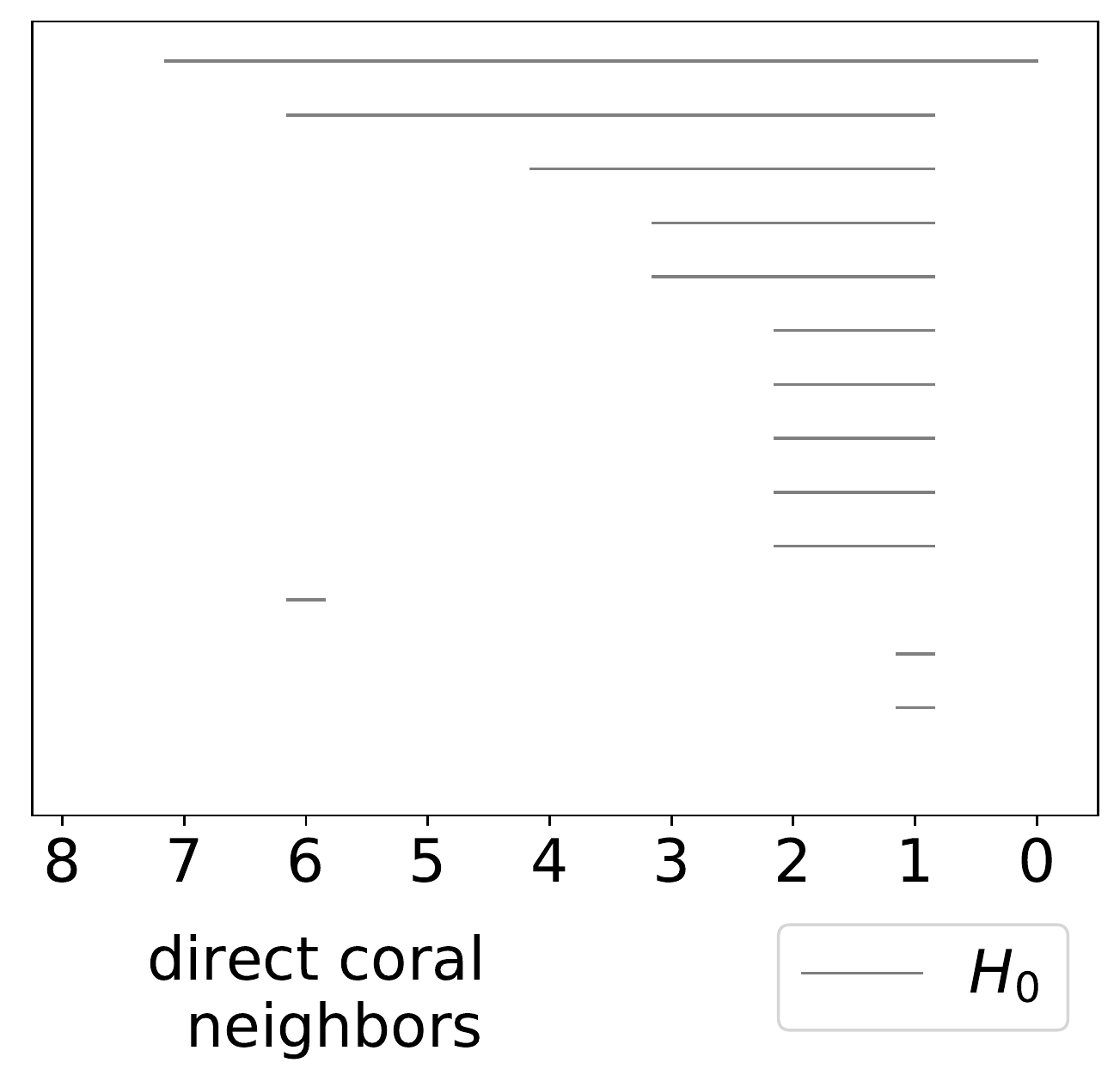}
\end{center}
& \begin{center}
$t=100$\\
\includegraphics[width = 0.175\textwidth]{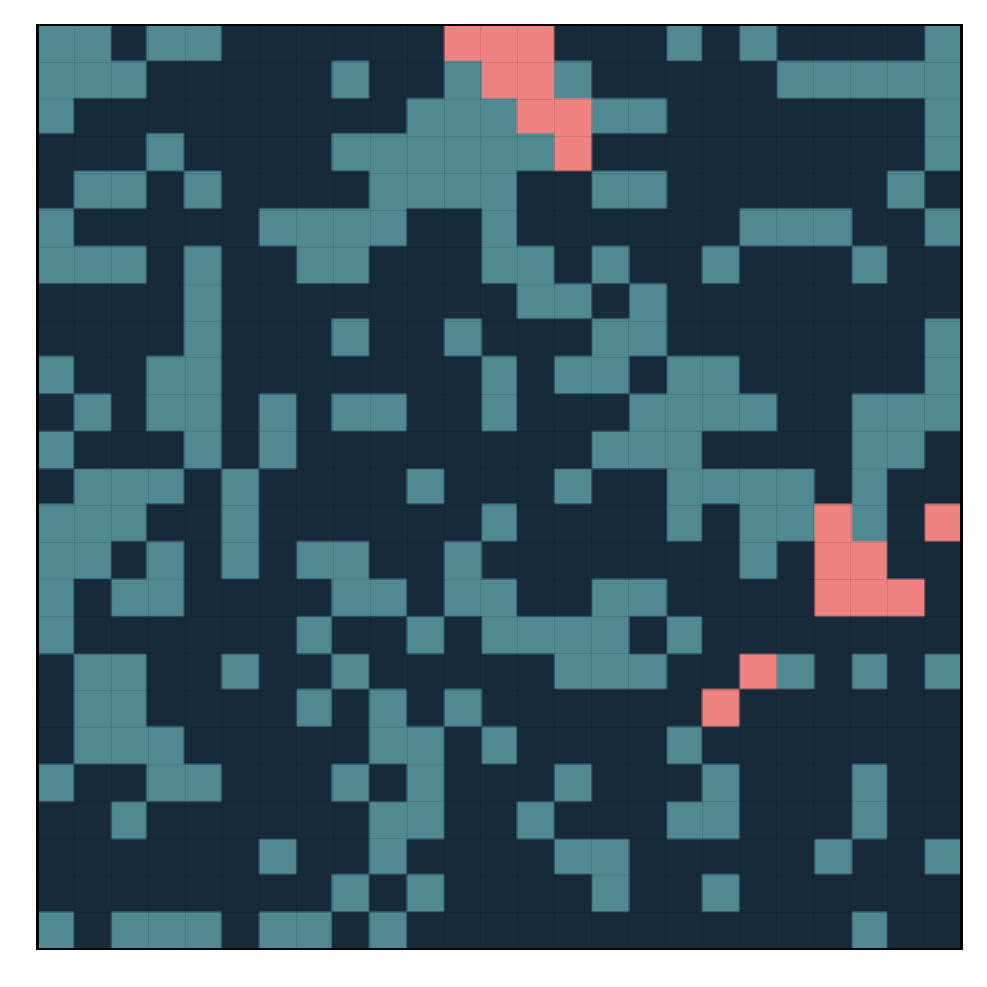}\\
\includegraphics[width = 0.175\textwidth]{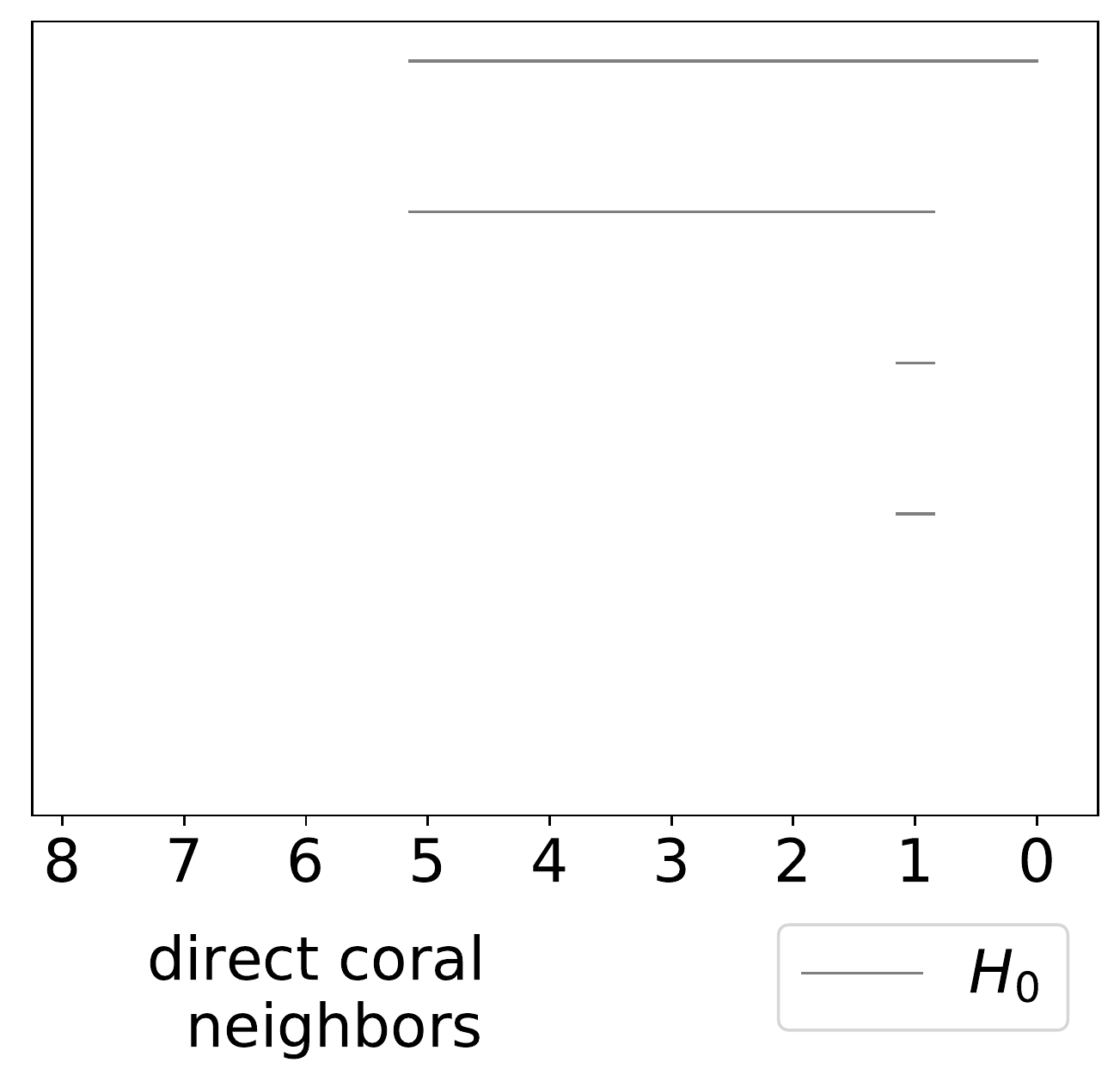}
\end{center} \\
\begin{center}
$t=120$\\
\includegraphics[width = 0.175\textwidth]{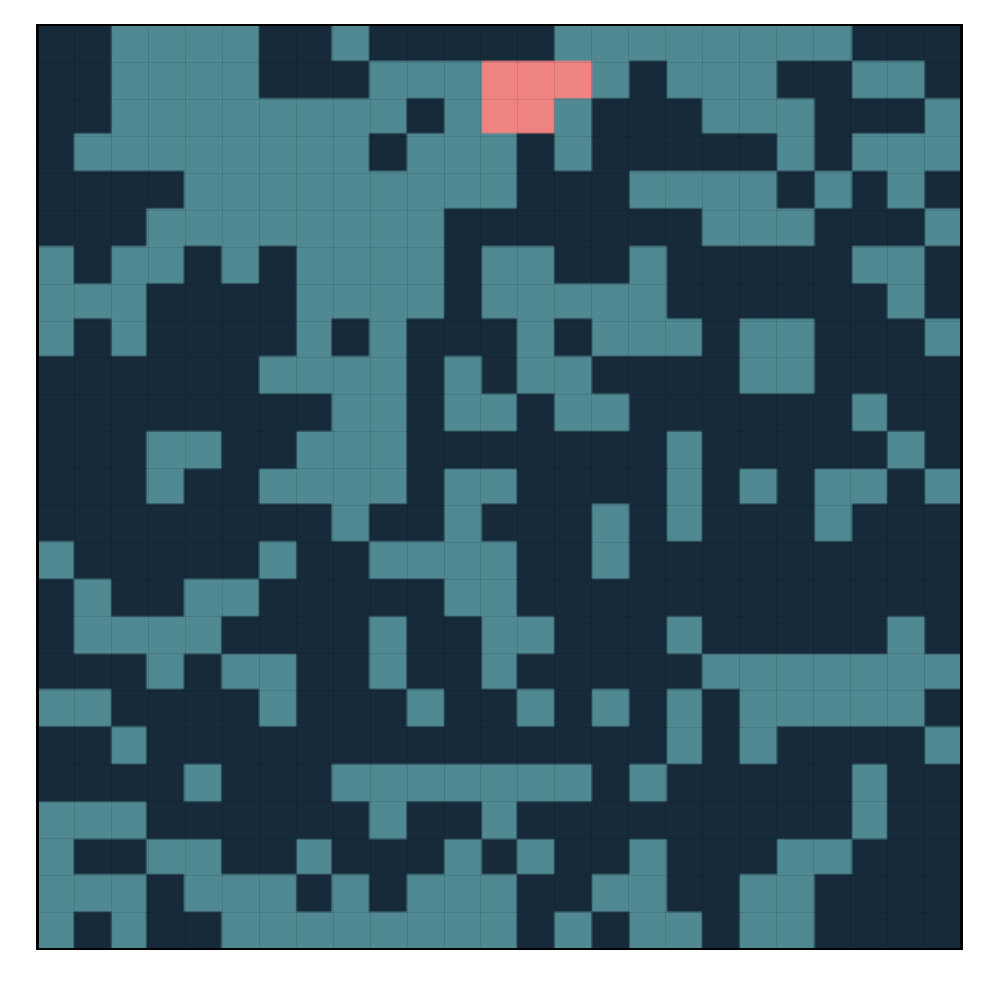}\\
\includegraphics[width = 0.175\textwidth]{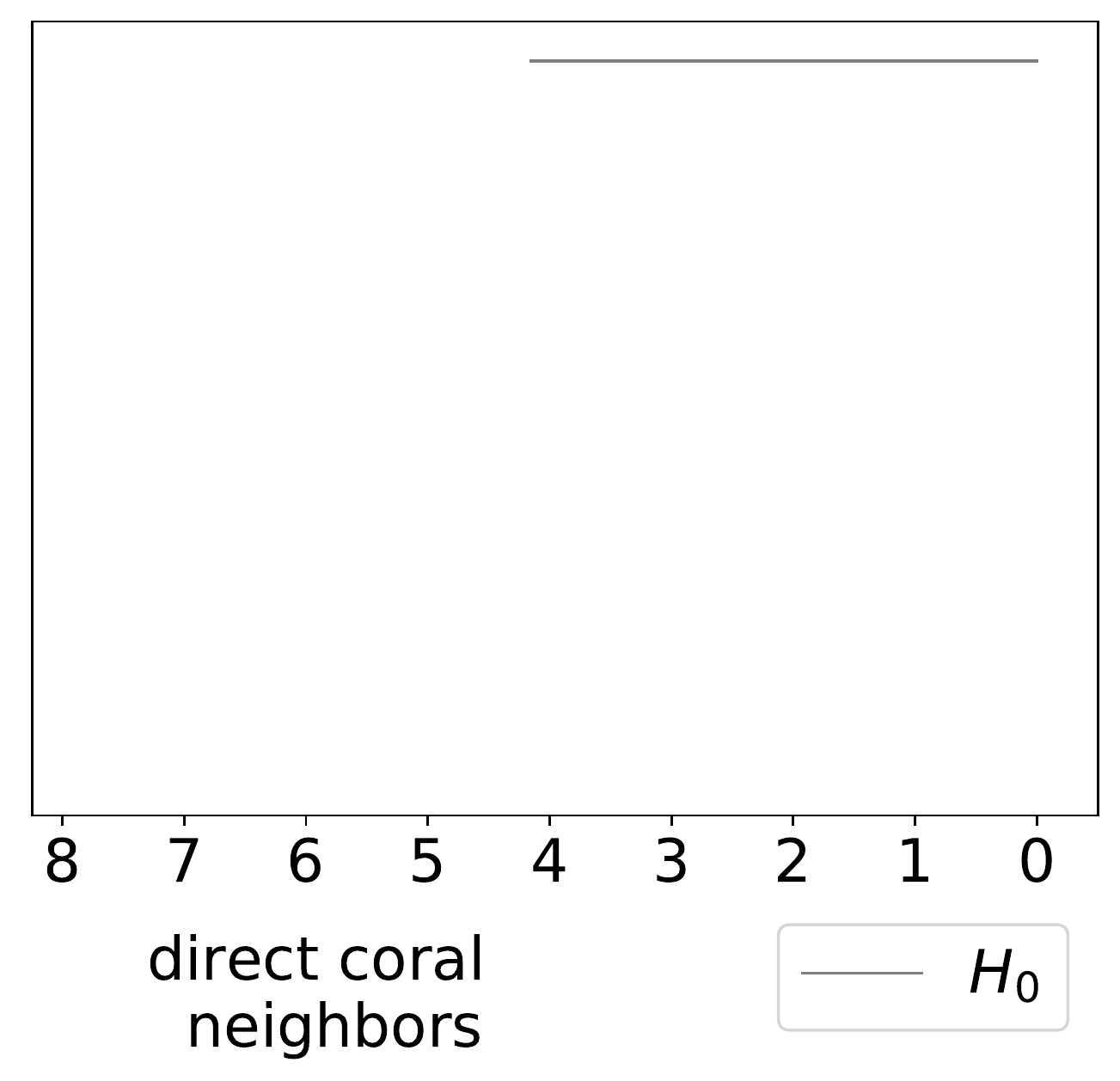}
\end{center} 
& \begin{center}
$t=140$\\
\includegraphics[width = 0.175\textwidth]{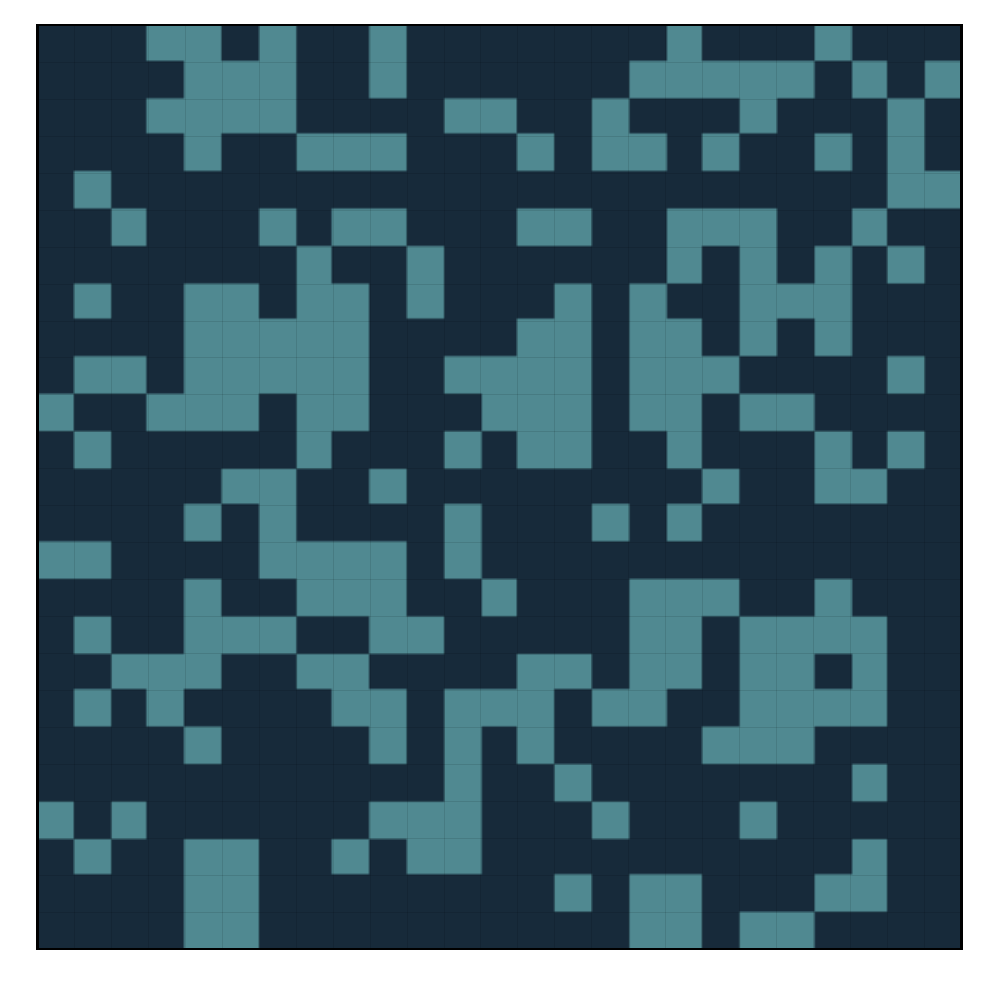} \\
\includegraphics[width = 0.175\textwidth]{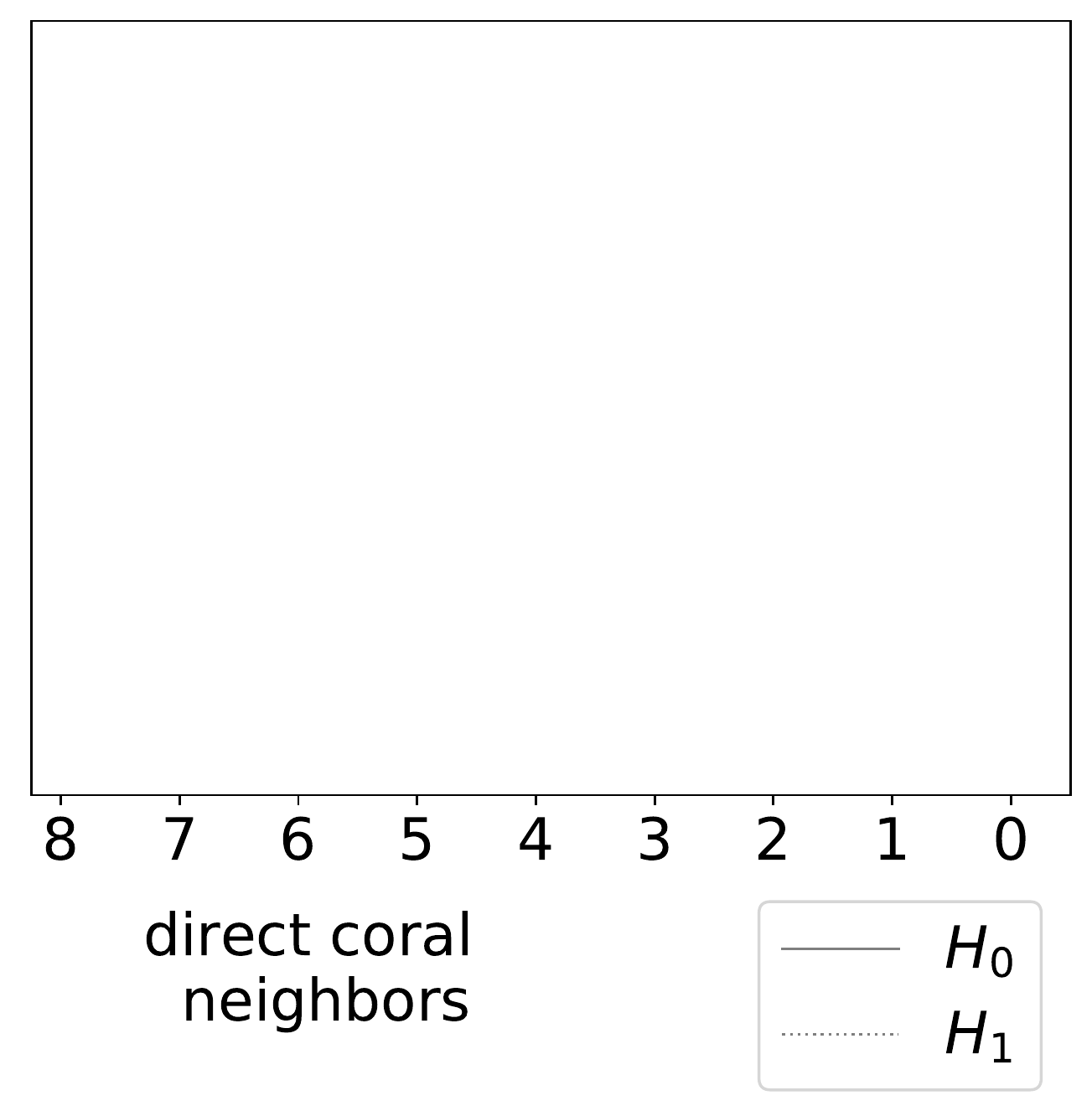}
\end{center}
& \begin{center}
$t=160$\\
\includegraphics[width = 0.175\textwidth]{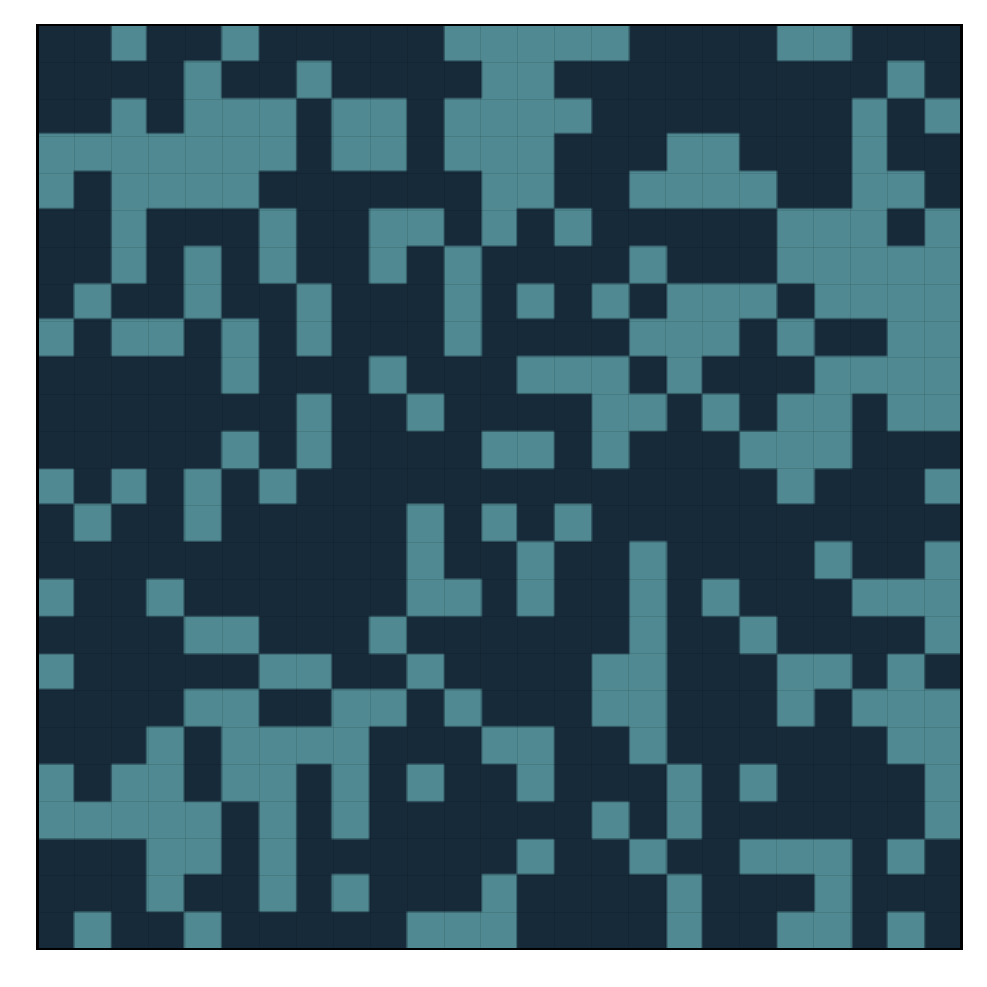}\\
\includegraphics[width = 0.175\textwidth]{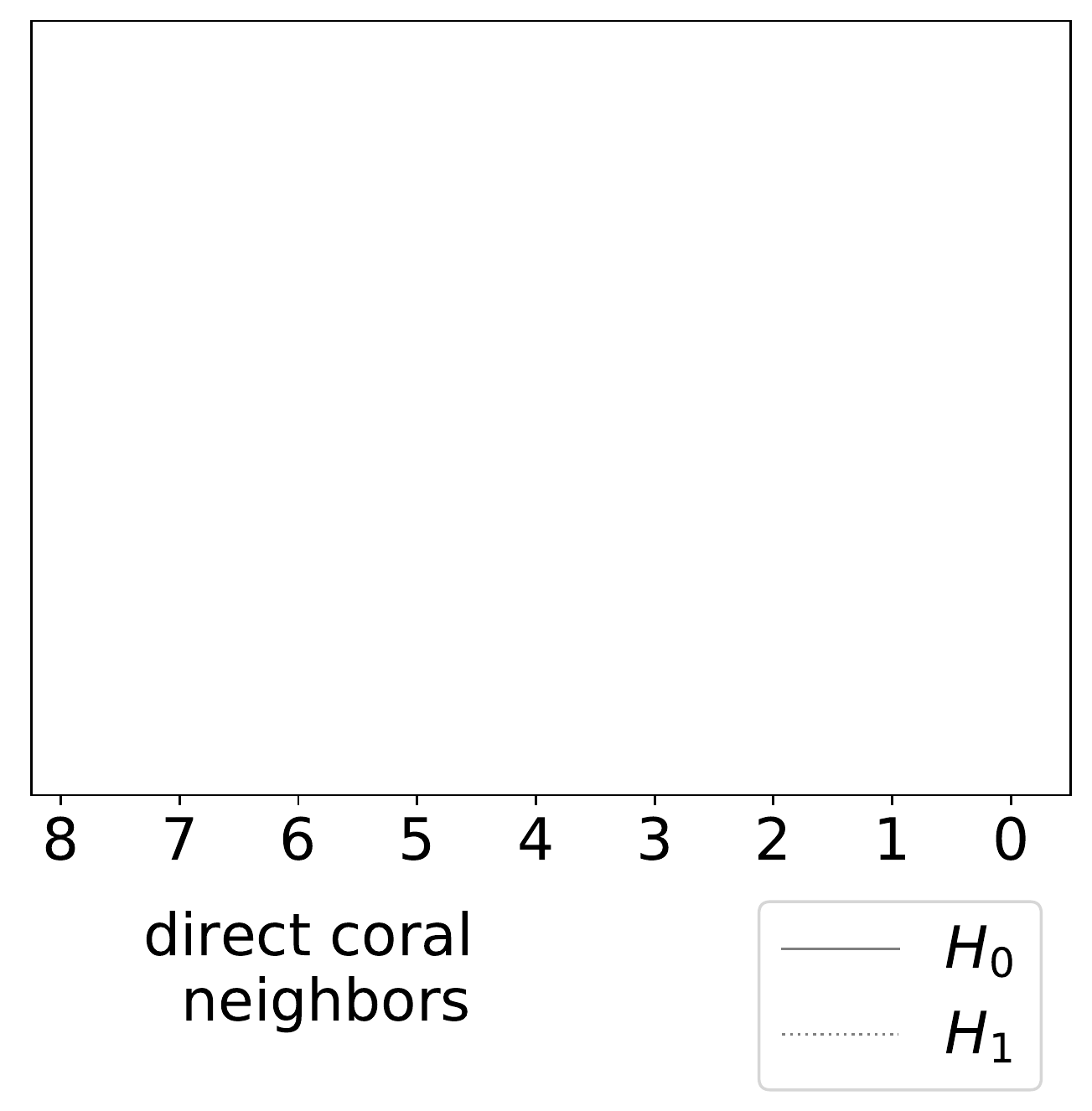}
\end{center} 
\end{tabular} 
}
\caption{Nine snapshots of the \ssMHE{} model, with persistent homology barcodes computed at each instant. The coral-cluster initial configuration is used with equal numbers of coral, turf, and macroalgae nodes. The grazing rate is $g=0.5$, which favours macroalgae.}
\label{SIfig:PHexample}
\end{center}
\end{figure}

To investigate the topological properties of the \ssMHE{} model, we could consider a single simulation of the model and produce a separate barcode at each snapshot.
\begin{ex}
Fig.~\ref{SIfig:PHexample} gives nine iterations of the \ssMHE{} model and plots the $\beta_0$ and $\beta_1$ barcodes. The number of bars gives the number of connected components in the snapshot, and the bars' lengths give the sizes of these components. A sequence of barcodes--one for each snapshot--gives a topological description of the simulation. While visual inspection of snapshots in this simulation shows that a single component of coral slowly shrinks and eventually disappears, one can not infer this from the barcodes. Although each barcode has one long bar, we can not generally determine whether the long bars in different barcodes represent the same component in different snapshots, or new components in each snapshot. 
\end{ex}

\clearpage
\section{Zigzag persistence}
\label{SIsec:zigzag}
In this section, we describe in detail how we use zigzag persistence to analyse \ssMHE{} model data. Subsection \ref{SIsubsec:zzD} demonstrates how we use a zigzag diagram to represent a time-evolving simulation of the \ssMHE{} model. Topological statistics are used to average our results over many such simulations in Subsection \ref{SIsubsec:TDAstats}. Finally, we introduce a pre-processing step that we perform before computing zigzag persistence in Subsection \ref{SIsubsec:preprocess}. For a comprehensive treatment of zigzag persistence, see the original work \cite{originalzigzag} by Carlsson and De Silva.

\subsection{Zigzag diagrams}
\label{SIsubsec:zzD}
Zigzag diagrams \cite{bradzzalgorithm} generalise the notion of a filtration in Definition \ref{SIdef:filtration}. In a filtration, the cubical complexes must be nested, so they allow the definition of the essential inclusion maps in \eqref{SIeq:filtrationinclusion}, which enable the computation of persistence. Zigzag diagrams still require inclusion maps, which may point left or right.
\begin{definition}[Zigzag diagram] Given a collection $\{\mathcal{Q}^z\}_{z\in \{0, 1, \dots, Z\}}$ of cubical complexes, a zigzag diagram is a sequence:
\begin{align}
    \mathcal{Q}^0\longleftrightarrow \mathcal{Q}^1 \longleftrightarrow \mathcal{Q}^2 &\longleftrightarrow \dots \longleftrightarrow \mathcal{Q}^Z,
\end{align}
where the symbol $\longleftrightarrow$ means that either $\mathcal{Q}^z \subset \mathcal{Q}^{z+1}$ or $\mathcal{Q}^z \supset \mathcal{Q}^{z+1}$, with inclusion maps defined in the relevant direction. \end{definition}

To construct a zigzag diagram from a simulation of the \ssMHE{} model, we must choose a single representative cubical complex $K_{\eta^*}^t$ from the filtration \ref{SIdef:filtration} (Fig. \ref{SIfig:coralPH}B) at each snapshot. For results in the main text, we choose $\eta^*=1$ and take $K_1^t$ as a cubical complex for each snapshot. The final panel in Fig.~\ref{SIfig:coralPH}B gives an example of $K_1^t$ for the \ssMHE{} snapshot in Fig.~\ref{SIfig:coralPH}A. This cubical complex represents all coral nodes with at least one direct coral neighbour, and zigzag diagrams allow tracking such components. To consider the time evolution of only large components of coral, we could take $\eta^*=8$ (for example, the first panel of Fig.~\ref{SIfig:coralPH}A). An result demonstrating the use $\eta^*=8$ over $\eta^*=1$ is given in Fig.~\ref{SIfig:fig3supplement}. 

\begin{figure}[tbhp]
\hspace{40mm} \textbf{A.} \hfill
\begin{center}
    \includegraphics[width = 0.5\textwidth]{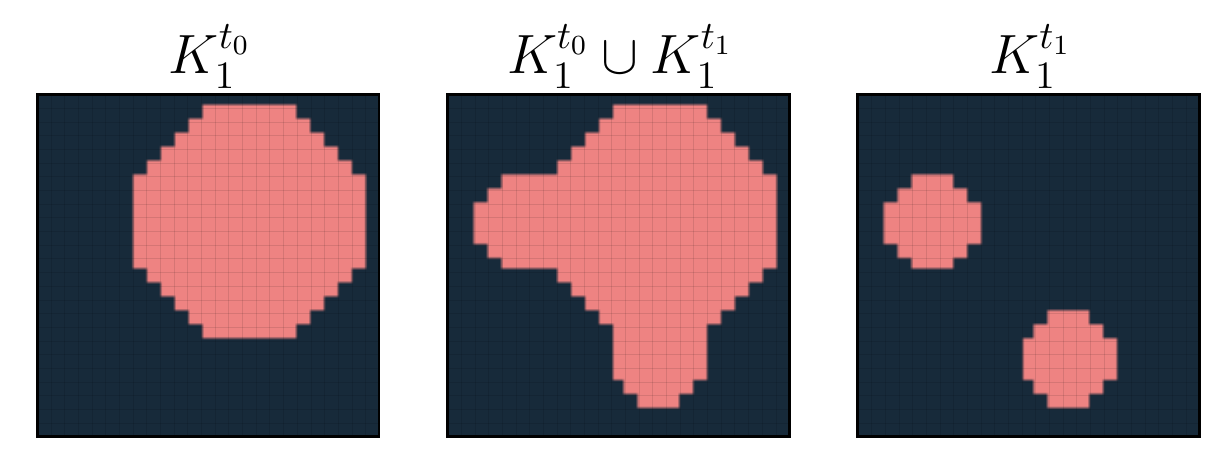}
\end{center}
\hspace{40mm} \textbf{B.} \hfill
\begin{center}
   \includegraphics[width = 0.5\textwidth]{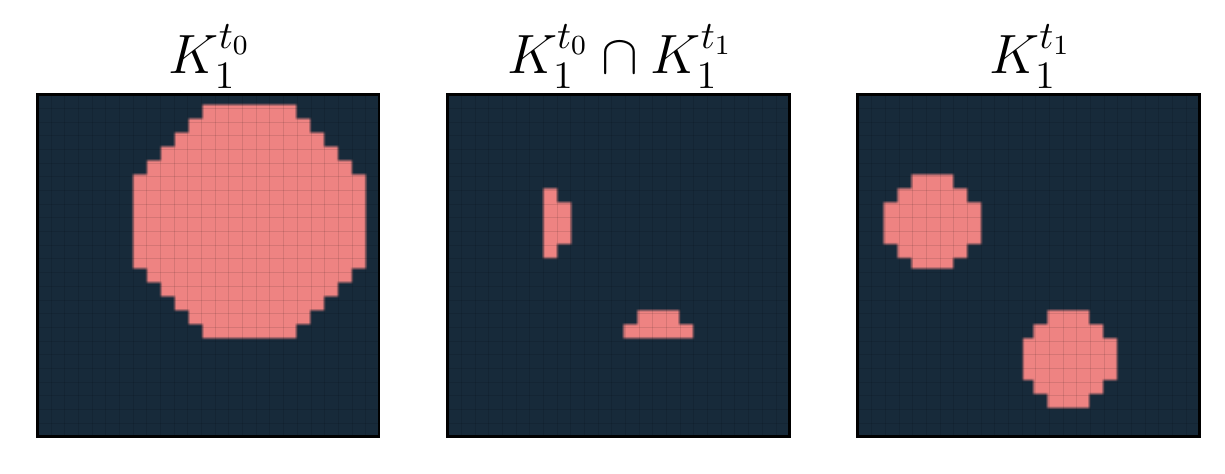}
\end{center}
     \caption{The two possibilities of intermediate complexes when defining a zigzag diagram. \textbf{A.} The union of two consecutive snapshots of the \ssMHE{} model is inserted in sequence between them. \textbf{B.} The intersection of two snapshots of the \ssMHE{} model is inserted.}
        \label{SIfig:zzoptions}
\end{figure}

The notion of a zigzag diagram allows the crucial extension that spaces no longer need to be nested for us to track topological features. Since we are interested in how topological features change from one model snapshot to the next, we insert an `intermediate' complex between each pair of complexes generated from successive snapshots. The two options for intermediate complexes from which we can define inclusion maps are the union and intersection of consecutive cubical complexes. These lead to two possibilities for zigzag diagrams (\eqref{SIeq:zzunion} or \eqref{SIeq:zzintersection}) that we could use to describe a simulation of the \ssMHE{} model.
\begin{subequations}
\begin{align}
    &K_1^{t_0}\longrightarrow K_1^{t_0} \cup K_1^{t_1} \longleftarrow K^{t_1} \longrightarrow K_1^{t_1} \cup K_1^{t_2} \longleftarrow K_1^{t_2} \longrightarrow K_1^{t_2} \cup K_1^{t_3} \cdots
    \label{SIeq:zzunion},\\
    &K_1^{t_0}\longleftarrow K_1^{t_0} \cap K_1^{t_1} \longrightarrow K_1^{t_1} \longleftarrow K_1^{t_1} \cap K_1^{t_2} \longrightarrow K_1^{t_2} \longleftarrow K_1^{t_2} \cap K_1^{t_3} \cdots.
    \label{SIeq:zzintersection}
\end{align}
\end{subequations}
Fig.~\ref{SIfig:zzoptions} shows the two options for zigzag diagrams. All non-coral nodes are dark-coloured.

If coral occupies the same nodes in two consecutive snapshots of the \ssMHE{} model, we assume that the same component of coral exists during the time interval between the two snapshots. In real coral data, for example, if coral is present in the same location in two photographs taken a short time apart, it is reasonable to assume it is the same piece of coral. 
To encode this assumption, we choose the second option (\eqref{SIeq:zzintersection}) and insert the intersection of adjacent cubical complexes into the sequence.

We now calculate zigzag persistence by applying the homology functor to the sequence \eqref{SIeq:zzintersection}, giving the following sequence of homology groups.
\begin{align}
    H_0(K_1^{t_0})\longleftarrow H_0(K_1^{t_0} \cap K_1^{t_1}) \longrightarrow H_0(K_1^{t_1}) \longleftarrow H_0(K_1^{t_1} \cap K_1^{t_2}) \longrightarrow   \cdots.
    \label{SIeq:zzHom}
\end{align}
We compute zigzag persistence using the \textsc{BATS} Python package \cite{bradzzalgorithm} developed by Carlsson, Dwaraknath, and Nelson.

As with PH, zigzag persistence may be visualised as a barcode (Fig.~\ref{SIfig:ZZBCtoZZLS}A). The start and end points of bars give the birth and death time of a component of coral within a time-evolving simulation. Fig.~\ref{SIfig:ZZBCtoZZLS}C gives additional information to Fig.~\ref{SIfig:PHexample} since we can now see a single persistent bar---which disappears only at time $t=120$. While, computationally, zigzag is more expensive to compute than PH, it can track topological features across multiple timesteps. In the simulation in Figs.~\ref{SIfig:PHexample} and \ref{SIfig:ZZBCtoZZLS}, zigzag persistence demonstrates that a single cluster of coral gradually disintegrates over time---which PH alone can not conclude. The shorter bars in the zigzag persistence barcode in Fig.~\ref{SIfig:ZZBCtoZZLS}A represent small fragments of coral appearing and disappearing, which can also be observed visually in Fig.~\ref{SIfig:PHexample}, but can not be inferred from the PH barcodes.

\subsection{Statistical TDA}
\label{SIsubsec:TDAstats}
We use persistence diagrams and landscapes to make statistical comparisons between the topological summaries of the \ssMHE{} model. Recall that a zigzag persistence barcode is a multiset of intervals $\{ [\mathbbm{b}_j,\mathbbm{d}_j]\}$ representing the birth ($\mathbbm{b}_j$) and death ($\mathbbm{d}_j$) times of components of coral. A persistence diagram is then simply the multiset of points $\{(\mathbbm{b}_j,\mathbbm{d}_j)\}$ in the plane, together with the line $\{(\mathbbm{t},\mathbbm{t}): \mathbbm{t} \in \R_{+}\}$. Persistence diagrams allow the definition of persistence landscapes.

For any point in a persistence diagram $(\mathbbm{b}_j,\mathbbm{d}_j) \in \R^2$, define a piece-wise linear function $g _{(\mathbbm{b}_j,\mathbbm{d}_j)}:  [0,T] \to \R$ by \eqref{SIeq:landscapeF}, where $T$ is the duration of a simulation of the \ssMHE{} model.
\begin{equation}
    g_{(\mathbbm{b}_j,\mathbbm{d}_j)}(t) = 
    \begin{cases} 
      0 & t\notin (\mathbbm{b}_j,\mathbbm{d}_j), \\
      t-\mathbbm{b}_j & t \in (\mathbbm{b}_j,\frac{\mathbbm{b}_j+\mathbbm{d}_j}{2}), \\
      -t + \mathbbm{d}_j & x \in (\frac{\mathbbm{b}_j+\mathbbm{d}_j}{2},\mathbbm{d}_j).
   \end{cases}
   \label{SIeq:landscapeF}
    \end{equation}
    For each point $(\mathbbm{b}_j, \mathbbm{d}_j)$ in the zigzag persistence diagram of a time-evolving simulation of the \ssMHE{} model, define $g_{(\mathbbm{b}_j,\mathbbm{d}_j)}$ as above. These functions together define persistence landscapes.
    \begin{definition}[Persistence landscapes]

     For $k \in \N$, the $k$th persistence landscape $\lambda_k(t):  \R \to [0,\infty)$ is defined by \eqref{SIeq:landscape}.
\begin{equation}
\lambda_k(t) = \textnormal{the }k\textnormal{th}\textnormal{ largest value of }\{g_{(\mathbbm{b}_j,\mathbbm{d}_j)}(t)\},
\label{SIeq:landscape}
\end{equation}
where $[\mathbbm{b}_j, \mathbbm{d}_j]$ range over all bars in the zigzag persistence barcode.
\label{SIdef:persistencelandscapes}
\end{definition}

\begin{figure}[tbh]
\textbf{A. \ssMHE{} model snapshots} \hfill

\begin{tabular}{p{0.3\textwidth}p{0.3\textwidth}p{0.3\textwidth}}
\begin{center}
$t=0$\\
\includegraphics[width = 0.175\textwidth]{SI_figs/PH_example_1_image.pdf}\\
\end{center} 
& \begin{center}
$t=60$\\
\includegraphics[width = 0.175\textwidth]{SI_figs/PH_example_4_image.pdf} \\
\end{center}
& \begin{center}
$t=120$\\
\includegraphics[width = 0.175\textwidth]{SI_figs/PH_example_7_image.pdf}\\
\end{center} 
\end{tabular}
    \begin{center}
       \pmb{ \Bigg\downarrow}
    \end{center}
    
\vspace{-2mm}

\textbf{B. Zigzag persistence} \hfill

\begin{center}
    
    \includegraphics[width=0.34\textwidth]{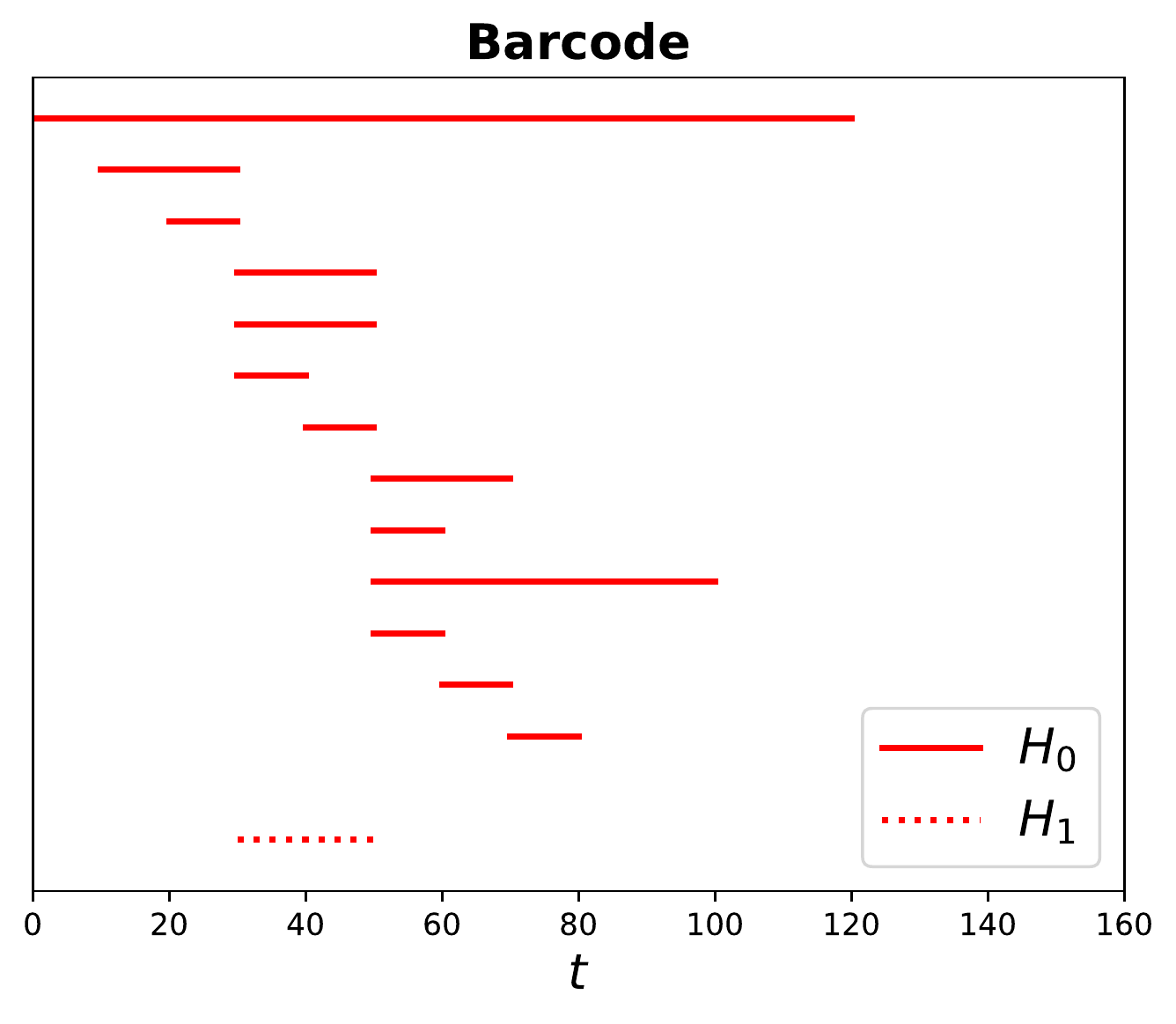} 
     \includegraphics[width=0.3\textwidth]{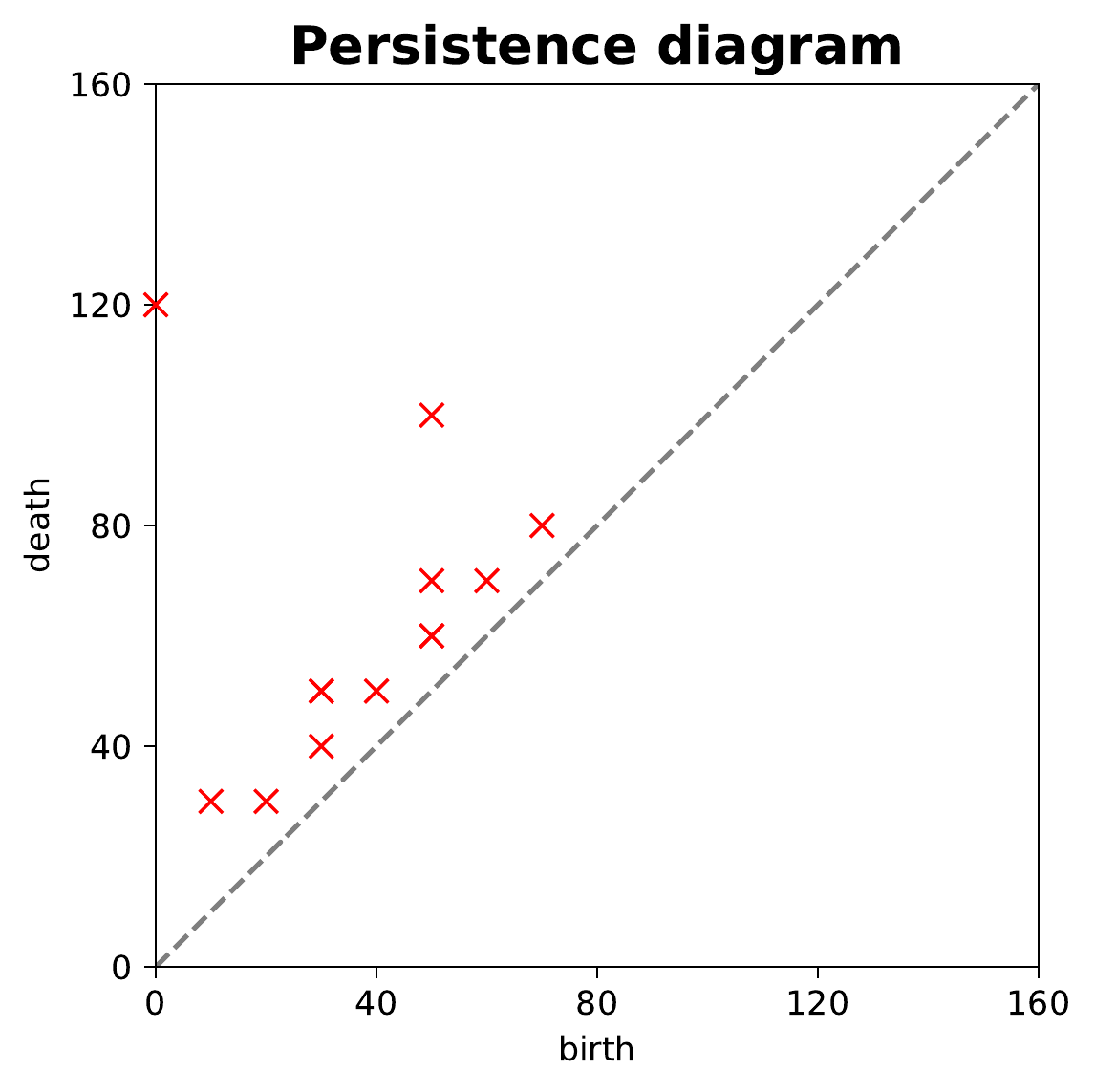}
     \includegraphics[width=0.3\textwidth]{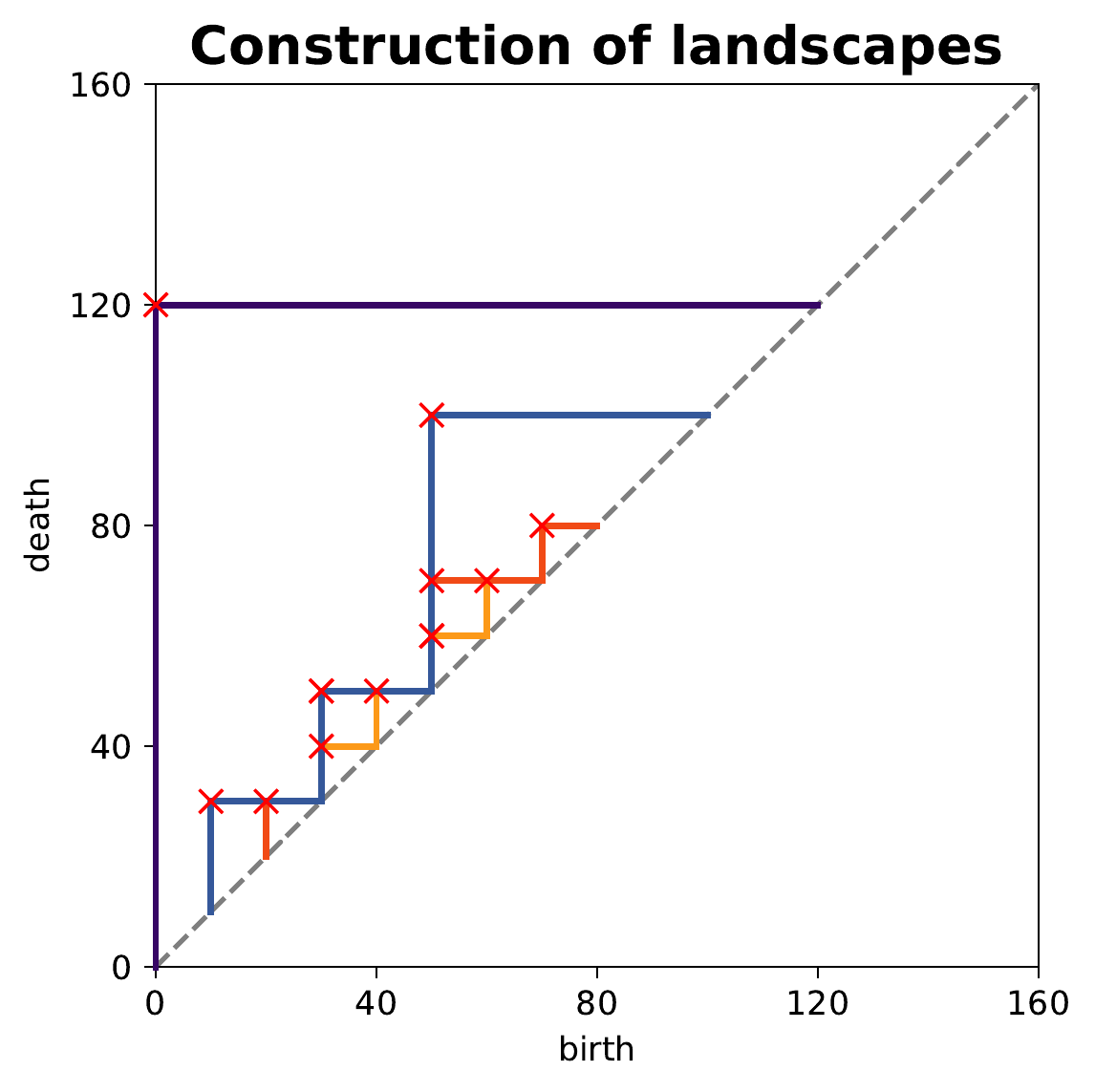}
     \end{center}
         \begin{center}
       \pmb{ \Bigg\downarrow}
    \end{center}
    
     \vspace{2mm}
\textbf{C. Landscapes} \hfill

     \centering 
     \includegraphics[width=0.6\textwidth]{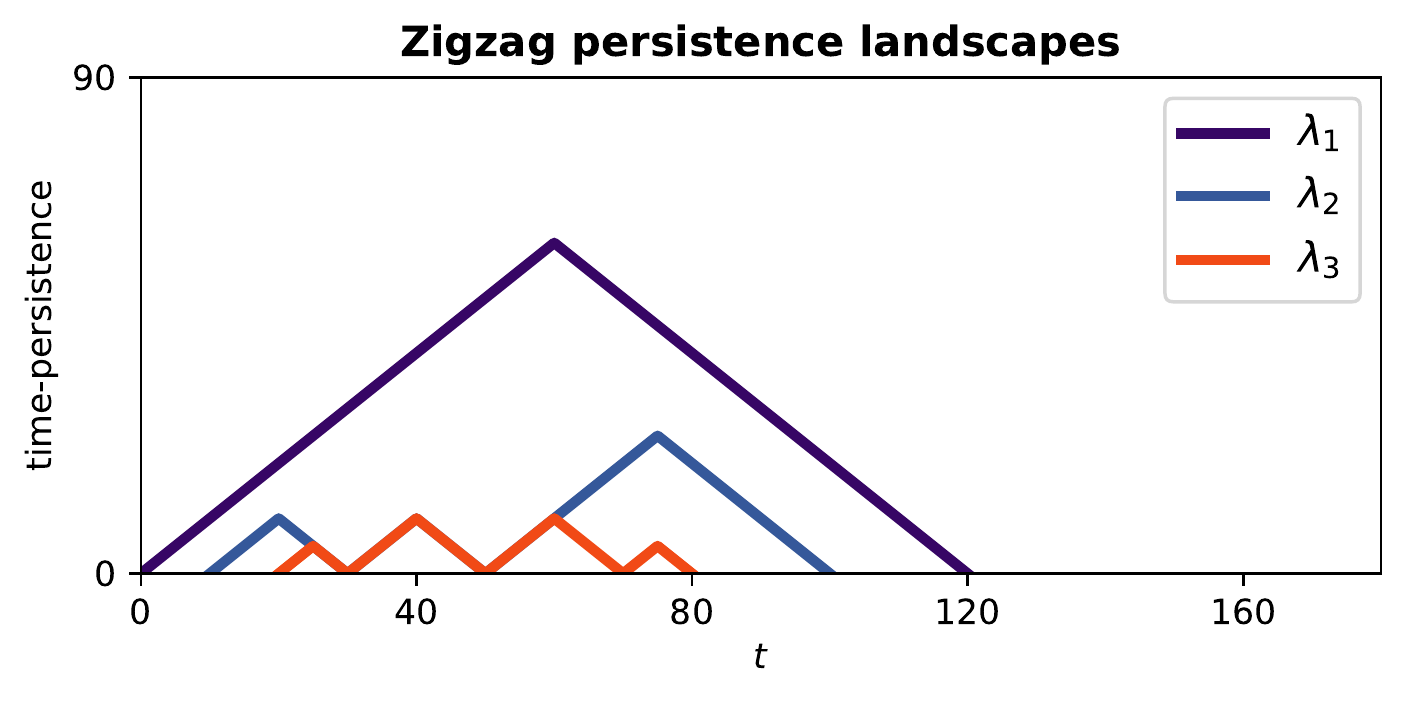}
\caption{Workflow for the computation of zigzag persistence from a time-evolving simulation of the \ssMHE{} model. \textbf{A. \ssMHE{} model snapshots} Three snapshots of the simulation from Fig.~\ref{SIfig:PHexample}. \textbf{B.} Left: The zigzag persistence barcode computed from the simulation of the \ssMHE{} model in Fig.~\ref{SIfig:PHexample}. Solid red bars represent clusters of coral, and dotted red bars represent loops enclosed by coral. The start- and end-points of the bars represent the times they emerge and disappear in the simulation. Middle: The zigzag persistence diagram converted from the zigzag persistence barcode in B. Solid bars $[\mathbbm{b}, \mathbbm{d}]$ are represented by points $(\mathbbm{b}, \mathbbm{d})$ in the plane. Right: An explanation of how a zigzag persistence barcode is converted into a zigzag persistence landscape. \textbf{C.} The zigzag persistence landscape generated from the zigzag persistence barcode of this simulation.}  
\label{SIfig:ZZBCtoZZLS}

\end{figure}

\begin{ex}
Fig.~\ref{SIfig:ZZBCtoZZLS} shows an example of the conversion of a zigzag persistence barcode to a zigzag persistence diagram to a persistence landscape.
\end{ex}

A fundamental property of persistence landscapes is that they live in a Banach space \cite{persland}. This property allows the computation of an average persistence landscape, $\bar \lambda_k$, which is the average function taken over many $\lambda_k$. We calculate and average persistence landscapes using the \textsc{Persistence representations} module by Dlotko in \textsc{GUDHI} \cite{gudhi:PersistenceRepresentations}. 

\subsection{Pre-processing simulated data for zigzag persistence}
\label{SIsubsec:preprocess}

Before converting a snapshot of the \ssMHE{} model into a cubical complex (for example, $K_1^t$) for the computation of zigzag persistence, we perform a pre-processing step. The coral-dominated stationary state of the \ssMHE{} model consists of around 80\% coral nodes and 20\% turf nodes--see the top row of Fig.~\ref{SIfig:preprocessZZ}A. We want our persistence landscapes to reflect that this is a stationary state. The average persistence landscapes in Fig.~\ref{SIfig:preprocessZZ}A, however, show many small landscapes that track the minor components of turf that appear and disappear quickly at the coral-dominated stationary state. We preprocess snapshots of the \ssMHE{} model to remove this noise before computing the zigzag persistence of simulations.

\begin{figure}[tbh]
	\begin{center}
	Original model snapshot \hspace{2cm} Processed model snapshot\\
	\includegraphics[width = 0.3\textwidth]{SI_figs/initial_profile_4.pdf} \includegraphics[width = 0.3\textwidth]{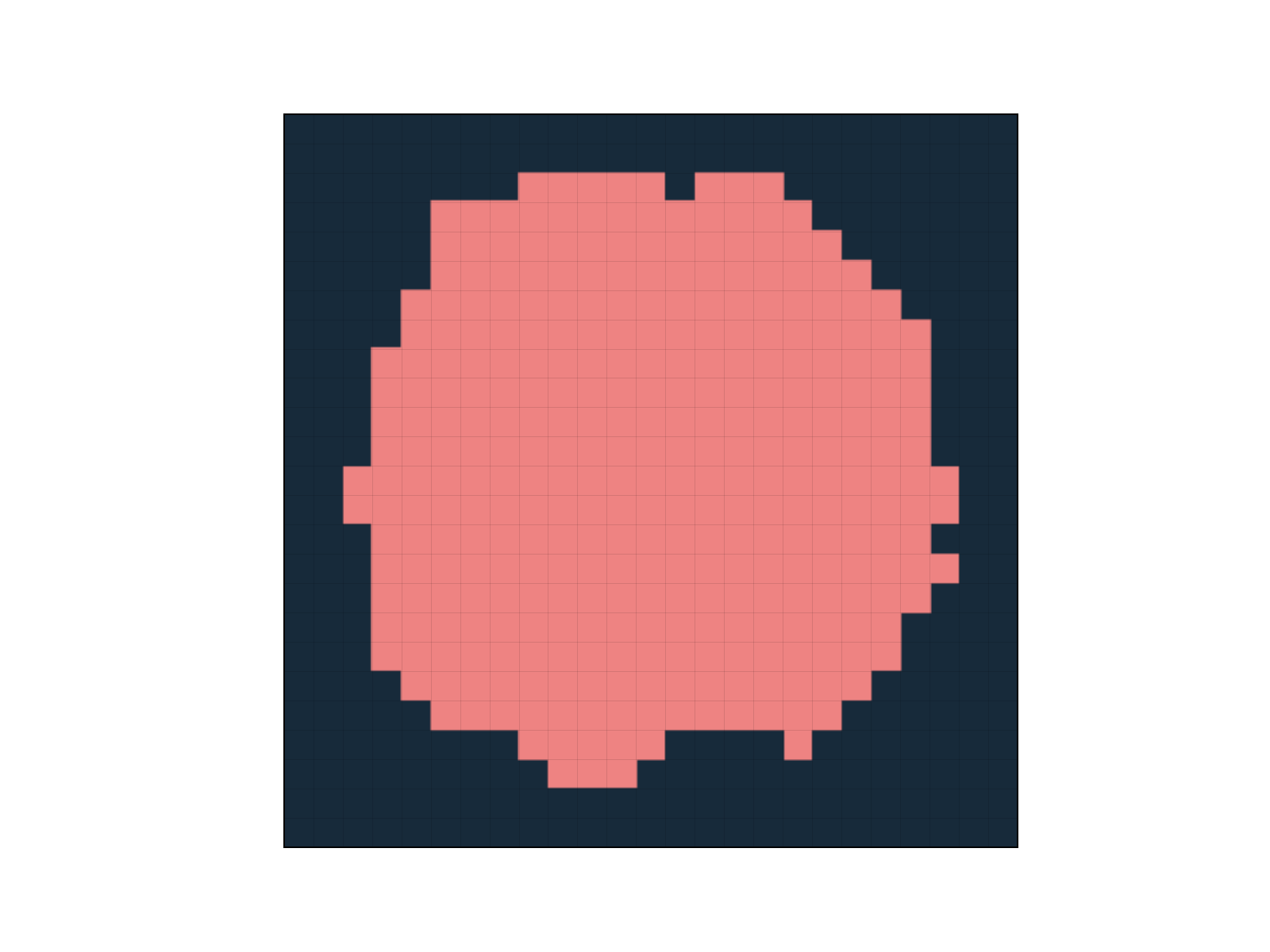} \\
	\includegraphics[width = 0.3\textwidth]{SI_figs/initial_profile_2.pdf} \includegraphics[width = 0.3\textwidth]{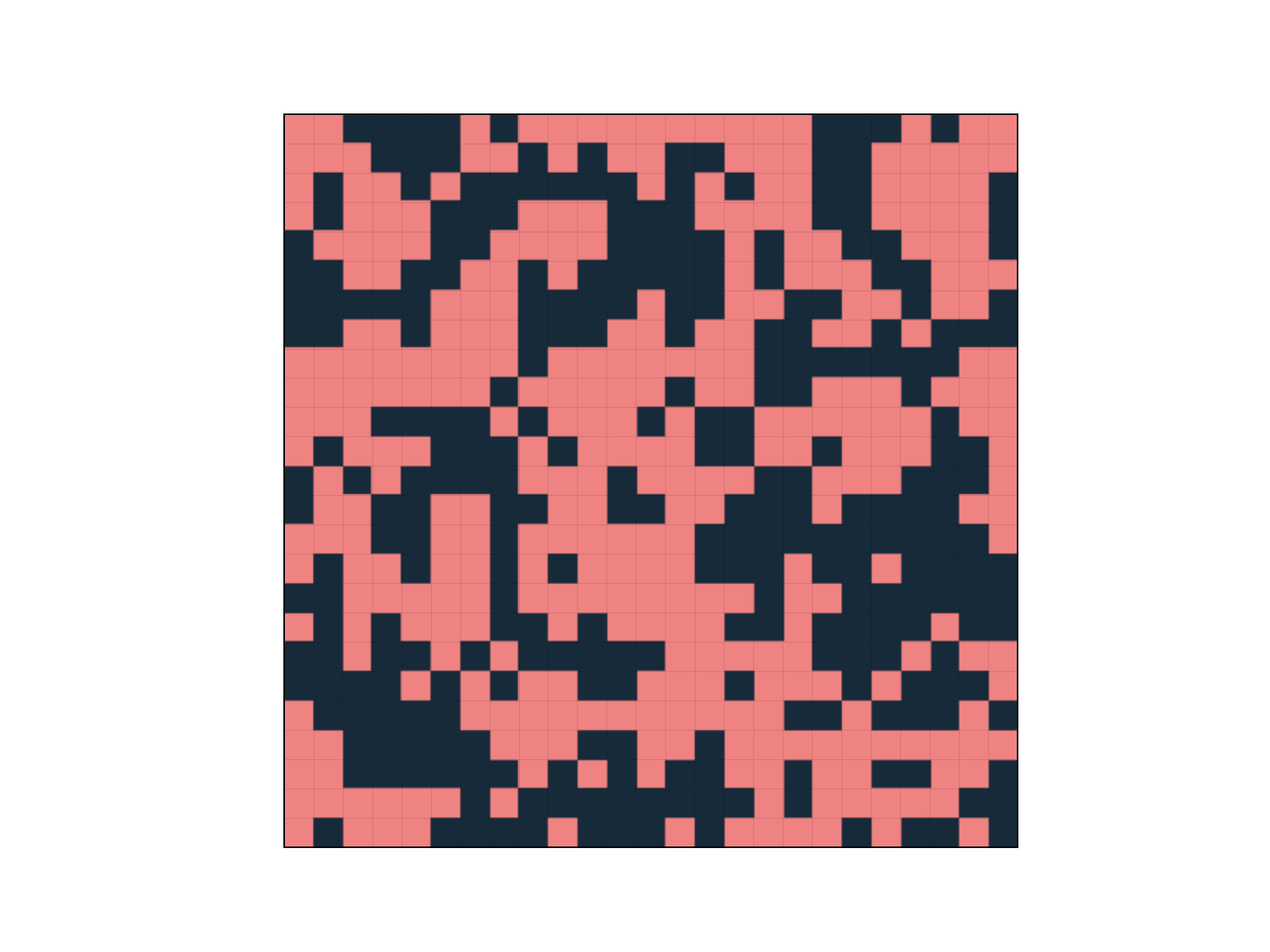} 
	\end{center}
	\caption{Two examples of snapshots of the \ssMHE{} model and the result of our pre-processing step.}
\label{SIfig:preprocess}
\end{figure}

\begin{figure}[tbh]
\hspace{10mm} \textbf{A. } \hfill
\begin{center}
        \includegraphics[width=0.2\textwidth]{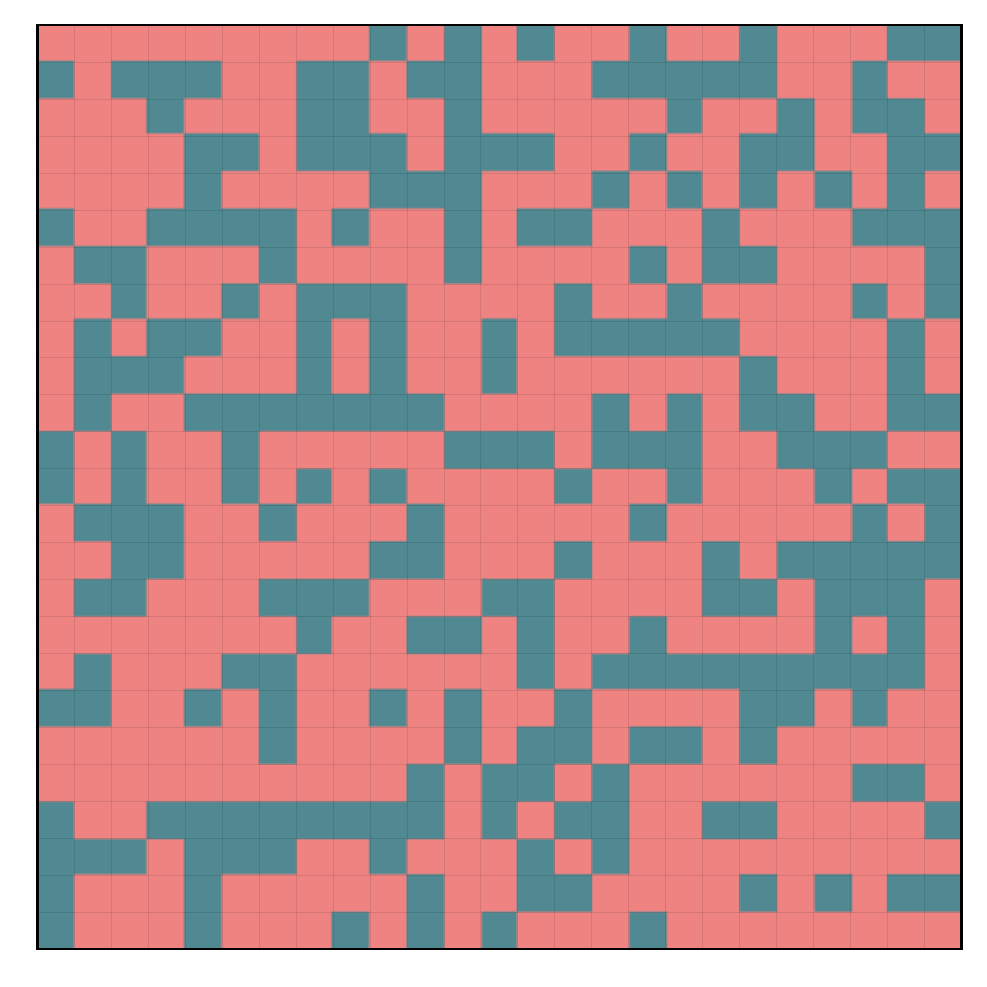}
    \includegraphics[width=0.2\textwidth]{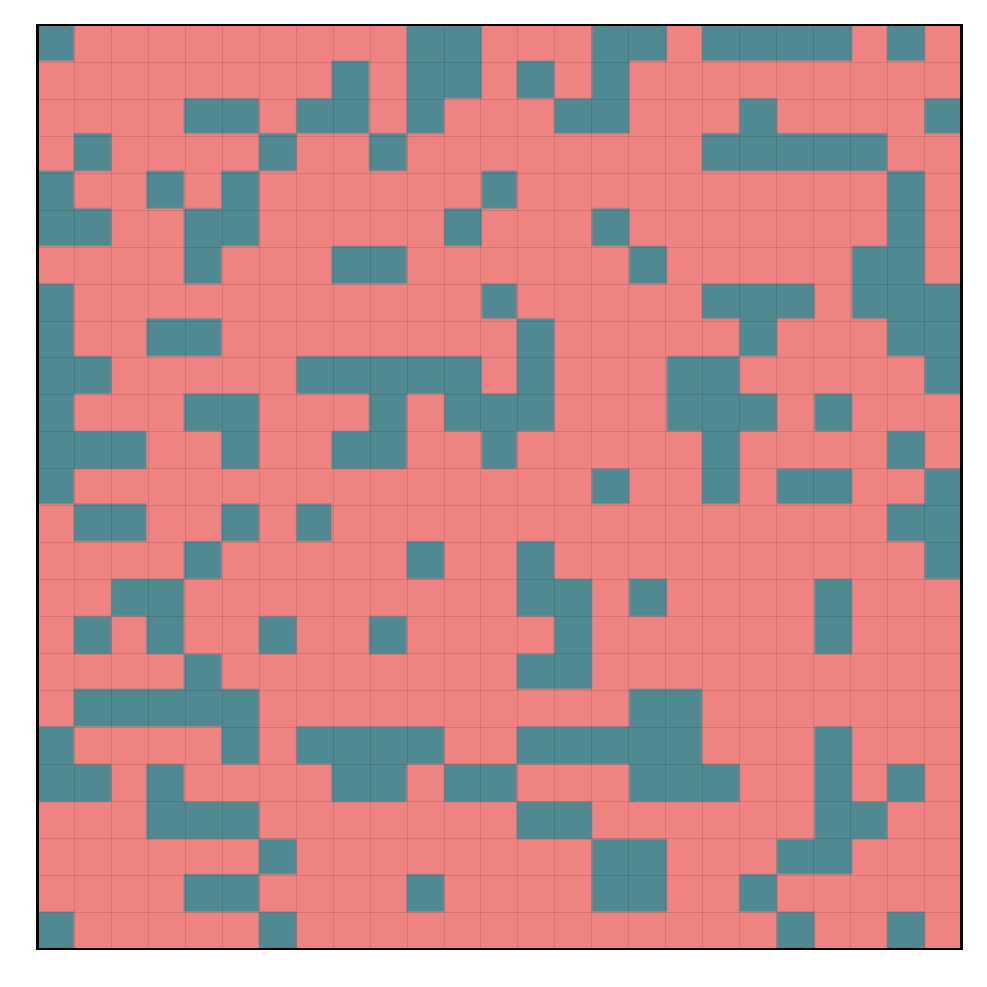}
    \includegraphics[width=0.2\textwidth]{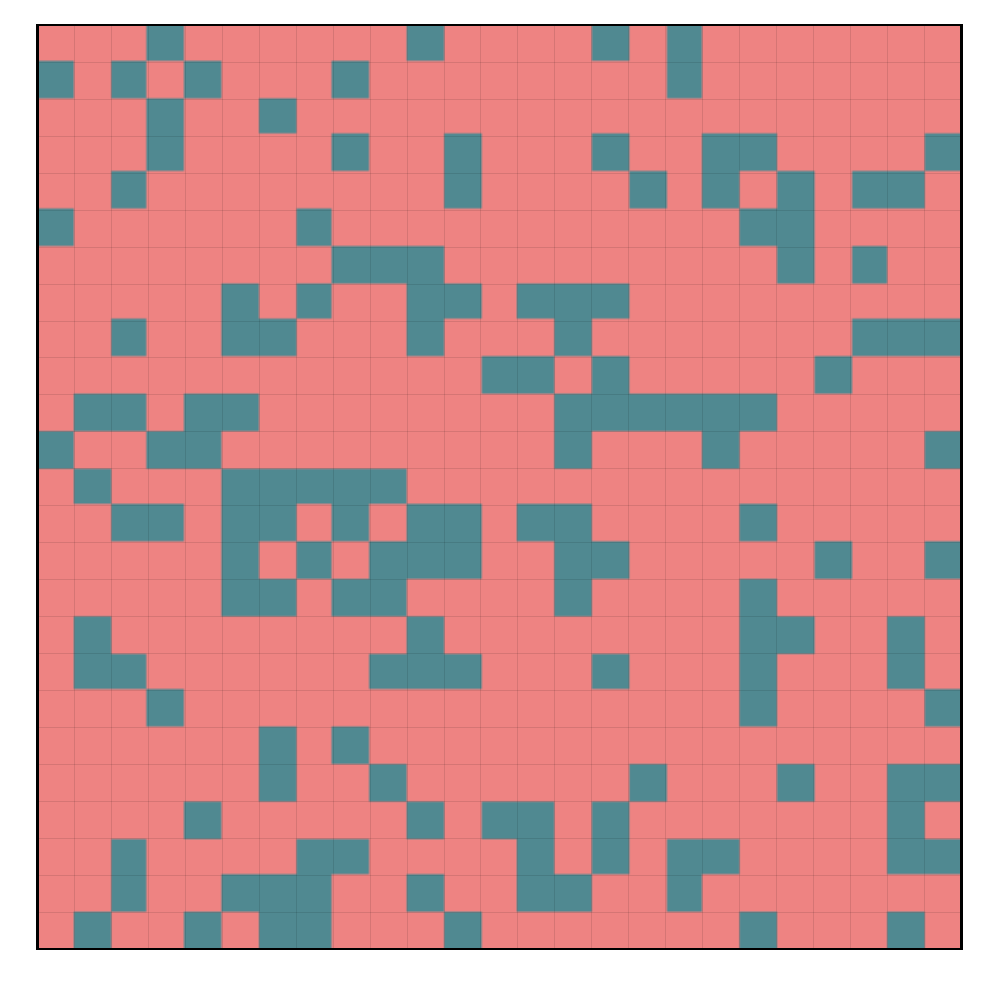}
    \includegraphics[width=0.2\textwidth]{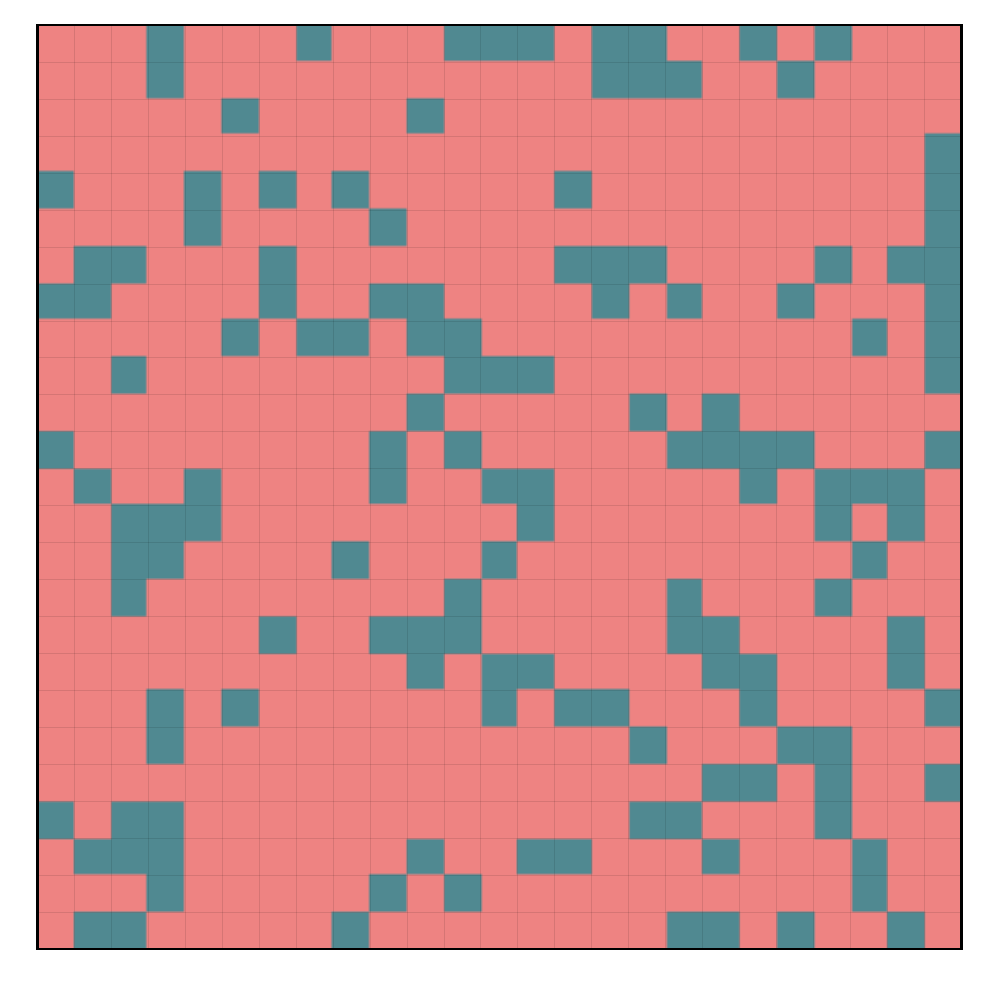}\\ 
   \includegraphics[width=0.75\textwidth]{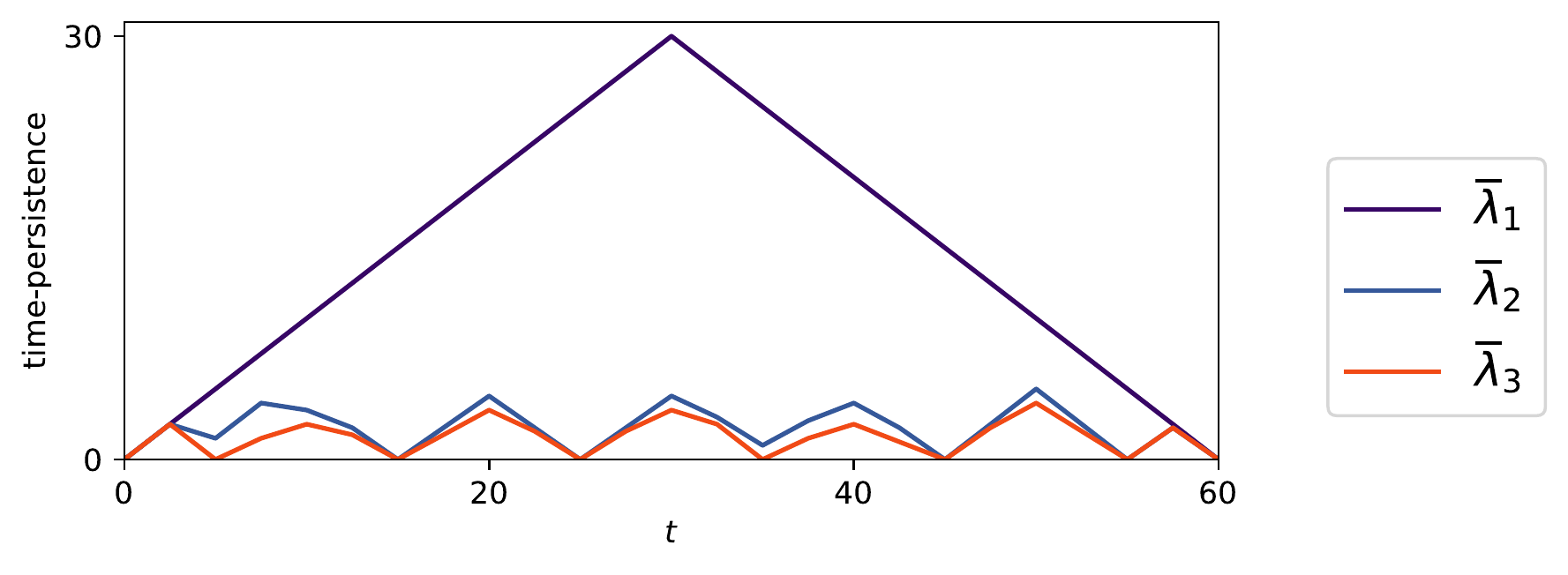}
   \end{center}
\hspace{10mm} \textbf{B. } \hfill
\begin{center}
  \includegraphics[width=0.2\textwidth]{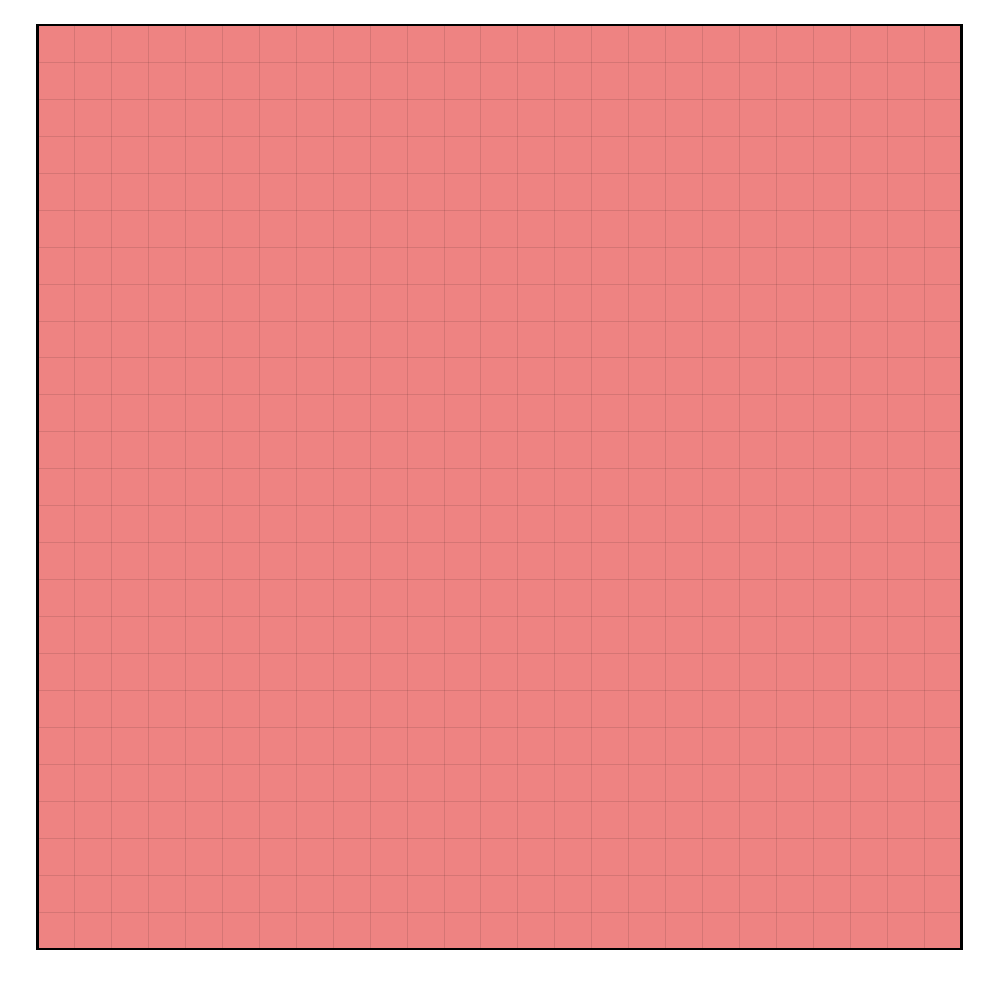}
    \includegraphics[width=0.2\textwidth]{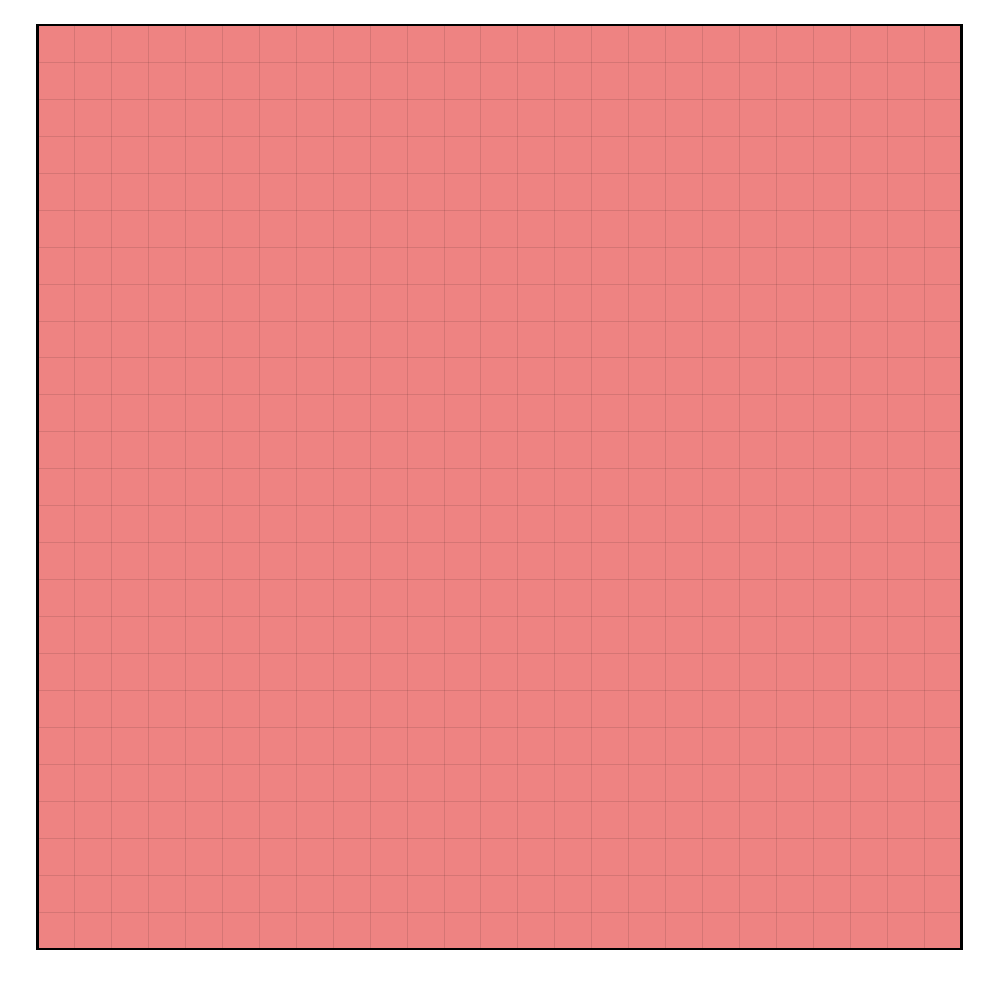}
    \includegraphics[width=0.2\textwidth]{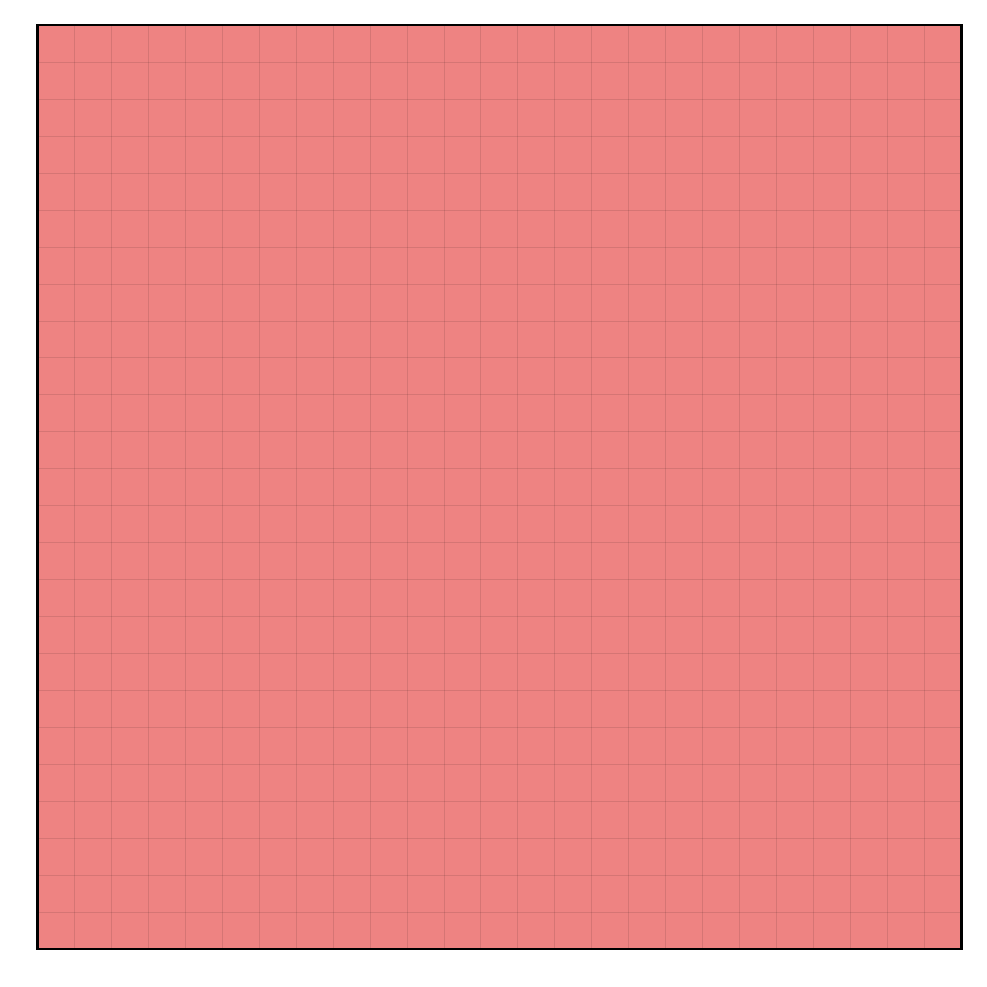}
    \includegraphics[width=0.2\textwidth]{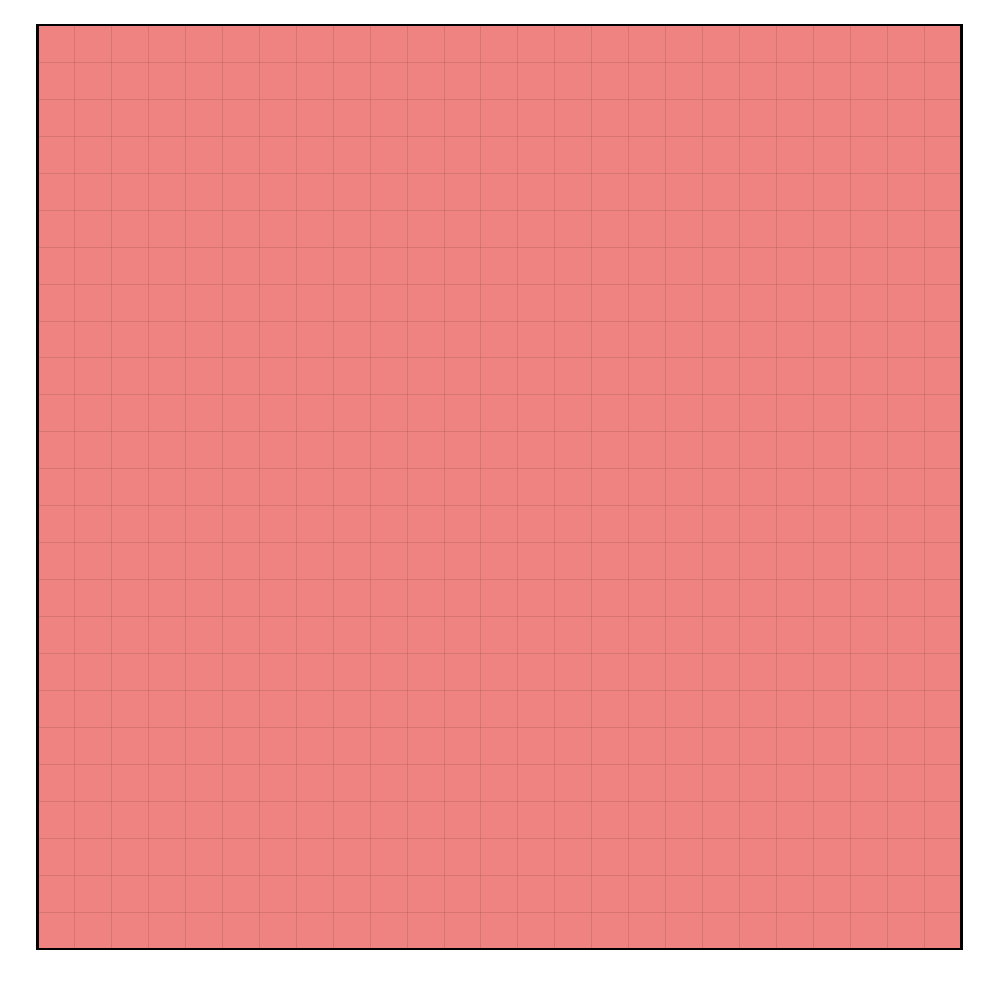}\\ 
      \includegraphics[width=0.75\textwidth]{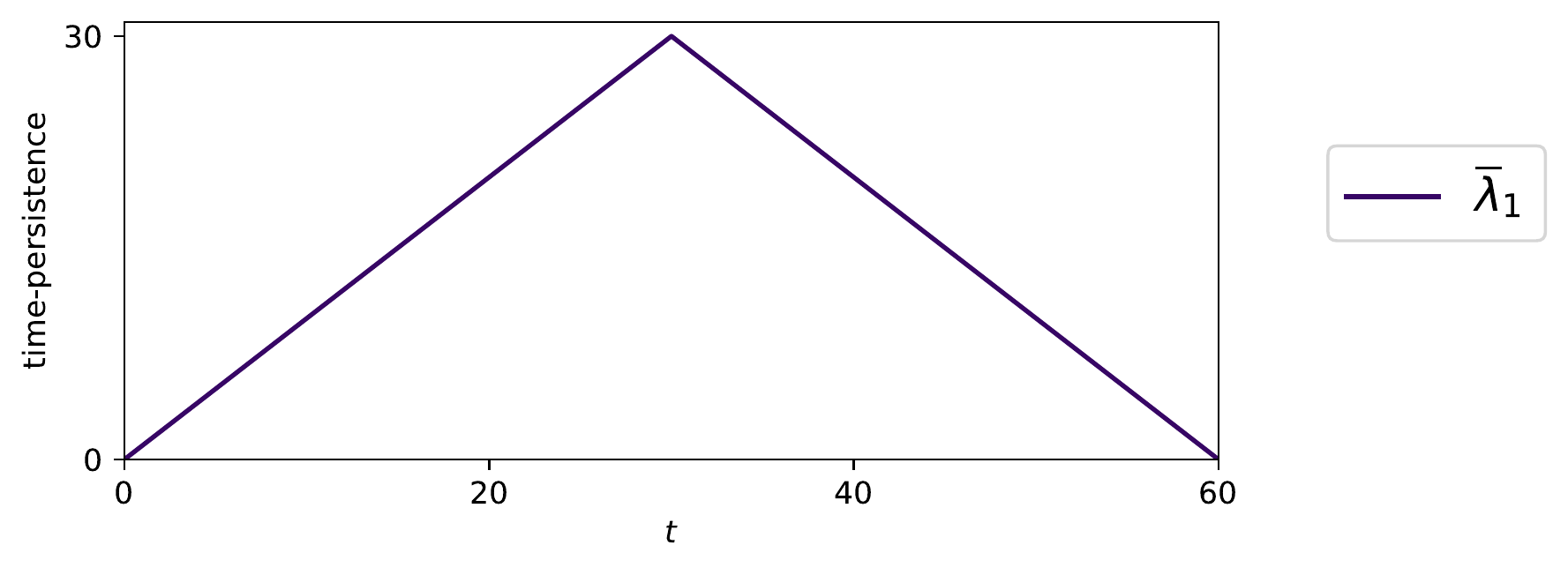}
\end{center}
\caption{\textbf{A. } Top: a simulation of the \ssMHE{} model at a coral-dominated stationary state. Bottom: the average persistence landscape of 10 such simulations without pre-processing. Small turf components are detected by $\bar \lambda_2$ and $\bar \lambda_3$, which represent only noise as the system is already at a stationary state. \textbf{B. } Top: The same simulation of the \ssMHE{} model at a coral-dominated stationary state, but now the pre-processing step has been performed. Snapshots of the \ssMHE{} model at the stationary state are viewed as 100\% coral. There is now a single landscape of maximum time-persistence. The average landscape $\bar \lambda_1$ indicates that a single component of coral lasts throughout each simulation. The pre-processing step, therefore, means that our zigzag persistence does not detect any spatial changes in a simulation of the \ssMHE{} model once it has reached the stationary state.}
    \label{SIfig:preprocessZZ}
\end{figure}

At each snapshot of the \ssMHE{} model, we remove nodes of turf by re-assigning them to either coral or macroalgae. If the neighbourhood of a turf node contains more coral than macroalgae, the turf node is re-assigned to coral. If not, the turf node becomes macroalgae. In the notation of Section \ref{SIsec:sMHE}, given $T_i=1$, if $\lC > \lM$, set $T_i=0$ and $C_i=1$. Otherwise, set $T_i=0$ and $M_i=1$. Fig.~\ref{SIfig:preprocess} shows two examples of such pre-processing of our simulated data.

\clearpage
  
\section{Supplementary figures}
\label{SIsec:supplementaryResults}
\subsection{Supplementary result to Fig. 3 of the main text}
\label{SIsubsec:supplementaryFig3}
\begin{figure}[h!]
\hspace{10mm}
\textbf{A.} \hspace{9cm} \textbf{B.} \\

\begin{center}
    \includegraphics[width=0.47\textwidth]{figs/fig3.pdf}   
    \hfill
    \includegraphics[width=0.41\textwidth]{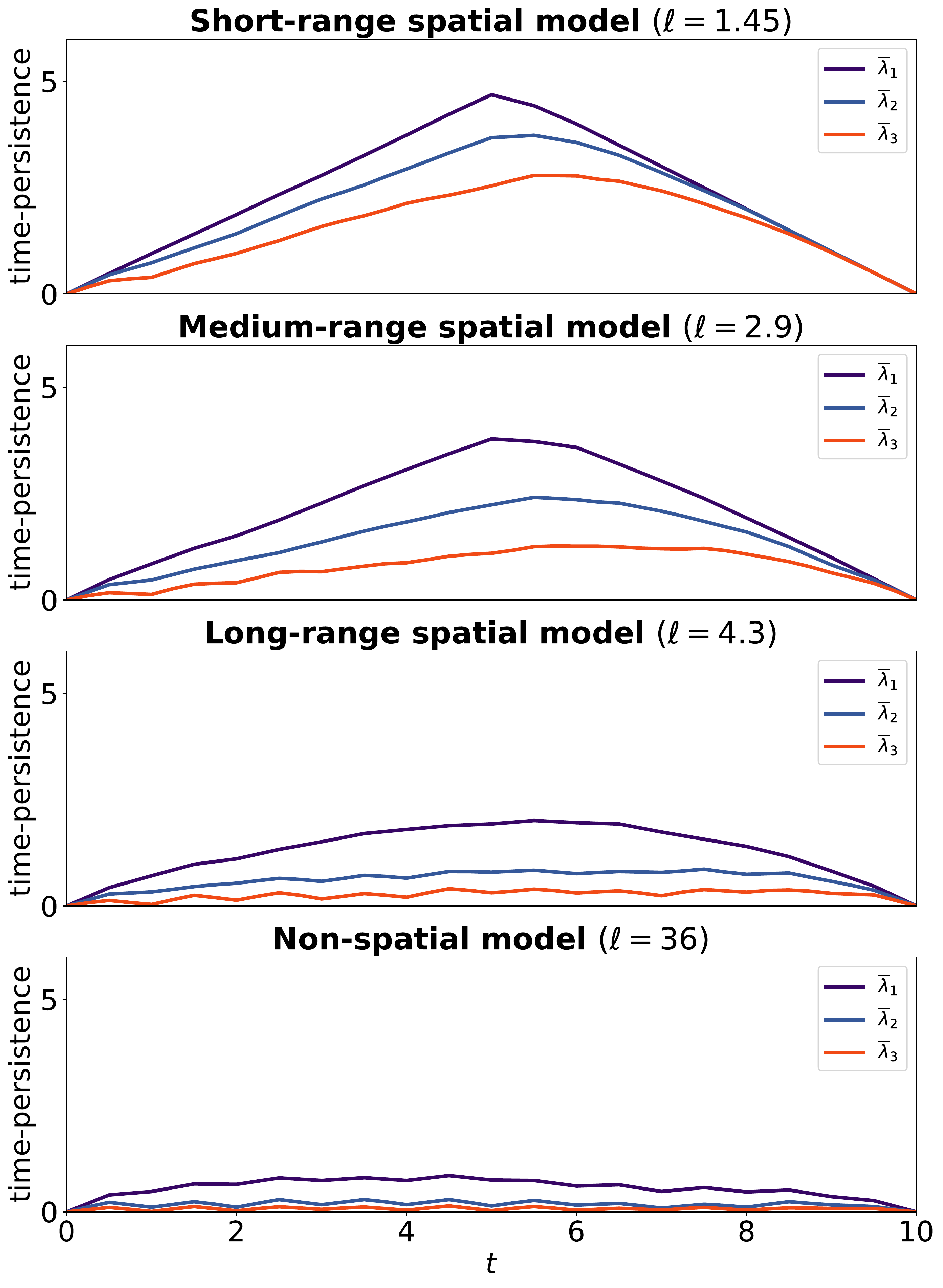}
\end{center}
\begin{center}
    \hspace{-5cm}\textbf{C.} \\
 \includegraphics[width=0.41\textwidth]{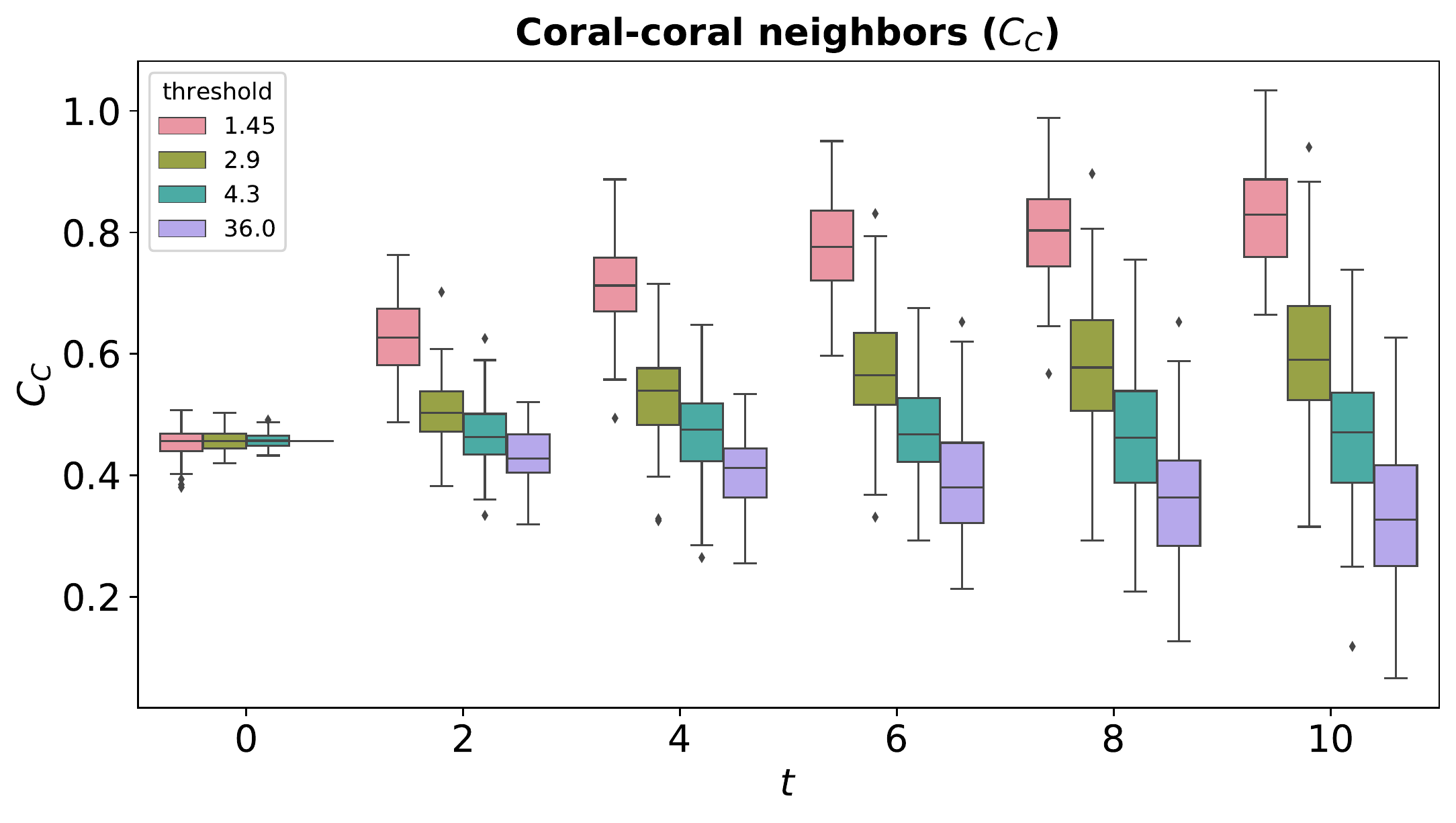}
  \end{center}
 

    
    \caption{\textbf{A.} Fig.~\ref{fig:fig3} from the main text, which shows three simulations of the \ssMHE{} model. We initialise the model with equal numbers of coral, turf, and macroalgae nodes in the random configuration and run the model for $10$ timesteps at four different local neighbourhood radii $\ell$. \textbf{B.} Average persistence landscapes for 100 simulations of the \ssMHE{} model using the same parameters as in Fig.~\ref{fig:fig3} of the main text. The size of the average landscape $\bar \lambda_1$ decreases as $\ell$ increases, indicating that the most time-persistent component persists for a shorter time for larger values of $\ell$. Note that the cubical complex $K_8^t$ was used to compute zigzag persistence here (as opposed to $K_1^t$ which is used in the rest of this work, see Subsection \ref{SIsubsec:preprocess}) Since we are investigating larger clusters of coral, we remove the noise of smaller components by taking $\eta_*=8$ in Subsection \ref{SIsubsec:zzD}. 
    \textbf{C.} Average plots of $C_C$ for $\ell=1.45, 2.9, 4.3, 36$ for 100 simulations using the same parameters used in Fig.~\ref{fig:fig3} of the main text. The average value of $C_C$ (at time $t=10$) decreases as $\ell$ increases. Together, \textbf{B} and \textbf{C} confirm that the results shown in \textbf{A} are representative behaviour of the \ssMHE{} model at these parameters.}
    \label{SIfig:fig3supplement}
\end{figure}

As shown in Fig.~\ref{fig:fig3} of the main text, we found that increasing the local neighbourhood radius $\ell$ (which corresponds to reducing the impact of spatial configuration on the behaviour of the \ssMHE{} model) led to decreased clustering of coral. For smaller values of $\ell$, we found larger values of $C_C$ and longer bars in the PH barcode, quantifying this clustering. We reprint Fig.~\ref{fig:fig3} of the main text in Fig.~\ref{SIfig:fig3supplement} and provide two supplementary plots. We compute averages of $C_C$ over 100 simulations of the \ssMHE{} model, showing that $C_C$ is, on average, higher for lower values of $\ell$. Then we compute average zigzag persistence landscapes for these 100 simulations. The average landscape $\bar \lambda_1$ is large for $\ell=1.45$, but then decreases in size as $\ell$ is increased. This average landscape confirms that the clusters we found for the lower values of $\ell$ persist throughout simulations and that no clusters persist over time for $\ell=36$. In this figure, we used the complex $K_8^t$ to represent a snapshot of the \ssMHE{} model at time $t$ (while we usually use $K_1^t$). See Subsection \ref{SIsubsec:preprocess} for a discussion of this choice. Including only coral nodes with eight direct coral neighbours ensures that we only track large components of coral in Fig.~\ref{SIfig:fig3supplement}--we want to track the persistence of large clusters in this case.

\clearpage

\subsection{Supplementary result to Fig. 5 of the main text}
\label{SIsubsec:supplementaryFig5}
Fig.~\ref{fig:fig5} of the main text demonstrates how coral dynamics change as the grazing parameter $g$ is increased. We initialised snapshots of the \ssMHE{} model with equal covers of coral, turf, and macroalgae with the random initial configuration and ran simulations for $100$ timesteps at different grazing rates.

\begin{figure}[h!]
\hspace{10mm}
\textbf{A.} \hspace{9cm} \textbf{B.} \\
\begin{center}
    \includegraphics[width=0.49\textwidth]{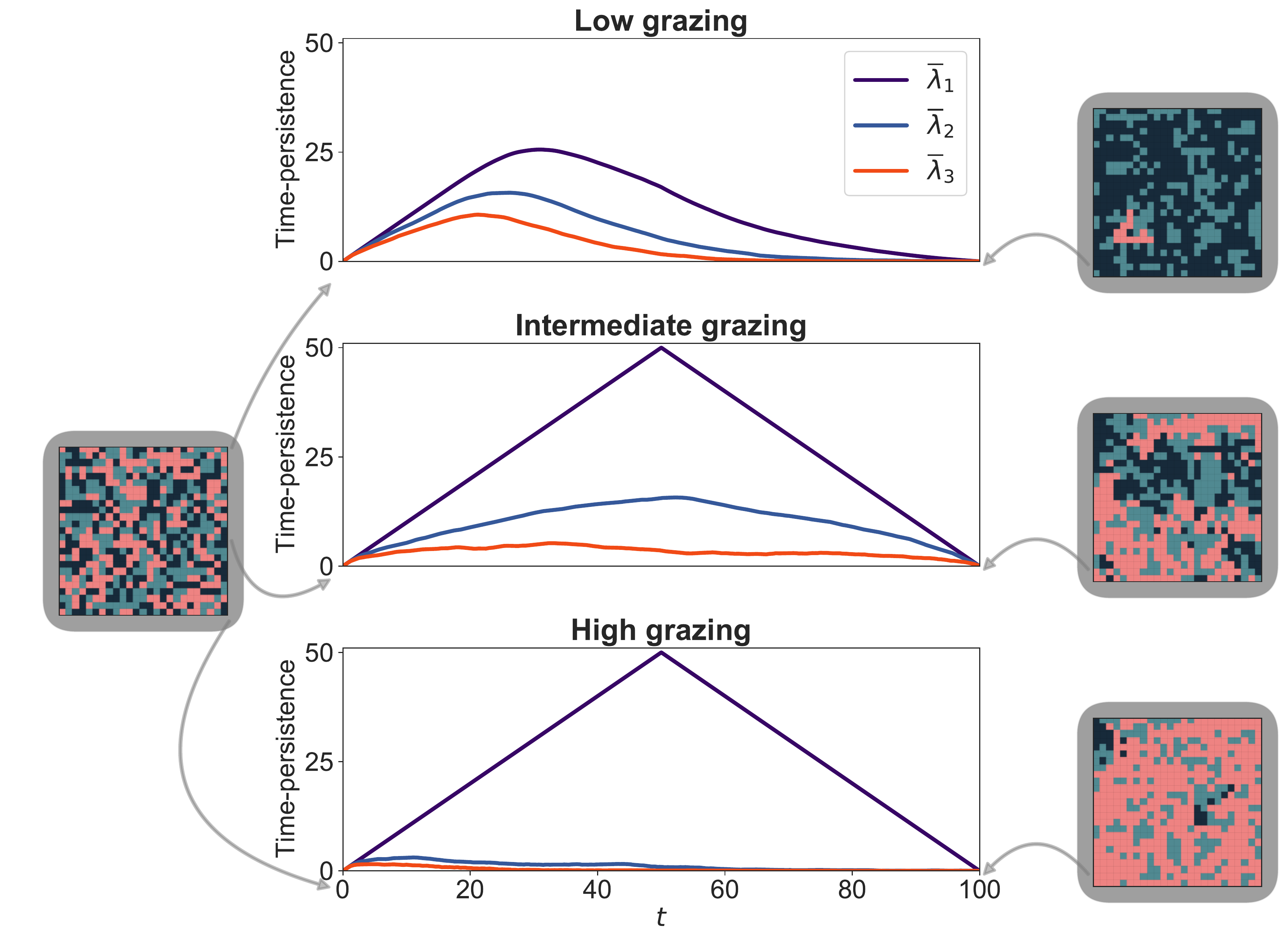}
    \hfill
    \includegraphics[width=0.49\textwidth]{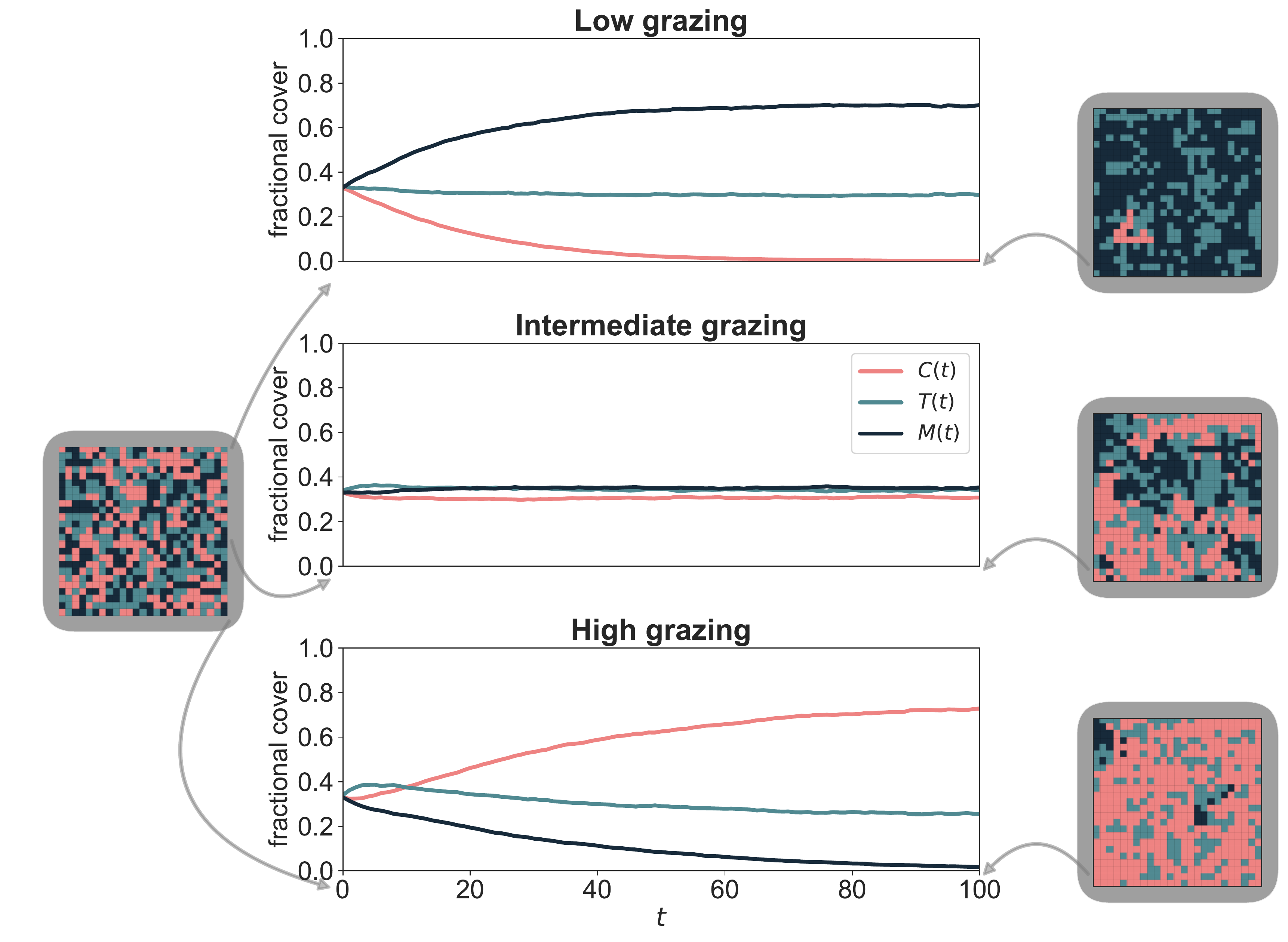}
    \end{center}
   \caption{\textbf{A.} Fig.~\ref{fig:fig5} of the main text, top, showing a spatial summary (zigzag persistence landscapes) of the \ssMHE{} model at three characteristic grazing rates $g=0.42, g=0.53, g=0.62$, averaged over 100 simulations. \textbf{B.} Plots of the fractional cover of coral, turf, and macroalgae at the same three grazing rates as in \textbf{A}, averaged at each time point over the 100 simulations. The average fractional cover of coral decreases for low $g$ and increases for high $g$. At the intermediate grazing rate, all average fractional covers remain constant since in half of the simulations, $C(t)$ is decreasing, and in the other half, $C(t)$ is increasing.}
\label{SIfig:grazingspatial}
\end{figure}

\begin{figure}[h!]
\textbf{A.} \hfill
\begin{center}
     \includegraphics[width=\textwidth]{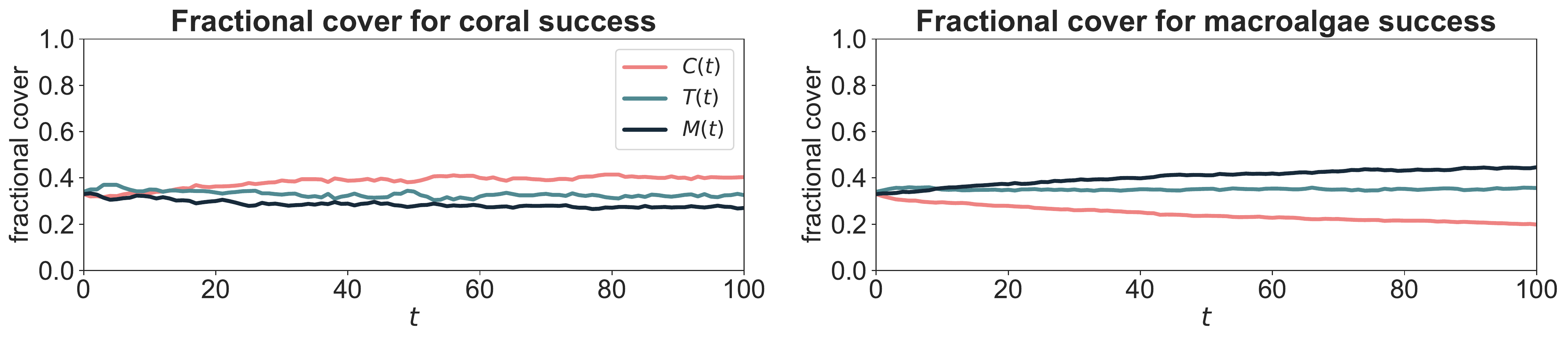}
\end{center}
\textbf{B.} \hfill
\begin{center}
    \includegraphics[width=\textwidth]{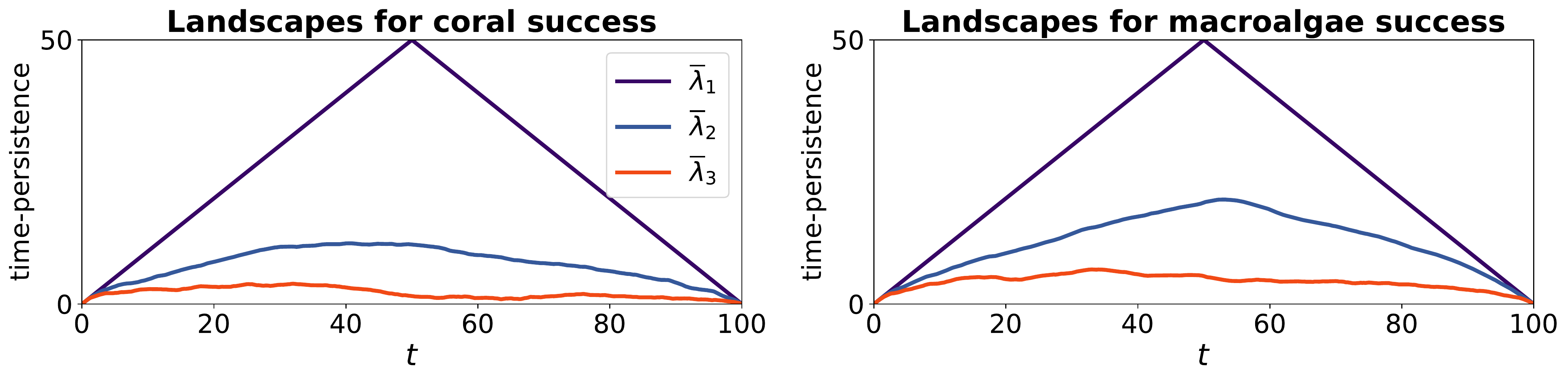}
\end{center}

   \caption{Analysis of simulations of the \ssMHE{} model at the intermediate grazing level ($g=0.53$), plotted separately for simulations where coral goes on to dominate and simulations where coral dies out.  \textbf{A.} The fractional covers of simulations of the \ssMHE{} model plotted for the first $100$ timesteps. We partition simulations according to which species dominate at $t=1000$. In the first $100$ timesteps, coral increases slowly on average in simulations where it will go on to dominate and decreases slowly otherwise. \textbf{B.} The first three zigzag persistence landscapes, averaged separately for those simulations that converge to one stationary state or the other. In coral-dominating simulations, $\bar \lambda_2$ and $\bar \lambda_3$ are smaller than in macroalgae-dominating simulations, indicating that coral survives in one time-persistent component or dies out via many short-lived components.}
    \label{SIfig:splitmeta}
\end{figure}

\begin{figure}[h!]
\hspace{40mm} \textbf{A.} \hfill
\begin{center}
 \includegraphics[width=0.7\textwidth]{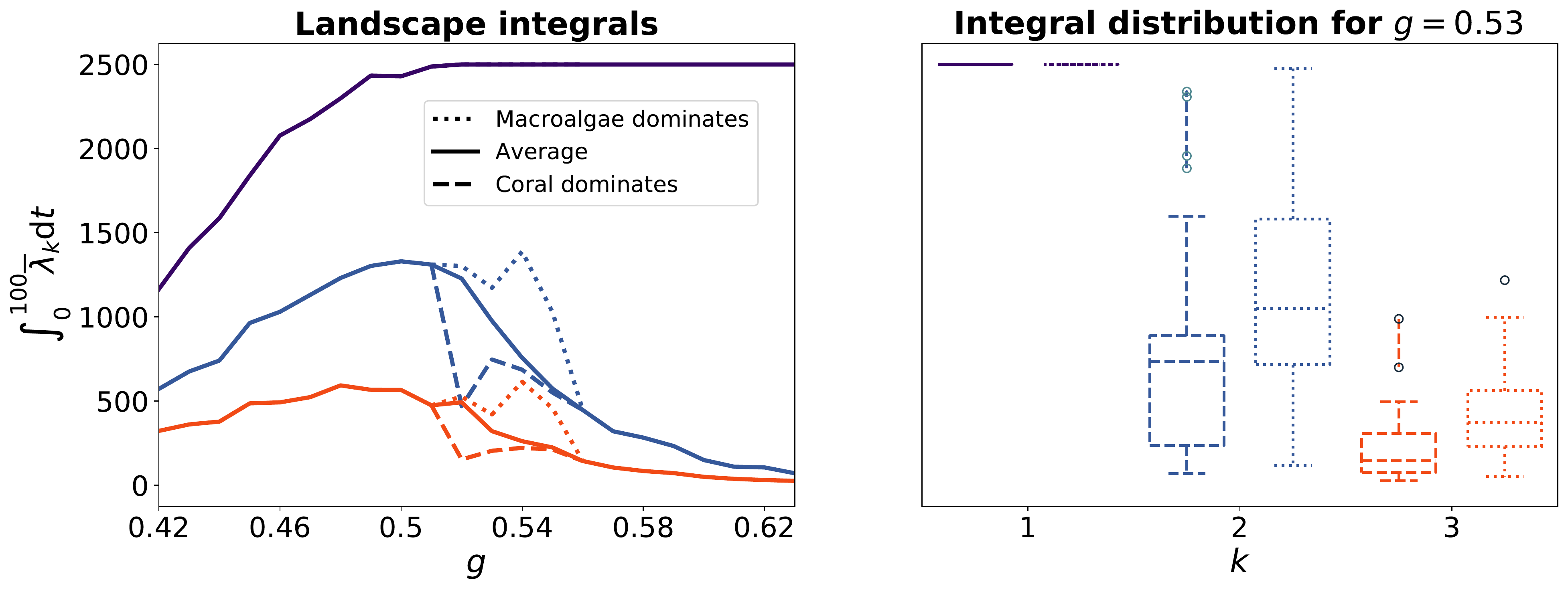}
\end{center}
\hspace{40mm} \textbf{B.} \hfill
\begin{center}
\includegraphics[width=0.7\textwidth]{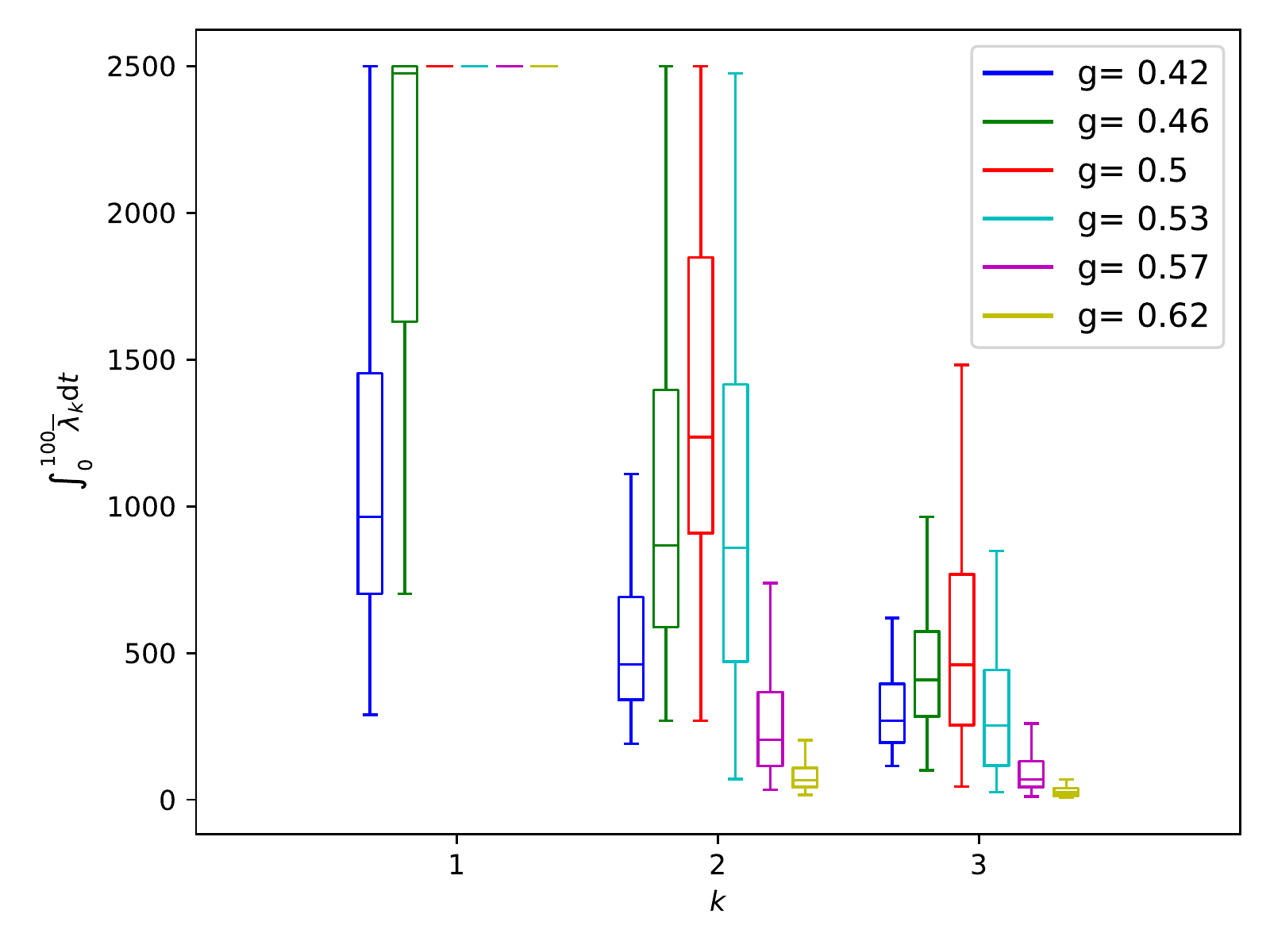}

\end{center}
\caption{\textbf{A.} Fig.~\ref{fig:fig5} of the main text, bottom, showing integrals of landscapes $\bar \lambda_1$, $\bar \lambda_2$, and $\bar \lambda_3$ for 100 simulations of the \ssMHE{} model at different grazing rates. At the intermediate grazing rate $g=0.53$, the integrals are plotted separately for simulations where coral dominates eventually from those where coral dies out. Error bars are shown to show significance. \textbf{B.} Integrals of the first three landscapes for more grazing rates, including error bars. The average integrals of $\lambda_2$ and $\lambda_3$ for the low ($g=0.42$), intermediate ($g=0.53$) and high ($g=0.62$) grazing rates show a significant difference.}
\label{SIfig:grazingboxplots}
\end{figure}

For the same three grazing rates given in Fig.~\ref{fig:fig5} of the main text, we give the non-spatial summaries $C(t), T(t), M(t)$ in Fig.~\ref{SIfig:grazingspatial}---also averaged over 100 simulations. The average coral fractional cover decreases to zero at the low grazing rate ($g=0.42$), whereas macroalgae cover reaches zero at high grazing ($g=0.62$). At the intermediate grazing rate of $g=0.53$, approximately half of the simulations evolve to a coral-dominated stationary state, with the other half evolving to a macroalgae-dominated stationary state. When averaged, the fractional covers $C(t)$ and $M(T)$ remain constant. However, when we average these separately for simulations where coral dies out and for those where coral ends up dominating, we see a small separation in the fractional covers (Fig.~\ref{SIfig:splitmeta}, bottom). Similarly, when we plot the average landscapes $\bar \lambda_k, k=1, 2, 3$ separately according to the stationary state reached after $1000$ timesteps, we see a difference in the sizes of $\bar \lambda_2$ and $\bar \lambda_3$.
Fig.~\ref{fig:fig5}B of the main text, right, gives the distribution of the integrals of the first three landscapes at $g=0.53$. Fig.~\ref{SIfig:grazingboxplots} provides this information for more grazing rates and includes error bars. We see a significant difference between the average integrals of $\lambda_2$ and $\lambda_3$ at the low, intermediate, and high grazing rates.

\newpage

\newpage

\clearpage
\nomenclature[M,01]{\(\ell\)}{local neighborhood radius used to define interactions between nodes}
\nomenclature[M,02]{\(J\)}{the total number of nodes in the lattice (in this work we use $J=25\times25$)}
\nomenclature[M,03]{\(\lN\)}{set of nodes that are neighbors of node $i$}

\nomenclature[M,04]{\(C \ (T, M)\)}{global fractional cover of coral (turf, macroalgae) within the reef}
\nomenclature[M,05]{\(C_i \ (T_i,M_i)\)}{presence of coral (turf, macroalgae) at node $i$ (can either take value 0 or 1)}
\nomenclature[M,06]{\(C_i^\ell \ (\lT,\lM)\)}{number of coral (turf, macroalgae) nodes in the neighborhood of radius $\ell$ of node $i$}
\nomenclature[M,07]{\(C_C \ (C_T, C_M) \)}{the average number of coral neighbors of coral (macroalgae, turf) nodes in the reef}
\nomenclature[M,08]{\(M_C \ (M_T, M_M)\)}{the average number of macroalgae neighbors of coral (turf, macroalgae) nodes in the reef}
\nomenclature[M,09]{\(T_C \ (T_T, T_M)\)}{the average number of turf neighbors of coral (turf, macroalgae) nodes in the reef}

\nomenclature[M]{\(r\)}{rate at which coral overgrows turf}
\nomenclature[M]{\(d\)}{base rate at which coral dies and is replaced by turf}
\nomenclature[M]{\(a\)}{rate at which macroalgae overgrows coral}
\nomenclature[M]{\(g\)}{fish grazing rate}
\nomenclature[M]{\(\gamma\)}{rate at which macroalgae overgrows turf}
\nomenclature[M]{\(t\)}{time}

\nomenclature[T,00]{\(E\)}{elementary interval}
\nomenclature[T,01]{\(e, v, w\)}{endpoints of elementary intervals}
\nomenclature[T,01]{\(q, q', \tilde{q}\)}{elementary cubes}
\nomenclature[T,02]{\(\iota\)}{embedding number of an elementary cube}
\nomenclature[T,03]{\(\mathcal{Q}\)}{general cubical complex}
\nomenclature[T,03]{\(\epsilon\)}{edge within a cubical complex}
\nomenclature[T,03]{\(\zeta\)}{square within a cubical complex}
\nomenclature[T,04]{\(\mathcal{I}\)}{set of nodes in a reef}
\nomenclature[T,05]{\(N\)}{set $\{0, 1, \dots, 7, 8\}$}
\nomenclature[T,06]{\(\eta\)}{number of direct coral neighbors of a node, $\eta \in N$}
\nomenclature[T,07]{\(f\)}{function that returns the number of direct coral nodes $\eta$ of a node in a reef}
\nomenclature[T,08]{\(K_\eta^t\)}{a cubical complex containing vertices for each coral node that has at least $\eta$ direct coral neighbors}
\nomenclature[T,09]{\(\{K_8^t, K_7^t, \dots, K_1^t\}\)}{the \ssMHE{} snapshot filtration of a snapshot of the \ssMHE{} model at time $t$}
\nomenclature[T,10]{\(D_0, D_1, D_2\)}{free Abelian groups generated by elementary cubes of dimension 0, 1 and 2}
\nomenclature[T,11]{\(\partial_0, \partial_1, \partial_2\)}{boundary homomorphisms between free Abelian groups $D_0$, $D_1$, $D_2$}

\nomenclature[T,12]{\(H_0(\mathcal{Q}), H_1(\mathcal{Q})\)}{zeroth and first homology groups of a cubical complex $\mathcal{Q}$}
\nomenclature[T,13]{\(\lambda_k\)}{$k$th persistence landscape constructed from the zigzag persistence of a simulation of the \ssMHE{} model}
\nomenclature[T,14]{\(\bar \lambda_k\)}{average of $\lambda_k$ over multiple simulations of the \ssMHE{} model}
\nomenclature[T]{\(\mathbbm{b} \ (\mathbbm{d})\)}{birth (death) time of a component of coral in a simulation of the \ssMHE{} model}
\label{sec:symbols}
\printnomenclature

\end{document}